\documentclass[amsmath,amssymb,aps,prd,11pt,tightenlines,superscriptaddress,
nofootinbib,preprintnumbers,notitlepage]{revtex4-1}

\usepackage[T1]{fontenc}
\usepackage[utf8]{inputenc}
\usepackage[british]{babel}

\usepackage{color,xcolor}

\definecolor{dark-red}{rgb}{0.8, 0.0, 0.1803921568627451}
\definecolor{dark-blue}{rgb}{0.0, 0.0, 0.803921568627451}
\definecolor{dark-green}{rgb}{0.0, 0.39215686274509803, 0.0}
\definecolor{dark-orange}{rgb}{0.8, 0.4, 0.0}

\usepackage{amsmath,dsfont,mathtools}
\usepackage{siunitx}
\usepackage{graphicx}
\usepackage{multirow,booktabs,tabularx,float}
\usepackage[font=normalsize]{subfig}
\usepackage{slashed,braket}
\usepackage{feynmp-auto}
\usepackage[sort&compress]{natbib}
\PassOptionsToPackage{hyphens}{url}\usepackage[pdftex,hidelinks,linkcolor={dark-blue}, citecolor={dark-green}, urlcolor={dark-red},bookmarksopen,linktoc=all]{hyperref}
\usepackage{enumitem} 
\usepackage[stretch=15,shrink=15]{microtype} 
\usepackage{tabto}
\usepackage{accents} 
\usepackage{tikz}
\usepackage{xspace} 
\usetikzlibrary{arrows, decorations.markings, arrows.meta}

\makeatletter
\def\l@subsubsection#1#2{}
\makeatother

\newcommand{\toolfont}[1]{\textsc{#1}}

\newcommand{\im}{\mathrm{i}}
\newcommand{\diff}{\mathrm{d}}
\newcommand{\eg}{{e.\,g.}~}
\newcommand{\ie}{{i.\,e.}~}

\newcommand{\yep}{\ensuremath{\color{dark-green}{  \checkmark}}}
\newcommand{\kindof}{\ensuremath{\color{dark-blue}{  \ast}}}

\newcommand{\qqquad}{\qquad \qquad}
\newcommand{\qqqquad}{\qqquad \qqquad}

\newcommand{\ope}[1]{\mathcal{O}_{#1}}
\newcommand{\lgr}[1]{\mathcal{L}_\text{#1}}
\newcommand{\ord}[1]{\mathcal{O}\left(#1\right)}

\newcommand{\phisq}{\phi^\dagger \phi}

\newcommand{\gev}{\si{GeV}}

\newcommand{\normal}[2]{\mathcal{N}\!\!\left(#1 \middle| #2 \right)}

\newcommand{\tder}[2]{\frac {\diff #1} {\diff #2}}

\newcommand{\fder}[2]{\frac {\delta #1} {\delta #2}}

\newcommand{\boldtheta}{{\ensuremath \boldsymbol{\theta}}}

\newcommand{\abc}{\textsc{Abc}\xspace}
\newcommand{\afc}{\textsc{Afc}\xspace}
\newcommand{\nde}{\textsc{Nde}\xspace}
\newcommand{\carl}{\textsc{Carl}\xspace}
\newcommand{\rolr}{\textsc{Rolr}\xspace}
\newcommand{\sally}{\textsc{Sally}\xspace}
\newcommand{\sallino}{\textsc{Sallino}\xspace}
\newcommand{\cascal}{\textsc{Cascal}\xspace}
\newcommand{\rascal}{\textsc{Rascal}\xspace}
\newcommand{\scandal}{\textsc{Scandal}\xspace}

\newcommand{\intractablep}{\ensuremath {\textcolor{dark-red}{p}}}
\newcommand{\intractabler}{\ensuremath {\textcolor{dark-red}{r}}}
\newcommand{\intractables}{\ensuremath {\textcolor{dark-red}{s}}}
\newcommand{\intractablet}{\ensuremath {\textcolor{dark-red}{t}}}
\newcommand{\intractablez}{\ensuremath {\textcolor{dark-red}{Z}}}
\newcommand{\localmodel}{\ensuremath {\textcolor{dark-red}{p_{\text{local}}}}}

\newcommand{\intx}{\int \! \diff x\;}

\newcommand{\intz}{\int \! \diff z\;}

\newcommand{\intxz}{\int \! \diff x \, \diff z\;}

\newlength{\hhatheight}
\newcommand{\hhat}[1]{%
    \settoheight{\hhatheight}{\ensuremath{\hat{#1}}}%
    \addtolength{\hhatheight}{-0.35ex}%
    \hat{\vphantom{\rule{1pt}{\hhatheight}}%
    \smash{\hat{#1}}}}

\DeclareMathOperator{\Real}{Re}

\DeclareMathOperator{\var}{var}
\DeclareMathOperator{\expectation}{\mathbb{E}}

\DeclareMathOperator*{\argmax}{arg\,max}

\newcolumntype{R}{>{\raggedleft\arraybackslash}X}%
\newcolumntype{L}{>{\raggedright\arraybackslash}X}%

\newcommand{\feynmansetup}{%
      \fmfpen{0.8pt}%
      \fmfset{arrow_len}{2mm}%
}

\AtBeginDocument{
  \heavyrulewidth=.08em
  \lightrulewidth=.05em
  \cmidrulewidth=.03em
  \belowrulesep=.65ex
  \belowbottomsep=0pt
  \aboverulesep=.4ex
  \abovetopsep=0pt
  \cmidrulesep=\doublerulesep
  \cmidrulekern=.5em
  \defaultaddspace= .5em
  \setlength{\tabcolsep}{0.5em}
}

\setlist[itemize]{itemsep=1pt,parsep=1pt, topsep=1pt}

\def\nnnodesize{0.75cm}
\def\nnlayersep{2cm}
\def\nnlabelsep{0.75cm}
\def\nnbordersep{0.5cm}
\def\nnboxshift{0.2cm}

\begin{document}

\count\footins = 1000 

\begin{fmffile}{diagrams}


\title{A Guide to Constraining Effective Field Theories with Machine Learning}

\author{Johann Brehmer}
\affiliation{New York University, USA}

\author{Kyle Cranmer}
\affiliation{New York University, USA}

\author{Gilles Louppe}
\affiliation{University of Li\`{e}ge, Belgium}

\author{Juan Pavez}
\affiliation{Federico Santa Mar\'ia Technical University,  Chile}

\date{\today}

\begin{abstract}
  We develop, discuss, and compare several inference techniques to constrain theory parameters in collider experiments. By harnessing the latent-space structure of particle physics processes, we extract extra information from the simulator. This augmented data can be used to train neural networks that precisely estimate the likelihood ratio. The new methods scale well to many observables and high-dimensional parameter spaces, do not require any approximations of the parton shower and detector response, and can be evaluated in microseconds. Using weak-boson-fusion Higgs production as an example process, we compare the performance of several techniques. The best results are found for likelihood ratio estimators trained with extra information about the score, the gradient of the log likelihood function with respect to the theory parameters. The score also provides sufficient statistics that contain all the information needed for inference in the neighborhood of the Standard Model. These methods enable us to put significantly stronger bounds on effective dimension-six operators than the traditional approach based on histograms. They also outperform generic machine learning methods that do not make use of the particle physics structure, demonstrating their potential to substantially improve the new physics reach of the LHC legacy results.
\end{abstract}

\maketitle

\tableofcontents

\section{Introduction}
\label{sec:intro}

An important aspect of the legacy of the Large Hadron Collider (LHC) experiments will be precise constraints on indirect signatures of physics beyond the Standard Model (SM), parameterized for instance by the dimension-six operators of the Standard Model effective field theory (SMEFT). The relevant measurements can easily involve tens of different parameters that predict subtle kinematic signatures in the high-dimensional space of the data. Traditional analysis techniques do not scale well to this complex problem, motivating the development of more powerful techniques.

The analysis of high-energy-physics data is based on an impressive suite of simulation tools that model the hard interaction, parton shower, hadronization, and detector response. The community has invested a tremendous amount of effort into developing these tools, yielding the high-fidelity modeling of LHC data needed for precision measurements. Simulators such as \toolfont{Pythia}~\cite{Sjostrand:2007gs} and \toolfont{Geant4}~\cite{Agostinelli:2002hh} use Monte-Carlo techniques to sample the multitudinous paths through which a particular hard scattering might develop. A single event's path through the simulation can easily involve many millions of random variables. While Monte-Carlo techniques can efficiently sample from the distributions implicitly defined by the simulators, it is not feasible to calculate the likelihood for a particular observation because doing so would require integrating over all the possible histories leading to that observation. Clearly it is infeasible to explicitly calculate a numerical integral over this enormous latent space. While this problem is ubiquitous in high energy physics, it is rarely acknowledged explicitly. 

Traditionally, particle physicists have approached this problem by restricting the analysis to one or two well-motivated discriminating variables, discarding the information contained in the remaining observables. The probability density for the restricted set of discriminating variables is then estimated with explicit functions or non-parametric approaches such as template histograms, kernel density estimates, or Gaussian Processes~\cite{Cranmer:2000du,*Cranmer:2012sba,*Frate:2017mai}. These low-dimensional density estimates are constructed and validated using Monte-Carlo samples from the simulation. While well-chosen variables may yield precise bounds along individual directions of the parameter space, they often lead to weak constraints in other directions in the parameter space~\cite{Brehmer:2016nyr}.  The sensitivity to multiple parameters can be substantially improved by using the fully differential cross section. This is the forte of the Matrix Element Method~\cite{Kondo:1988yd, Abazov:2004cs, Artoisenet:2008zz, Gao:2010qx, Alwall:2010cq, Bolognesi:2012mm, Avery:2012um, Andersen:2012kn, Campbell:2013hz, Artoisenet:2013vfa, Gainer:2013iya, Schouten:2014yza, Martini:2015fsa, Gritsan:2016hjl, Martini:2017ydu} and Optimal Observables~\cite{Atwood:1991ka, Davier:1992nw, Diehl:1993br} techniques, which are based on the parton-level structure of a given process. Shower and event deconstruction~\cite{Soper:2011cr, Soper:2012pb, Soper:2014rya, Englert:2015dlp} extend this approach to the parton shower. But all these methods still require some level of approximations on the parton shower and either neglect or crudely approximate the detector response. Moreover, even a simplified description of the detector effects requires the numerically expensive evaluation of complicated integrals for each observed event. None of these established approaches scales well to high-dimensional problems with many parameters and observables, such as the SMEFT measurements.

In recent years there has been increased appreciation that several real-world phenomena are better described by simulators that do not admit a tractable likelihood. This appears in fields as diverse as ecology, phylogenetics, epidemiology, cardiac simulators, quantum chemistry, and particle physics. Inference in this setting is often referred to as \emph{likelihood-free inference}, where the inference strategy is restricted to samples generated from the simulator. Implicitly, these techniques aim to estimate the likelihood. A particularly ubiquitous technique is Approximate Bayesian Computation (\abc)~\cite{rubin1984, beaumont2002approximate, marjoram2003markov, sisson2007sequential, sisson2011likelihood, marin2012approximate, Charnock:2018ogm}.  \abc is closely related to the traditional template histogram and kernel density estimation approach used by physicists. More recently, approximate inference techniques based on machine learning and neural networks have been proposed~\cite{Cranmer:2015bka, Cranmer:2016lzt, 2012arXiv1212.1479F, NIPS2016_6084, 2016arXiv160206701P, 2016arXiv161110242D, gutmann2017likelihood, 2017arXiv170208896T, 2017arXiv170707113L, 2014arXiv1410.8516D, 2015arXiv150505770J, 2016arXiv160508803D, 2017arXiv170507057P, 2016arXiv160502226U, 2016arXiv160903499V, 2016arXiv160605328V, 2016arXiv160106759V, 2018arXiv180507226P}. All these techniques have in common that they only take into account simulated samples similar to the actual observables\,---\,they do not exploit the structure of the process that generates them.

We develop new \emph{simulation-based inference} techniques that are tailored to the structure of particle physics processes. The key insight behind these methods is that we can extract more information than just samples from the simulations, and that this additional information can be used to efficiently train neural networks that precisely estimate likelihood ratios, the preferred test statistics for LHC measurements. These methods are designed for scalability to both high-dimensional parameter spaces as well as to many observables. They do not require any simplifying assumptions to the underlying physics: they support state-of-the-art event generators with parton shower, reducible and irreducible backgrounds, and full detector simulations. After an upfront training phase, they are very efficient to evaluate. Our tools directly provide an estimator for the likelihood ratio, an intuitive and easily interpretable quantity. Finally, limits derived from these tools with toy experiments have the reassuring property that even if they might not be optimal, they are never wrong, \ie no points are said to be excluded that should not be excluded at a given confidence level.

In Ref.~\cite{companion_short}, the companion paper of this publication, we focus on the key ideas and sensitivity enabled by these techniques. Reference~\cite{companion_nips} presents the methods in a more abstract setting. Here we describe the actual algorithms in detail, developing several different methods side by side. Given the number of discussed variations, this publication might have the look and feel of a review article and we present it as a guide to the interested practitioner. We focus on the main ideas and differences between the approaches and postpone many technical details until the appendices.

We evaluate the performance of these different methods on a specific example problem, the measurement of two dimension-six operators in Higgs production in weak boson fusion (WBF) in the four-lepton mode at the LHC. For part of this analysis, we work in an idealized setting in which we can access the true likelihood function, providing us with a ground truth for the comparison of the different analysis methods. After establishing the precision of the likelihood ratio estimation, we turn towards the more physical question of how strongly the two operators can be constrained with the different techniques. We repeat the analysis with a simplified detector response where the ground-truth likelihood is no longer tractable. 

We begin by laying out the problem in Sec.~\ref{sec:problem}: we summarize the effective field theory idea, list the challenges posed by EFT measurements, translate the problem from a physics perspective into the language of statistics, and discuss its important structural properties. We also set up the example process used throughout the rest of the paper. The description of the analysis methods are split in two parts: in Sec.~\ref{sec:modeling} we define the different techniques to estimate the likelihood ratio, which includes most of the conceptual work presented here. Section~\ref{sec:inference} then explains how to set limits on the EFT parameters based on these tools. In Sec.~\ref{sec:results} we evaluate the performance of the different tools in our example process. Finally, in Sec.~\ref{sec:conclusions} we summarize our findings and give recommendations for practitioners. The appendices describe the different algorithms in more detail and provide additional results. The code and data used for this paper are available online at Ref.~\cite{repository}.

\section{The EFT measurement problem}
\label{sec:problem}

\subsection{Effective field theory}
\label{sec:eft}

Effective field theories (EFTs)~\cite{Coleman:1969sm, Callan:1969sn, Weinberg:1980wa}  parameterize the effects of physics at an energy scale $\Lambda$ on observables at smaller energies $E \ll \Lambda$ as a set of local operators. The form of these operators is fixed by the light particles and the symmetry structure of the theory and is entirely independent of the high-energy model. Systematically expanding the Lagrangian in $1/\Lambda$, equivalent to ordering the operators by their canonical dimension, leaves us with a finite set of operators weighted by Wilson coefficients that describe all possible new physics effects up to some order in $E/\Lambda$.

In the absence of new particles at the TeV scale, and assuming the symmetry structure of the SM, we can thus describe any new physics signature in LHC processes in terms of a set of higher-dimensional operators~\cite{Burges:1983zg, Leung:1984ni, Buchmuller:1985jz, Arzt:1994gp, Hagiwara:1993ck, Grzadkowski:2010es}. In this SM Effective Field Theory (SMEFT), the leading effects beyond the SM come from 59 independent dimension-six operators $\ope{o}$ with Wilson coefficients $f_o$,
\begin{equation}
  \lgr{D6} = \lgr{SM} + \sum_o \frac {f_o} {\Lambda^2} \, \ope{o} \,,
\end{equation}
where the SM corresponds to all $f_o = 0$ and any measurement of a deviation hints at new physics.

The dimension-six Wilson coefficients are perfectly suited as an interface between experimental measurements and theory interpretations. They are largely model-independent, can parameterize a wide range of observables, including novel kinematic features, and are theoretically consistent beyond tree level. On the technical side, dimension-six operators are implemented in standard Monte-Carlo event generators~\cite{Alwall:2014hca}, allowing us to generate predictions for rates and kinematic observables for any combination of Wilson coefficients. Measured values of $f_o / \Lambda^2$ can easily be translated to the parameters of specific models through well-established matching procedures~\cite{Henning:2014wua}. All in all, SMEFT measurements will likely be a key part of the legacy of the LHC experiments~\cite{deFlorian:2016spz}.

Let us briefly comment on the question of EFT validity. A hierarchy of energy scales $E \ll \Lambda$ is the key assumption behind the EFT construction, but in a bottom-up approach the cutoff scale $\Lambda$ cannot be known without additional model assumptions. From a measurement $f_o / \Lambda^2 \neq 0$ we can estimate the new physics scale $\Lambda$ only by assuming a characteristic size of the new physics couplings $\sqrt{f_o}$, and compare it to the energy scale $E$ of the experiment. It has been found that dimension-six operators often capture the dominant effects of new physics even when there is only a moderate scale separation $E \lesssim \Lambda$~\cite{Brehmer:2015rna}. All these concerns are not primarily of interest for the measurement of Wilson coefficients, but rather important for the interpretation of the results in specific UV theories.

\subsection{Physics challenges and traditional methods}
\label{sec:challenge}

EFT measurements at the LHC face three fundamental challenges:
\begin{enumerate}
  \item Individual scattering processes at the LHC are sensitive to several operators and require simultaneous inference over a multi-dimensional parameter space. While a naive parameter scan works well for one or two dimensions, it becomes prohibitively expensive for more than a few parameters.
\item Most operators introduce new coupling structures and predict non-trivial kinematic features. These do not translate one-to-one to traditional kinematic observables such as transverse momenta, invariant masses or angular correlations. An analysis based on only one kinematic variable typically cannot constrain the full parameter space efficiently. Instead, most of the operator effects only become fully apparent when multiple such variables including their correlations are analysed~\cite{Brehmer:2016nyr, Brehmer:2017lrt}.
\item The likelihood function of the observables is intractable, making this the setting of ``likelihood-free inference'' or ``simulator-based inference''. There are simulators for the high-energy interactions, the parton shower, and detector effects that can generate events samples for any theory parameter values, but they can only be run in the forward mode. Given a set of reconstruction-level observables, it is not possible to evaluate the likelihood of this observation given different theory parameters. The reason is that this likelihood includes the integral over all possible different parton shower histories and particle trajectories through the detector as a normalizing constant, which is infeasible to calculate in realistic situations. We will discuss this property in more detail in the following section.
\end{enumerate}

The last two issues are typically addressed in one of three ways. Most commonly, a small set of discriminating variables (also referred to as summary statistics or engineered features) is handpicked for a given problem. The likelihood in this low-dimensional space is then estimated, for instance, by filling histograms from simulations. While well-chosen variables may lead to good constraints along individual directions of the parameter space, there are typically  directions in the parameter space with limited sensitivity~\cite{Brehmer:2016nyr, Brehmer:2017lrt}.

The Matrix Element Method~\cite{Kondo:1988yd, Abazov:2004cs, Gao:2010qx, Alwall:2010cq, Bolognesi:2012mm, Avery:2012um, Andersen:2012kn, Campbell:2013hz, Artoisenet:2013vfa, Gainer:2013iya, Martini:2015fsa, Gritsan:2016hjl, Martini:2017ydu} or Optimal Observables~\cite{Atwood:1991ka, Davier:1992nw, Diehl:1993br} go beyond a few specific discriminating variables and use the matrix element for a particular process to estimate the likelihood ratio. While these techniques can be very powerful, they suffer from two serious limitations. The parton shower and detector response are either entirely neglected or approximated through ad-hoc transfer function. Shower and event deconstruction~\cite{Soper:2011cr, Soper:2012pb, Soper:2014rya, Englert:2015dlp} allow for the calculation of likelihood ratios at the level of the parton shower, but still rely on transfer functions to describe the detector response. Finally, even with such a simple description of the shower and detector, the evaluation of the likelihood ratio estimator requires the numerically expensive computation of large integrals for each observed event. 

Finally, there is a class of generic methods for likelihood-free inference. For Bayesian inference, the best-known approach is Approximate Bayesian Computation (\abc)~\cite{rubin1984, beaumont2002approximate, marjoram2003markov, sisson2007sequential, sisson2011likelihood, marin2012approximate}. Similar to the histogram approach, it relies on the choice of  appropriate low-dimensional summary statistics, which can severely limit the sensitivity of the analysis. Different techniques based on machine learning have been developed recently. In particle physics, the most common example are discriminative classifiers between two discrete hypotheses, such as a signal and a background process. This approach has recently been extended to parameter measurements~\cite{Cranmer:2015bka, Cranmer:2016lzt}. More generally, many techniques based on the idea of using a classification model, such as neural networks, for inference in the absence of a tractable likelihood function have been introduced in the machine learning community~\cite{2012arXiv1212.1479F, NIPS2016_6084, 2016arXiv160206701P, 2016arXiv161110242D, gutmann2017likelihood, 2017arXiv170208896T, 2017arXiv170707113L, 2014arXiv1410.8516D, 2015arXiv150505770J, 2016arXiv160508803D, 2017arXiv170507057P, 2016arXiv160502226U, 2016arXiv160903499V, 2016arXiv160605328V, 2016arXiv160106759V, 2018arXiv180507226P}. All of these methods only require samples of events trained according to different parameter points. They do not make use of the structure of the particle physics processes, and thus do not use all available information.

All of these methods come with a price. We develop new techniques that
\begin{itemize}
  \item are tailored to particle physics measurements and leverage their structural properties,
  \item scale well to high-dimensional parameter spaces,
  \item can accommodate many observables,
  \item capture the information in the fully differential cross sections, including all correlations between observables,
  \item fully support state-of-the art simulators with parton showers and full detector simulations, and
  \item are very efficient to evaluate after an upfront training phase.
\end{itemize}

\subsection{Structural properties of EFT measurements}
\label{sec:structure}

\subsubsection{Particle-physics structure}
\label{sec:latent_structure}

One essential step to finding the optimal measurement strategy is identifying the structures and symmetries of the problem. Particle physics processes, in particular those described by effective field theories, typically have two key properties that we can exploit.

First, any high-energy particle physics process factorizes into the parton-level process, which contains the matrix element and in it the entire dependence on the EFT coefficients, and a residual part describing the parton shower and detector effects. In many plausible scenarios of new physics neither the strong interactions in the parton shower nor the electromagnetic and strong interactions in the detector are affected by the parameters of interest. The likelihood function can then be written as
\begin{equation}
  \intractablep(x | \theta) = \intz \intractablep (x,z | \theta) = \intz \intractablep (x | z) \, p (z | \theta) \,.
  \label{eq:latent_structure}
\end{equation}
Here and in the following $x$ are the actual observables after the shower, detector, and reconstruction; $\theta$ are the theory parameters of interest; and $z$ are the parton-level momenta (a subset of the latent variables). Table~\ref{tbl:dictionary} provides a dictionary of these and other important symbols that we use.

The first ingredient to this likelihood function is the distribution of parton-level four-momenta
\begin{equation}
  p(z | \theta) = \frac 1 {\sigma(\theta)} \, \tder {\sigma(\theta)} {z} \,,
\end{equation}
where $\sigma(\theta)$ and $\diff\sigma (\theta)/ \diff z$ are the total and differential cross sections, respectively. Crucially, this function is tractable: the matrix element and the parton density functions can be evaluated for arbitrary four-momenta $z$ and parameter values $\theta$. In practice this means that matrix-element codes such as \toolfont{MadGraph}~\cite{Alwall:2014hca} can not only be run in a forward, generative mode, but also define functions that return the squared matrix element for a given phase-space point $z$. Unfortunately, there is typically no user-friendly interface to these functions, so evaluating it requires some work.

Second, the conditional density $\intractablep (x|z)$ describes the probabilistic evolution from the parton-level four-momenta to observable particle properties. While this symbol looks innocuous, it represents the full parton shower, the interaction of particles with the detector material, the sensor response and readout, and the reconstruction of observables. Different simulators such as \toolfont{Pythia}~\cite{Sjostrand:2007gs}, \toolfont{Geant4}~\cite{Agostinelli:2002hh}, or \toolfont{Delphes}~\cite{deFavereau:2013fsa} are often used to generate samples $\{x\} \sim \intractablep (x |z)$ for given parton-level momenta $z$. This sampling involves the Monte-Carlo integration over the possible shower histories and detector interactions,
\begin{equation}
  \intractablep (x | z) = \int  \! \diff z_{\text{detector}} \, \int \! \diff z_{\text{shower}}\; \intractablep(x | z_{\text{detector}}) \; \intractablep(z_{\text{detector}} | z_{\text{shower}}) \; \intractablep(z_{\text{shower}} | z) \,.
  \label{eq:intractable_integrals}
\end{equation}
This enormous latent space can easily involve many millions of random numbers, and these integrals are clearly intractable, which we denote with the red symbol $\intractablep$. In other words, given a set of  reconstruction-level observables $x$, we cannot calculate the likelihood function $\intractablep(x | z)$ that describes the compatibility of parton-level momenta $z$ with the observation. By extension, we also cannot evaluate $\intractablep(x | \theta)$, the likelihood function of the theory parameters given the observation. The intractable integrals in Eq.~\eqref{eq:intractable_integrals} are the crux of the EFT measurement problem.

The factorization of Eq.~\ref{eq:latent_structure} together with the tractability of the parton-level likelihood $p(z | \theta)$ is immensely important. We will refer to the combination of these two properties as \emph{particle-physics structure}. The far-reaching consequences of this structure for EFT measurements will be the topic of Sec.~\ref{sec:availability}. Many (but not all) of the inference strategies we discuss will rely on this condition.

Note that this Markov property holds even with reducible and irreducible backgrounds and when a matching scheme is used to combine different parton-level multiplicities. In these situations there may be different disjoint parts of $z$ space, even with different dimensionalities, for instance when events with $n$ and $n+1$ partons in the final state can lead to the same configuration of observed jets. The integral over $z$ then has to be replaced with a sum over ``$z_n$ spaces'' and an integral over each $z_n$, but the logic remains unchanged.

\begin{table}
  \small
  \begin{tabularx}{\linewidth}{lLL}
    \toprule
    Symbol & Physics meaning & Machine learning abstraction \\
    \midrule
    $x$ & Set of all observables & Features \\
    $v$ & One or two kinematic variables & Low-dimensional summary statistics\,/engineered feature \\
    $z \equiv z_{\text{parton}}$ & Parton-level four-momenta & Latent variables \\
    $z_{\text{shower}}$ & Parton shower trajectories & Latent variables \\
    $z_{\text{detector}}$ & Detector interactions & Latent variables \\
    $z_{\text{all}} = (z_{\text{parton}}, z_{\text{shower}}, z_{\text{detector}}) $ & Full simulation history of event & All latent variables \\
    $\theta$ & Theory parameters (Wilson coefficients) & Parameters of interest \\
    $\hat{\theta}$ & Best fit for theory parameters & Estimator for parameters of interest \\
    \midrule
    $\intractablep(x | \theta)$ & Distributions of observables given theory parameters & Intractable likelihood \\
    $p(z | \theta)$ & Parton-level distributions from matrix element & Tractable likelihood of latent variables \\
    $\intractablep(x | z)$ & Effect of shower, detector, reconstruction & Intractable density defined through stochastic generative process \\
    \midrule
    $\intractabler(x | \theta_0, \theta_1)$ & \multicolumn{2}{c}{Likelihood ratio between hypotheses $\theta_0$, $\theta_1$, see Eq.~\eqref{eq:likelihood_ratio}.} \\
    $\hat{r}(x | \theta_0, \theta_1)$ & \multicolumn{2}{c}{Estimator for likelihood ratio} \\
    $\intractablet(x | \theta)$ & \multicolumn{2}{c}{Score, see Eq.~\eqref{eq:score}.} \\
    $\hat{t}(x | \theta)$ & \multicolumn{2}{c}{Estimator for score} \\
    \midrule
    $x_e$, $z_e$ & Event & Data point \\
    $\theta_o$ & Wilson coefficient for one operator & Individual parameter of interest \\
    $\theta_c$, $w_c(z)$, $p_c(x)$ & \multicolumn{2}{c}{Morphing basis points, coefficients, densities, see Eq.~\eqref{eq:morphing_structure}.} \\
    \bottomrule
  \end{tabularx}
  \caption{Dictionary defining many symbols that appear in this paper. Red symbols denote intractable likelihood functions. The last three rows explain our conventions for indices.}
  \label{tbl:dictionary}
\end{table}

\subsubsection{Operator morphing}
\label{sec:morphing}

Effective field theories (and other parameterisations of indirect signatures of new physics) typically contribute a finite number of amplitudes to a given process, each of which is multiplied by a function of the Wilson coefficients.\footnote{Exceptions can arise for instance when particle masses or widths depend on the parameters of interest. But in an EFT setting one can expand these quantities in $1 / \Lambda$, restoring the factorization.} In this case the likelihood can be written as
\begin{equation}
  p(z | \theta) = \sum_{c'} \tilde{w}_{c'}(\theta) \, f_{c'}(z)
  \label{eq:morphing_naive}
\end{equation}
where $c'$ labels the different amplitude components, and the functions $f_{c'}(z)$ are not necessarily properly positive definite or normalized.

The simplest example is a process in which one SM amplitude $\mathcal{M}_0(z)$ interferes with one new physics amplitude $\mathcal{M}_{\text{BSM}}(z | \theta) = \theta \mathcal{M}_1 (z)$, which scales linearly with a new physics parameter $\theta$. The differential cross section, proportional to the squared matrix element, is then $\diff \sigma (z) \propto |\mathcal{M}_0(z)|^2 + 2 \theta \, \Real \mathcal{M}_0(z)^\dagger \mathcal{M}_1 (z) + \theta^2 \, |\mathcal{M}_1(z)|^2$. There are three components, representing the SM, interference, and pure BSM terms, each with their own parameter dependence $\tilde{w}_{c'}(z)$ and momentum dependence $f_{c'}(z)$.

We can then pick a number of basis\footnote{Note that the morphing basis points $\theta_c$ are unrelated to the choice of an operator basis for the effective field theory.} parameter points $\theta_{c}$ equal to the number of components $c'$ in Eq.~$\eqref{eq:morphing_naive}$. They can always be chosen such that the matrix $W_{cc'} = \tilde{w}_{c'} (\theta_c)$ is invertible, which allows us to rewrite $\eqref{eq:morphing_naive}$ as a mixture model
\begin{equation}
  p(z | \theta) = \sum_{c}  w_c(\theta)  \, p_c(z)
  \label{eq:morphing_structure}
\end{equation}
with weights $w_c(\theta) = \sum_{c'} \tilde{w}_c(\theta) \, W^{-1}_{\;cc'}$ and (now properly normalized) basis densities $p_c(z) = p(z | \theta_c)$. The weights $w_c(\theta)$ depend on the choice of basis points and are analytically known. This ``morphing'' procedure therefore allows us to extract the full likelihood function $p(z | \theta)$ from a finite set of evaluations of basis densities $p_c(z)$.

Calculating the full statistical model through morphing requires the likelihood $p(z | \theta)$ to be tractable, which is true for parton-level momenta as argued above. However, the same trick can be applied even when the exact likelihood is intractable, but we can estimate it. For instance, the marginal distribution of any individual kinematic variable $v(x)$ can be reliably estimated through histograms or other density estimation techniques, even when shower and detector effects are taken into account. The morphing procedure then lets us evaluate the full conditional distribution $p(v|\theta)$ based on a finite number of Monte-Carlo simulations~\cite{ATLAS:morphing}.

Finally, note that Eq.~\eqref{eq:morphing_structure} together with Eq.~\eqref{eq:latent_structure} imply
\begin{equation}
  \intractablep(x|\theta) = \sum_c w_c(\theta)  \, \intractablep_c(x) \,,
  \label{eq:morphing_structure_intractable}
\end{equation}
even if the likelihood function $\intractablep(x|\theta)$ and the components $\intractablep_c(x)$ are intractable. This will later allow us to impose the morphing structure on likelihood ratio estimators.

Not all EFT amplitudes satisfy the morphing structure in Eq.~\eqref{eq:morphing_naive}, so we discuss both measurement strategies that rely on and make use of this property as well as more general ones that do not require it to hold.

\subsection{Explicit example}
\label{sec:process}

\subsubsection{Weak-boson-fusion Higgs to four leptons}

\begin{figure}
  \fmfframe(0,15)(15,15){ 
    \begin{fmfgraph*}(150,70)
      \feynmansetup
      \fmfleft{i2,i1}
      \fmfright{o6,o5,o4,o3,o2,o1}
      \fmflabel{\small $q$}{i1}
      \fmflabel{\small $q$}{i2}
      \fmflabel{\small $q$}{o1}
      \fmflabel{\small $\ell^+$}{o2}
      \fmflabel{\small $\ell^-$}{o3}
      \fmflabel{\small $\ell^+$}{o4}
      \fmflabel{\small $\ell^-$}{o5}
      \fmflabel{\small $q$}{o6}
      \fmf{fermion,tension=4}{i1,v3}
      \fmf{fermion,tension=4}{i2,v4}
      \fmf{fermion,tension=2.5}{v3,o1}
      \fmf{fermion,tension=2.5}{v4,o6}
      \fmf{wiggly,label=\small $W$,, $Z$,label.side=right}{v3,v5}
      \fmf{wiggly,label=\small $W$,, $Z$,label.side=left}{v4,v5}
      \fmf{dashes,label=\small $h$,tension=0.5}{v5,v6}
      \fmf{wiggly,tension=0.3,label=\small $Z$,tension=0.3,label.side=right}{v7,v6}
      \fmf{wiggly,tension=0.3,label=\small $Z$,tension=0.3,label.side=right}{v6,v8}
      \fmf{fermion,tension=0.2}{o2,v7,o3}
      \fmf{fermion,tension=0.2}{o4,v8,o5}
      \fmfv{decoration.shape=circle,foreground=(0.8,,0.,,0.18),decoration.size=8}{v5,v6}
    \end{fmfgraph*}
  }
  \caption{Feynman diagram for Higgs production in weak boson fusion in the $4 \ell $
    mode. The red dots show the Higgs-gauge interactions affected by
    the dimension-six operators of our analysis.}
  \label{fig:feynman_diag}
\end{figure}
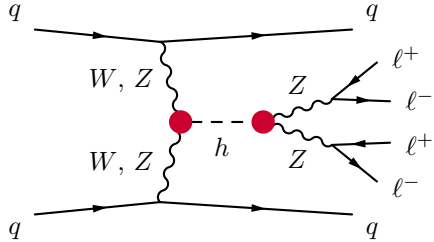

As an explicit example LHC process we consider Higgs production in weak boson fusion (WBF) with a decay of the Higgs into four leptons,
\begin{equation}
  q q \to q q \; h \to q q \; Z Z \to q q \; \ell^+ \ell^- \; \ell^+ \ell^-
\end{equation}
with $\ell = e, \mu$, as shown in Fig.~\ref{fig:feynman_diag}.

While this process is rare and is likely to only be observed during the high-luminosity run of the LHC, it has a few compelling features that make it a prime candidate to study the efficient extraction of information. First, the two jets from the quarks and in particular the four leptons can be reconstructed quite precisely in the LHC detectors. Even when assuming on-shell conditions and energy-momentum conservation, the final-state momenta span a 16-dimensional phase space, giving rise to many potentially informative observables.

Second, both the production of the Higgs boson in weak boson fusion as well as its decay into four leptons are highly sensitive to the effects of new physics in the Higgs-gauge sector. We parameterize these with dimension-six operators in the SMEFT, following the conventions of the Hagiwara-Ishihara-Szalapski-Zeppenfeld basis~\cite{Hagiwara:1993ck}. For simplicity, we limit our analysis to the two particularly relevant operators
\begin{equation}
  \lgr{} = \lgr{SM}
  + \frac {f_{W}} {\Lambda^2} \; \underbrace{\dfrac{\im g}{2} \, (D^\mu\phi)^\dagger \, \sigma^a \, D^\nu\phi \; W_{\mu\nu}^a}_{\ope{W}} \;
  {} - \frac {f_{WW}} {\Lambda^2} \;  \underbrace{\frac{g^2}{4} \, (\phisq) \; W^a_{\mu\nu} \, W^{\mu\nu\, a}}_{\ope{WW}} \,.
\end{equation}
For convenience, we rescale the Wilson coefficients to the dimensionless parameters of interest
\begin{equation}
  \theta = \left( \frac {f_W\, v^2} {\Lambda^2} \,, \quad  \frac {f_{WW}\, v^2} {\Lambda^2} \right)^T
\end{equation}
where $v =246~\gev$ is the electroweak vacuum expectation value. As alluded to above, the validity range of the EFT cannot be determined in a model-independent way. For moderately weakly to moderately strongly coupled underlying new physics models, one would naively expect $|f_o| \lesssim \ord{1}$ and the EFT description to be useful in the range $E \approx v \lesssim \Lambda$, or $-1 \lesssim \theta_o \lesssim 1$. This is the parameter range we analyse in this paper.

The interference between the Standard Model amplitudes and the dimension-six operators leads to an intricate relation between the observables and parameters in this process, which has been studied extensively. The precise measurement of the momenta of the four leptons provides access to a range of angular correlations that fully characterize the $h \to ZZ$ decay~\cite{DellAquila:1985mtb, Bolognesi:2012mm}. These variables are sensitive to the effects of dimension-six operators. But the momentum flow $p$ through the decay vertex is limited by the Higgs mass, and the relative effects of these dimension-six operators are suppressed by a factor $p^2 / \Lambda^2$. On the other hand, the Higgs production through two off-shell gauge bosons with potentially high virtuality does not suffer from this suppression. The properties of the two jets recoiling against them are highly sensitive to operator effects in this vertex~\cite{Plehn:2001nj, Hankele:2006ma, Hagiwara:2009wt, Englert:2012xt}.

\begin{figure}
  \includegraphics[width=0.89 \textwidth]{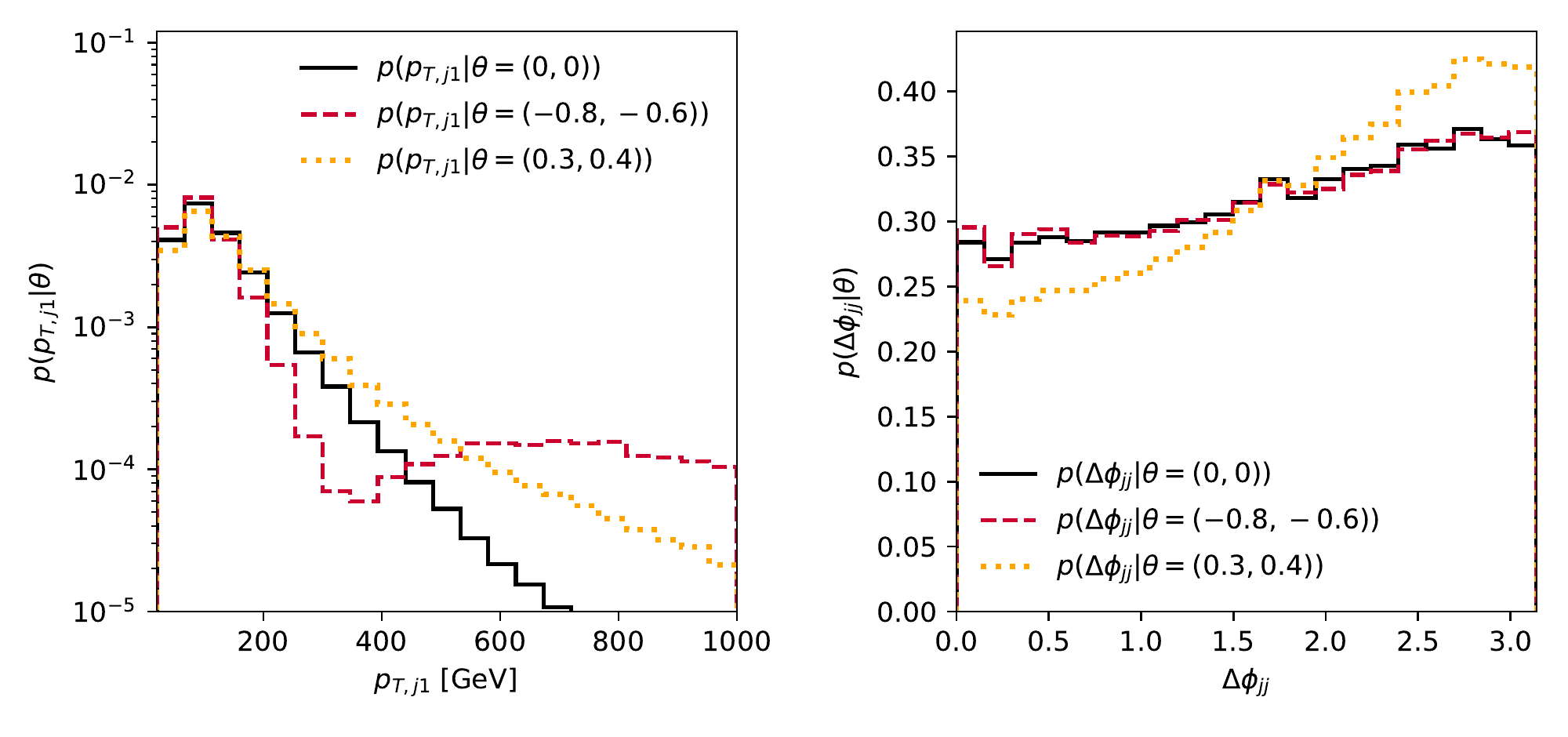}%
  \caption{Kinematic distributions in our example process for three example parameter points. We assume an idealized detector response to be discussed in Sec.~\ref{sec:event_generation}. Left: transverse momentum of the leading (higher-$p_T$) jet, a variable strongly correlated with the momentum transfer in the process. The dip around 350~\gev is a consequence of the amplitude being driven through zero, as discussed in the text. Right: separation in azimuthal angle between the two jets.}
  \label{fig:features}
\end{figure}

In Fig.~\ref{fig:features} we show example distributions of two particularly informative observables, the transverse momentum of the leading (higher-$p_T$) jet $p_{T,j1}$, and the azimuthal angle between the two jets, $\Delta \phi_{jj}$. The two quantities are sensitive to different directions in parameter space. Note also that the interference between the different amplitudes can give rise to non-trivial effects. The size of the dimension-six amplitudes grows with momentum transfer, which is strongly correlated with the transverse momentum of the leading jet. If the interference of new-physics amplitudes with the SM diagrams is destructive, this can drive the total amplitude through zero~\cite{Brehmer:2015rna}. The jet momentum distribution then dips and rises again with higher energies, as seen in the red curve in the left panel of Fig.~\ref{fig:features}. Such depleted regions of low probability can lead to very small or large likelihood ratios and potentially pose a challenge to inference methods.

By analysing the Fisher information in these distributions, it is possible to compare the discrimination power in these two observables to the information contained in the full multivariate distribution or to the information in the total rate. It turns out that the full multivariate distribution $p(z | \theta)$ contains significantly more information than the one-dimensional and two-dimensional marginal distributions of any standard kinematic variables~\cite{Brehmer:2016nyr}. The total rate is found to carry much less information on the two operators, in particular when systematic uncertainties on the cross sections are taken into account. In this study we therefore only analyse the kinematic distributions for a fixed number of observed events.

\subsubsection{Sample generation}
\label{sec:event_generation}

Already in the sample generation we can make use of the structural properties of the process discussed in Sec.~\ref{sec:structure}. The amplitude of this process factorizes into a sum of parameter-dependent factors times phase-space-dependent amplitudes, as given in Eq.~\eqref{eq:morphing_naive}. The effect of the operators $\ope{W}$ and $\ope{WW}$ on the total Higgs width breaks this decomposition, but this effect is tiny and in practice irrelevant when compared to the experimental resolution. The likelihood function of this process therefore follows the mixture model in Eq.~\eqref{eq:morphing_structure} to good approximation, and the weights $w_c(\theta)$ can be calculated. Since the parton-level likelihood function is tractable, we can reconstruct the entire likelihood function $p(z | \theta)$ based on a finite number of simulator runs, as described in Sec.~\ref{sec:morphing}.

To this end, we first generate a parton-level sample $\{z_e\}$ of $5.5  \cdot 10^6$ events with \toolfont{MadGraph~5}~\cite{Alwall:2014hca} and its add-on \toolfont{MadMax}~\cite{Cranmer:2006zs, Plehn:2013paa, Kling:2016lay}, using the setup described in Ref.~\cite{Brehmer:2016nyr}. With \toolfont{MadMax} we can evaluate the likelihood $p(z_e | \theta_c)$ for all events $z_e$ and for 15 different basis parameter points $\theta_c$. Calculating the morphing weights $w_c(\theta)$ finally gives us the true parton-level likelihood function $p(z_e | \theta)$ for each generated phase-space point $z_e$.

\begin{figure}
  \includegraphics[width=\textwidth]{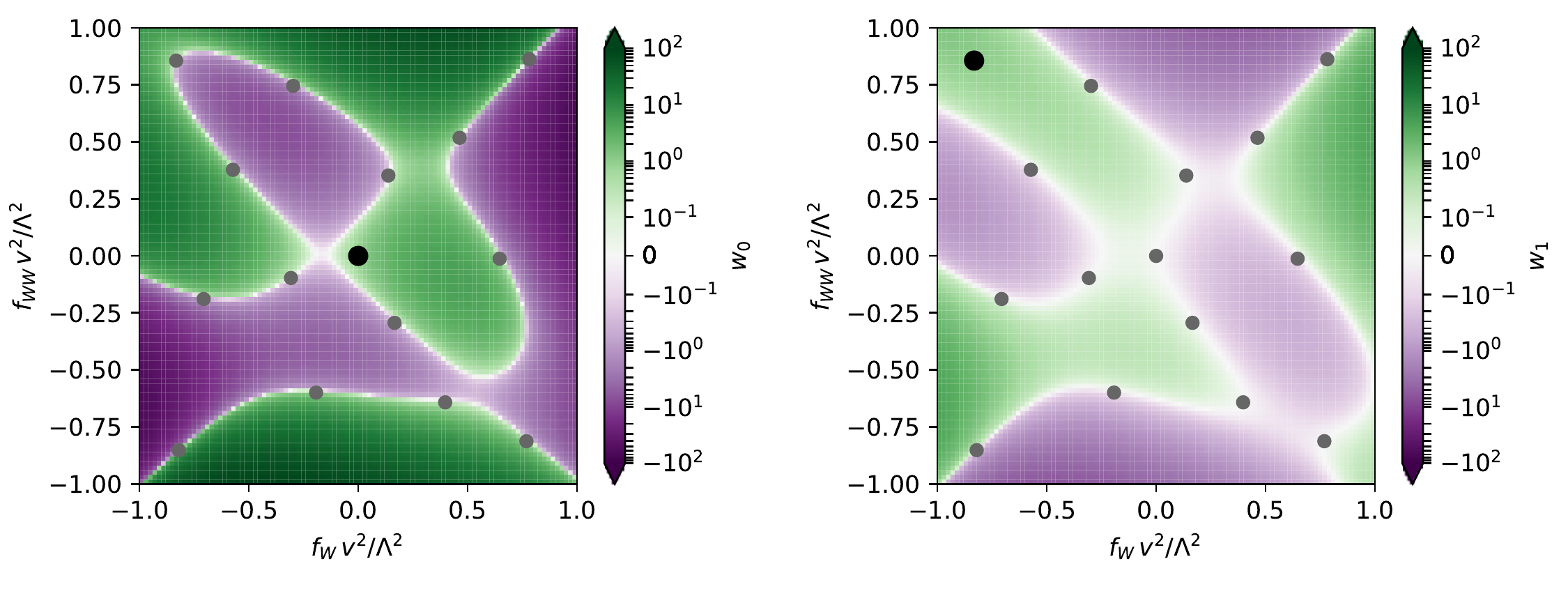}%
  \caption{Basis points $\theta_c$ and some of the morphing weights $w_c (\theta)$ for our example process. Each panel shows the morphing weight of one of the components $c$ as a function of parameter space. The weights of the remaining 13 components (not shown) follow qualitatively similar patterns. The dots show the position of the basis points $\theta_c$, the big black dot denotes the basis point corresponding to the morphing weight shown in that panel. Away from the morphing basis points, the morphing weights can easily reach $\ord{100}$, with large cancellations between different components.}
  \label{fig:morphing}
\end{figure}

In Fig.~\ref{fig:morphing} we show the basis points $\theta_c$ and two of the morphing weights $w_c(\theta)$ with their dependence on $\theta$. In some corners of parameter space the weights easily reach up to $|w_c| \lesssim \ord{100}$, and there are large cancellations between positive and negative weights. This will pose a challenge for the numerical stability of every inference algorithm that directly uses the morphing structure of the process, as we will discuss later. Other basis choices have led to comparable or larger morphing weights.

Parton shower and detector effects smear the observed particle properties $x$ with respect to the parton-level momenta $z$ and make the likelihood function in Eq.~\eqref{eq:latent_structure} intractable. We develop inference methods that can be applied exactly in this case and that do not require any simplifying assumptions on the shower and detector response. However, in this realistic scenario we cannot evaluate their performance by comparing them to the true likelihood ratio. We therefore test them first on an idealized scenario in which the four-momenta, flavor, and charges of the leptons, and the momenta of the partons, can be measured exactly, $\intractablep(x | z) \approx \delta (x-z)$. In this approximation we can evaluate the likelihood $\intractablep(x | \theta)$.

After establishing the performance of the various algorithms in this idealized setup, we will analyse the effect of parton shower and detector simulation on the results. We generate an approximate detector-level sample by drawing events from a smearing distribution $\intractablep(x | z)$ conditional on the parton-level momenta $z$. This smearing function is loosely motivated by the performance of the LHC experiments and is defined in Appendix~\ref{sec:appendix_smearing}.

\section{Likelihood ratio estimation}
\label{sec:modeling}

According to the Neyman-Pearson lemma, the likelihood ratio
\begin{equation}
  \intractabler(x | \theta_0, \theta_1)
  \equiv \frac {\intractablep(x | \theta_0)} {\intractablep(x | \theta_1)}
  = \frac {\intz \intractablep(x,z | \theta_0)} {\intz \intractablep(x,z | \theta_1)}
  \label{eq:likelihood_ratio}
\end{equation}
is the most powerful test statistic to discriminate between two hypotheses $\theta_0$ and $\theta_1$. Unfortunately, the integral over the latent space $z$ makes the likelihood function $\intractablep(x| \theta)$ as well as the likelihood ratio $\intractabler(x | \theta_0, \theta_1)$ intractable. The first and crucial stage of all our EFT measurement strategies is therefore the construction of a \emph{likelihood ratio estimator} $\hat{r}(x | \theta_0, \theta_1)$ that is as close to the true $\intractabler(x | \theta_0, \theta_1)$ as possible and thus maximizes the discrimination power between $\theta_0$ and $\theta_1$.

This estimation problem has several different aspects that we try to disentangle as much as possible. The first choice is the overall structure of the likelihood ratio estimator and its dependence on the theory parameters $\theta$. We discuss this in Sec.~\ref{sec:eft_modeling}. Section~\ref{sec:availability} analyses what information is available and useful to construct (train) the estimators for a given process. Here we will introduce the main ideas that harness the structure of the EFT to increase the information that is used in the training process.

These basic concepts are combined into concrete strategies for the estimation of the likelihood ratio in Sec.~\ref{sec:strategies}. After training the estimators, there is an optional additional calibration stage, which we introduce in Sec.~\ref{sec:calibration}. Section~\ref{sec:details} describes the technical implementation of these strategies in terms of neural networks. Finally, we discuss the challenges that the different algorithms face in Sec.~\ref{sec:challenges} and introduce diagnostic tools for the uncertainties.

\subsection{Modeling likelihood ratios}
\label{sec:eft_modeling}

\subsubsection{Likelihood ratios}

There are different approaches to the structure of this estimator, in particular to the dependence on the theory parameters $\theta$:
\begin{description}
\item[Point by point (PbP)]
A common strategy is to scan the parameter space, randomly or in a grid. To reduce the complexity of the scan one can keep the denominator $\theta_1$ fixed, while scanning only $\theta_0$. Likelihood ratios with other denominators can be extracted trivially as $\hat{r}(x | \theta_0, \theta_2) = \hat{r} (x | \theta_0, \theta_1) / \hat{r} (x | \theta_2, \theta_1)$.  Instead of a single reference value $\theta_1$, we can also use a composite reference hypothesis $p(x | \theta_1) \to p_{\text{ref}}(x) = \int \!\diff \theta_1 \; \pi(\theta_1) \, p(x | \theta_1)$ with some prior $\pi(\theta_1)$. This can reduce the regions in feature space with small reference likelihood $p(x|\theta_1)$ and improve the numerical stability.

For each pair $(\theta_0, \theta_1)$ separately, the likelihood ratio $\hat{r}(x | \theta_0, \theta_1)$ as a function of $x$ is estimated. Only the final results are interpolated between the scanned values of $\theta_0$.

This approach is particularly simple, but discards all information about the structure and smoothness of the parameter space. For high-dimensional parameter spaces, the parameter scan can become prohibitively expensive. The final interpolation may introduce additional uncertainties.
\item[Agnostic parameterized estimators]
Alternatively we can train one estimator as the full model $\hat{r}(x | \theta_0,  \theta_1)$ as a function of both $x$ and the parameter combination $(\theta_0, \theta_1)$~\cite{Cranmer:2015bka, Baldi:2016fzo}. A modification is again to leave $\theta_1$ at a fixed reference value (or fixed composite reference hypothesis with a prior $\pi(\theta_1)$) and only learn the dependence on $x$ and $\theta_0$.

This parameterized approach leaves it to the estimator to learn the typically smooth dependence of the likelihood ratio on the physics parameters and does not require any interpolation in the end. There are no assumptions on the form of the dependence of the likelihood on the ratios.
\item[Morphing-aware estimators]
For problems that satisfy the morphing condition of Eq.~\eqref{eq:morphing_structure} and thus also Eq.~\eqref{eq:morphing_structure_intractable}, we can impose this structure and the explicit knowledge of the weights $w_c(\theta)$ onto the estimator. Again, one option is to keep the denominator fixed at a reference value (or composite reference hypothesis), leading to
\begin{equation}
  \hat{r} (x | \theta_0, \theta_1) = \sum_c w_c(\theta_0) \, \hat{r}_c(x)
  \label{eq:morphing_aware}
\end{equation}
where the basis estimators $\hat{r}_c(x) = \hat{r} (x | \theta_c, \theta_1)$ only depend on $x$.

Alternatively, we can decompose both the numerator and denominator distributions to find~\cite{Cranmer:2015bka, Cranmer:2016swd}
\begin{equation}
  \hat{r} (x | \theta_0, \theta_1) = \sum_c \left[ \sum_{c'} \frac {w_{c'} (\theta_1) } {w_{c'} (\theta_0)} \, \hat{r}_{c',c} (x) \right]^{-1}
  \label{eq:decomposed_ratio}
\end{equation}
with pairwise estimators $\hat{r}_{c',c} (x) = \hat{r} (x | \theta_{c'}, \theta_c)$.
\end{description}

\subsubsection{Score and local model}
\label{sec:local_model}

One remarkably powerful quantity is the score, defined as the relative tangent vector
\begin{equation}
  \intractablet(x|\theta_0) = \nabla_\theta \log \intractablep(x|\theta) \Bigr|_{\theta_0} \,.
  \label{eq:score}
\end{equation}
It quantifies the relative change of the likelihood under infinitesimal changes in parameter space and can be seen as a local equivalent of the likelihood ratio.

In a small patch around $\theta_0$ in which we can approximate $\intractablet(x | \theta)$ as independent of $\theta$, Eq.~\eqref{eq:score} is solved by the \emph{local model}
\begin{equation}
  \localmodel(x|\theta) = \frac 1 {\intractablez(\theta)} \, \intractablep(t(x|\theta_0) | \theta_0) \, \exp[ \intractablet(x|\theta_0) \cdot (\theta - \theta_0) ]
  \label{eq:local_model}
\end{equation}
with a normalisation factor $\intractablez(\theta)$. The local model is in the exponential family. Note that the $\intractablet(x| \theta_0)$ are the sufficient statistics for $\localmodel(x|\theta)$. This is significant: if we can estimate the vector-valued function $\intractablet(x|\theta_0)$ (with one component per parameter of interest) of the high-dimensional $x$, we can reduce the dimensionality of our space dramatically without losing any information, at least in the local model approximation~\cite{Alsing:2018eau}.

In fact, ignoring the normalization factors and in the local model the likelihood ratio between $\theta_0$ and $\theta_1$ only depends on the scalar product between the score and $\theta_0 - \theta_1$, which will allow us to take this dimensionality reduction one step further and compress high-dimensional data $x$ into a scalar without loss of power.

\begin{figure}
  \includegraphics[width=\textwidth]{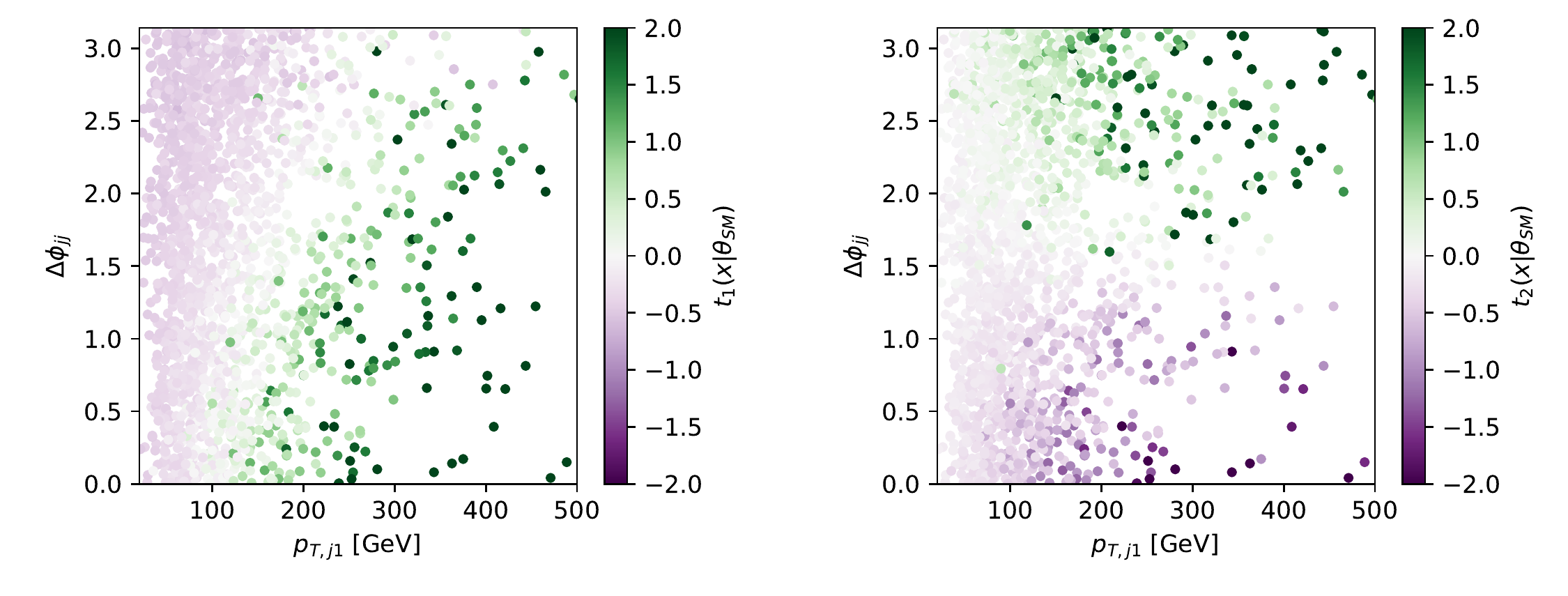}%
  \caption{Score vector as a function of kinematic observables in our example process. Left: first component of the score vector, representing the relative change of the likelihood with respect to small changes in $\ope{W}$ direction. Right: second component of the score vector, representing the relative change of the likelihood with respect to small changes in $\ope{WW}$ direction. In both panels, the axes show two important kinematic variables. We find that the score vector is clearly correlated with these two variables.}
  \label{fig:score_vs_observables}
\end{figure}

In our example process, we are interested in the Wilson coefficients of two dimension-six operators. The score vector therefore has two components. In Fig.~\ref{fig:score_vs_observables} we show the relation between these two score components and two informative kinematic variables, the jet $p_T$ and the azimuthal angle between the two jets, $\Delta \phi$. We find that the score vector is very closely related with these two kinematic quantities, but the relation is not quite one-to-one. Larger energy transfer, measured as larger jet $p_T$, increases the typical size of the score vector. The $\ope{WW}$ component of the score is particularly sensitive to the angular correlation variable, in agreement with detailed studies of this process~\cite{Brehmer:2016nyr}.

\subsection{Available information and its usefulness}
\label{sec:availability}

\subsubsection{General likelihood-free case}
\label{sec:availability_likelihood_free}

All measurement strategies have in common that the estimator $\hat{r}(x| \theta_0, \theta_1)$ is learned from data provided by Monte-Carlo simulations (the stochastic generative process). In the most general likelihood-free scenario, we can only generate samples of events $\{x_e\} $ with $x_e \sim \intractablep(x | \theta)$ through the simulator, and base an estimator $\hat{r}(x| \theta_0, \theta_1)$ on these generated samples.

\begin{figure}
  \includegraphics[width=0.90 \textwidth]{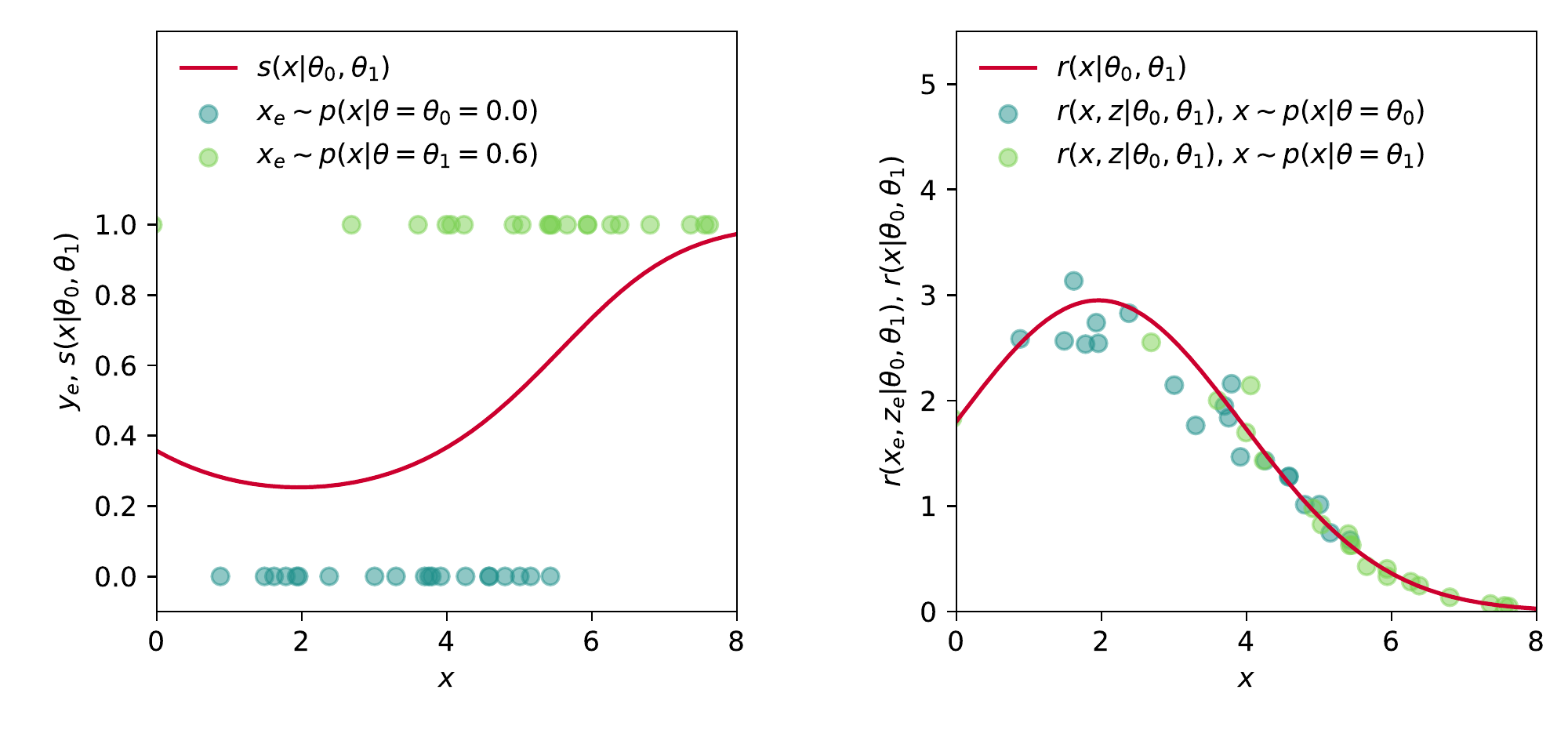}
  \caption{Illustration of some key concepts with a one-dimensional Gaussian toy example. Left: classifiers trained to distinguish two sets of events generated from different hypotheses (green dots) converge to an optimal decision function $\intractables(x | \theta_0, \theta_1)$ (in red) given in Eq.~\eqref{eq:carl_optimal_decision_function}. This lets us extract the likelihood ratio. Right: regression on the joint likelihood ratios $r(x_e, z_e | \theta_0, \theta_1)$ of the simulated events (green dots) converges to the likelihood ratio $\intractabler(x | \theta_0, \theta_1)$ (red line).}
  \label{fig:illustration_carl_ratio_regression}
\end{figure}

One strategy~\cite{Cranmer:2015bka} is based on training a classifier with decision function $\hat{s}(x)$ between two equal-sized samples $\{x_e\} \sim \intractablep(x | \theta_0)$, labelled $y_e=0$, and $\{x_e\} \sim \intractablep(x | \theta_1)$, labelled $y_e=1$. The cross-entropy loss functional
\begin{equation}
  L[\hat{s}] = - \frac 1 N \, \sum_e \left( y_e \, \log \hat{s}(x_e) + (1-y_e) \log (1- \hat{s}(x_e)) \right)
  \label{eq:cross-entropy}
\end{equation}
is minimized by the optimal decision function
\begin{equation}
  \intractables(x| \theta_0, \theta_1) = \frac {\intractablep(x | \theta_1)} {\intractablep(x | \theta_0) + \intractablep(x | \theta_1)} \,.
  \label{eq:carl_optimal_decision_function}
\end{equation}
From the decision function $\hat{s}(x)$ of a classifier we can therefore extract an estimator for the likelihood ratio as
\begin{equation}
  \hat{r}(x | \theta_0, \theta_1) = \frac {1 - \hat{s}(x| \theta_0, \theta_1)} {\hat{s}(x| \theta_0, \theta_1)} \,.
  \label{eq:r_from_s}
\end{equation}
This idea, sometimes called the likelihood ratio trick, is visualized in the left panel of Fig.~\ref{fig:illustration_carl_ratio_regression}.

As pointed out in Ref.~\cite{Cranmer:2015bka}, we can use the weaker assumption of any loss functional that is minimized by a decision function $s(x)$ that is a strictly monotonic function of the likelihood ratio. The underlying reason is that the likelihood ratio is invariant under any transformation $s(x)$ with this property. In practice, the output of any such classifier can be brought closer to the form of Eq.~\eqref{eq:carl_optimal_decision_function} through a calibration procedure, which we will discuss in Sec.~\ref{sec:calibration}.

\subsubsection{Particle-physics structure}

As we have argued in Sec.~\ref{sec:structure}, particle physics processes have a specific structure that allow us to extract additional information. Most processes satisfy the factorization of Eq.~\eqref{eq:latent_structure} with a tractable parton-level likelihood $p(z | \theta)$. The generators do not only provide samples $\{x_e\}$, but also the corresponding parton-level momenta (latent variables) $\{z_e\}$ with $(x_e, z_e) \sim \intractablep(x,z|\theta_0)$. By evaluating the matrix elements at the generated momenta $z_e$ for different hypotheses $\theta_0$ and $\theta_1$, we can extract the parton-level likelihood ratio $p(z_e|\theta_0) / p(z_e | \theta_1)$. Since the distribution of $x$ is conditionally independent of the theory parameters, this is the same as the joint likelihood ratio
\begin{align}
  r(x_e, z_{\text{all}\, e}| \theta_0 , \theta_1) &\equiv \frac {\intractablep(x_e, z_{\text{detector}\, e}, z_{\text{shower}\, e}, z_e | \theta_0)} {\intractablep(x_e, z_{\text{detector}\, e}, z_{\text{shower}\, e}, z_e | \theta_1)} \notag \\
&= \frac {\intractablep(x_e | z_{\text{detector}\, e})} {\intractablep(x_e | z_{\text{detector}\, e})} \, \frac {\intractablep(z_{\text{detector}\, e} | z_{\text{shower}\, e})}  {\intractablep(z_{\text{detector}\, e} | z_{\text{shower}\, e})} \, \frac {\intractablep(z_{\text{shower}\, e} | z_{e})}  {\intractablep(z_{\text{shower}\,e} | z_e)} \, \frac {p(z_e | \theta_0)} {p(z_e | \theta_1)} \notag \\
&= \frac {p(z_{e} | \theta_0)} {p(z_{e} | \theta_1)} \,.
\end{align}
So while we cannot directly evaluate the likelihood ratio at the level of measured observables $\intractabler(x | \theta_0, \theta_1)$, we \emph{can} calculate the likelihood ratio for a generated event conditional on the latent parton-level momenta.

\begin{figure}
  \includegraphics[width=0.90 \textwidth]{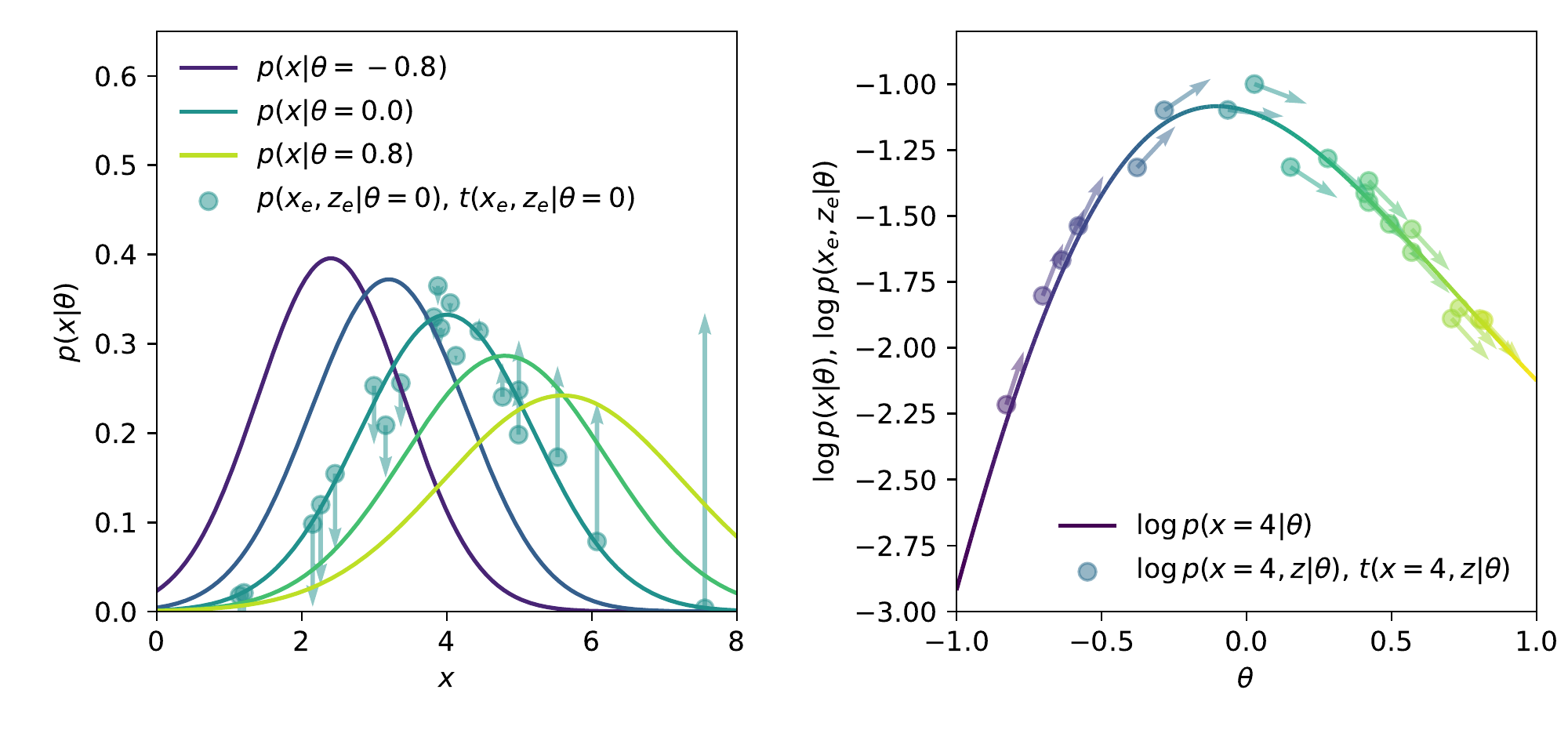}
  \caption{Illustration of some key concepts with a one-dimensional Gaussian toy example. Left: probability density functions for different values of $\theta$ and the scores $t(x_e, z_e | \theta)$ at generated events $(x_e, z_e)$. These tangent vectors measure the relative change of the density under infinitesimal changes of $\theta$. Right: dependence of $\log \intractablep (x | \theta)$ on $\theta$ for fixed $x=4$. The arrows again show the (tractable) scores $t(x_e, z_e | \theta)$.}
  \label{fig:illustration_score}
\end{figure}

\begin{figure}
  \includegraphics[width=0.90 \textwidth]{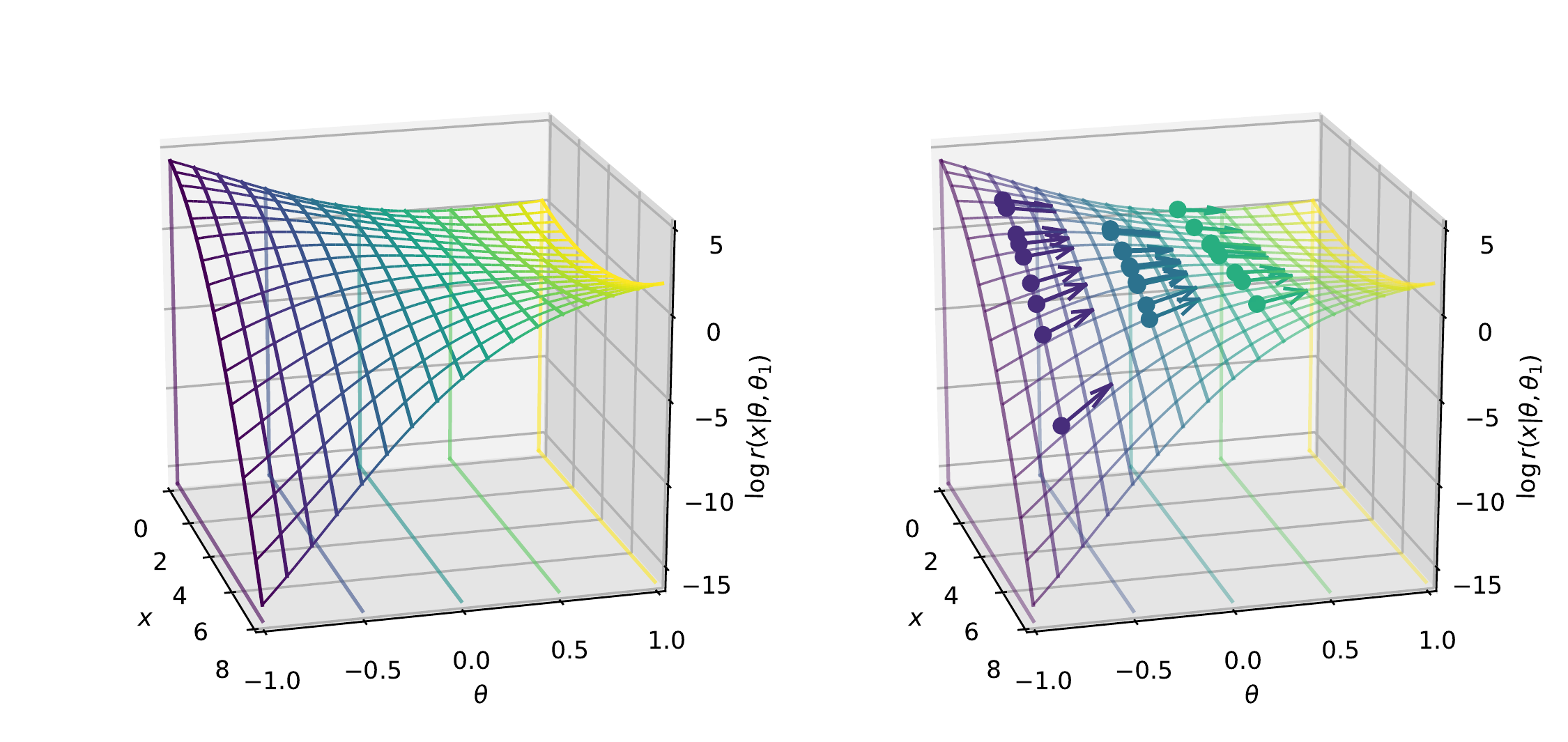}
  \caption{Illustration of some key concepts with a one-dimensional Gaussian toy example. Left: full statistical model $\log \intractabler (x | \theta, \theta_1)$ that we are trying to estimate. Right: available information at the generated events $(x_e, z_e)$.  The dots mark the joint likelihood ratios $\log r(x_e, z_e | \theta_0, \theta_1)$, the arrows the scores $t(x_e, z_e | \theta_0, \theta_1)$.}
  \label{fig:illustration_3d}
\end{figure}

The same is true for the score, \ie the tangent vectors or relative change of the (log) likelihood under infinitesimal changes of the parameters of interest. While the score $\intractablet(x_e | \theta_0) = \nabla_\theta \log \intractablep(x| \theta) |_{\theta_0}$ is intractable, we can extract the joint score
\begin{align}
  t(x_e, z_{\text{all}\, e} | \theta_0) &\equiv \nabla_\theta \log \intractablep(x_e, z_{\text{detector}\, e}, z_{\text{shower}\, e}, z_e | \theta_0)  \notag \\
&= \left. \frac {\intractablep(x_e | z_{\text{detector}\, e})} {\intractablep(x_e | z_{\text{detector}\, e})} \, \frac {\intractablep(z_{\text{detector}\, e} | z_{\text{shower}\, e})}  {\intractablep(z_{\text{detector}\, e} | z_{\text{shower}\, e})} \, \frac {\intractablep(z_{\text{shower}\, e} | z_{e})}  {\intractablep(z_{\text{shower}\, e} | z_e)} \, \frac {\nabla_\theta p(z_e | \theta)} {p(z_e |  \theta)} \right |_{\theta_0} \notag \\
&= \left. \frac {\nabla_\theta p(z_e | \theta)} {p(z_e |  \theta)} \right |_{\theta_0}
\end{align}
from the simulator. Again, all intractable parts of the likelihood cancel. We visualize the score in Fig.~\ref{fig:illustration_score} and all available information on the generated samples in Fig.~\ref{fig:illustration_3d}. It is worth repeating that we are not making any simplifying approximations about the process here, these statements are valid with reducible backgrounds, for state-of-the-art generators including higher-order matrix elements, matching of matrix element and parton shower, and with full detector simulations.

But how does the availability of the \emph{joint} likelihood ratio $r(x,z | \theta)$ and score $t(x,z | \theta)$ (which depend on the latent parton-level momenta $z$) help us to estimate the likelihood ratio $\intractabler(x | \theta)$, which is the one we are interested in?

Consider the $L^2$ squared loss functional for functions $\hat{g}(x)$ that only depend on $x$, but which are trying to approximate a function $g(x,z)$,
\begin{align}
  L[\hat{g}(x)]
  &= \intxz \intractablep(x,z|\theta) \;  |g(x,z) - \hat{g}(x)|^2 \notag \\
  &= \intx \, \underbrace{\left[ \hat{g}^2(x) \, \intz \intractablep(x,z | \theta) - 2 \hat{g}(x) \, \intz \intractablep(x,z | \theta) \, g(x,z) + \intz \intractablep(x,z | \theta) \, g^2(x,z)  \right ]}_{F(x)} \,.
\end{align}
Via calculus of variations we find that the function ${g}^*(x)$ that extremizes $L[\hat{g}]$ is given by~\cite{companion_nips}
\begin{equation}
  0 = \left. \fder{F}{\hat{g}} \right|_{g^*}
  = 2 \hat{g} \,\underbrace{\intz \intractablep(x,z | \theta)}_{= \intractablep(x|\theta)} - 2 \intz \intractablep(x,z | \theta) \, g(x,z) \,,
  \label{eq:squared_loss_general}
\end{equation}
therefore
\begin{equation}
  g^*(x) = \frac{1}{\intractablep(x|\theta)} \, \intz \intractablep(x,z | \theta) \, g(x,z) \,.
\end{equation}

We can make use of this general property in our problem in two ways. Identifying $g(x_e, z_e)$ with the joint likelihood ratios $r(x_e,  z_{\text{all}, e} | \theta_0, \theta_1)$ (which we can calculate!) and $\theta = \theta_1$, we find
\begin{equation}
  g^*(x) = \frac{1}{\intractablep(x|\theta_1)} \intz \intractablep(x,z | \theta_1)\, \frac{\intractablep(x,z|\theta_0) }{\intractablep(x,z|\theta_1)}
  = \intractabler(x | \theta_0, \theta_1) \,.
\end{equation}
By minimizing the squared loss
\begin{equation}
  L [ \hat{r} (x | \theta_0, \theta_1) ]
  =\frac 1 N \, \sum_{(x_e, z_e) \sim \intractablep(x,z | \theta_1)} | r(x_e,  z_{\text{all}, e} | \theta_0, \theta_1) - \hat{r}(x_e| \theta_0, \theta_1)|^2
\end{equation}
of a sufficiently expressive function $\hat{r}(x| \theta_0, \theta_1)$, we can therefore regress on the true likelihood ratio~\cite{companion_nips}! This is illustrated in the right panel of Fig.~\ref{fig:illustration_carl_ratio_regression}. Note that to get the correct minimum, the events $(x_e, z_e)$ have to be sampled according to the denominator hypothesis $\theta_1$.

We can also identify $g(x_e, z_e)$ in Eq.~\eqref{eq:squared_loss_general} with the scores $t(x_e,  z_{\text{all}, e} | \theta)$, which can also be extracted from the generator. In this case,
\begin{equation}
  g^*(x) = \frac{1}{\intractablep(x|\theta)} \intz \nabla_\theta \intractablep(x,z|\theta)
  = \intractablet(x | \theta) \,.
\end{equation}
Thus minimizing
\begin{equation}
  L [ \hat{t} (x | \theta) ]
  = \frac 1 N \, \sum_{(x_e, z_e) \sim \intractablep(x,z | \theta)} |t(x_e,  z_{\text{all}, e} | \theta) - \hat{t}(x_e| \theta)|^2
\end{equation}
of a sufficiently expressive function $\hat{t}(x| \theta)$ allows us to regress on the score $\intractablet(x | \theta)$~\cite{companion_nips}.\footnote{A similar loss function (with a non-standard use of the term ``score'') was used in Ref.~\cite{hyvarinen2005estimation}, though the derivative is taken with respect to $x$ and, critically, the model did not involve marginalization over the latent variable $z$.} Now the $(x_e, z_e)$ have to be sampled according to $\theta$. We summarize the availability of the (joint) likelihood, likelihood ratio, and score in the most general likelihood-free setup and in particle physics processes in Table~\ref{tbl:availability}.

\begin{table}
  \begin{tabular}{ll cc}
    \toprule
    \multicolumn{2}{l}{Quantity} & General likelihood-free & Particle physics\\
    \midrule
    Samples & $\{x_e\}$ & \yep & \yep \\
    Likelihood & $\intractablep(x_e|\theta)$ &  &  \\
    Likelihood ratio & $\intractabler(x_e | \theta_0, \theta_1)$ &  & \kindof \\
    Score & $\intractablet(x_e | \theta)$ &  & \kindof \\
    \midrule
    Latent state & $\{x_e, z_e\}$ &  & \yep \\
    Joint likelihood & $\intractablep(x_e,  z_{\text{all}, e}|\theta)$ &  &  \\
    Joint likelihood ratio & $r(x_e,  z_{\text{all}, e} | \theta_0, \theta_1)$ &  & \yep \\
    Joint score & $t(x_e,  z_{\text{all}, e} | \theta)$ &  & \yep \\
    \bottomrule
  \end{tabular}
  \caption{Availability of different quantities from the generative process in the most general likelihood-free setup vs.\ in the particle-physics scenario with the structure given in Eq.~\eqref{eq:latent_structure}. Asterisks (\kindof) denote quantities that are not immediately available, but can be regressed from the corresponding joint quantity, as shown in Sec.~\ref{sec:availability}.}
  \label{tbl:availability}
\end{table}

This is one of our key results and opens the door for powerful new inference methods. Particle physics processes involve the highly complex effects of parton shower, detector, and reconstruction, modelled by a generative process with a huge latent space and an intractable likelihood. Still, the specific structure of this class of processes allows us to calculate how much more or less likely a generated event becomes when we move in the parameter space of the theory. We have shown that by regressing on the joint likelihood ratios or scores extracted in this way, we can recover the actual likelihood ratio or score as a function of the observables!

\subsection{Strategies}
\label{sec:strategies}

Let us now combine the estimator structure discussed in Sec.~\ref{sec:eft_modeling} with the different quantities available during training discussed in Sec.~\ref{sec:availability} and define our strategies to estimate the likelihood ratio. Here we restrict ourselves to an overview over the main ideas of the different approaches. A more detailed explanation and technical details can be found in Appendix~\ref{sec:appendix_models}.

\subsubsection{General likelihood-free case}

Some approaches are designed for the most general likelihood-free scenario and only require the samples $\{x_e\}$ from the generator:
\begin{description}
\item[Histograms of observables]
  The traditional approach to kinematic analyses relies on one or two kinematic variables $v(x)$, manually chosen for a given process and set of parameters. Densities $\hat{p}(v(x) | \theta)$ are estimated by filling histograms with generated samples, leading to the likelihood ratio
  \begin{equation}
    \hat{r} (x | \theta_0, \theta_1) = \frac {\hat{p}(v(x) | \theta_0)} {\hat{p}(v(x) | \theta_1)} \,.
  \end{equation}

  We use this algorithm point by point in $\theta_0$, but a morphing-based setup is also possible (see Sec.~\ref{sec:morphing}). We discuss the histogram approach in more detail in Appendix~\ref{sec:appendix_models_histos}.

\item[Approximate Frequentist Computation (\afc)]
  Approximate Bayesian Computation (\abc) is currently the most widely used method for likelihood-free inference in a Bayesian setup. It allows to sample parameters from the intractable posterior, $\theta \sim \intractablep(\theta|x) = \intractablep(x | \theta) p(\theta) / \intractablep(x)$. Essentially, \abc relies on the approximation of the likelihood function through a rejection probability
  \begin{equation}
    p_{\text{rejection}}(x | \theta) = K_\epsilon (v(x), v(x_e)) \,,
    \label{eq:abc_rejection_probability}
  \end{equation}
  with $x_e \sim \intractablep(x|\theta)$, a kernel $K_\epsilon$ that depends on a bandwidth $\epsilon$, and a sufficiently low-dimensional summary statistics $v(x)$.

Inference in particle physics is usually performed in a frequentist setup, so this sampling mechanism is not immediately useful. But we can define a frequentist analogue, which we call ``Approximate Frequentist Computation'' (\afc). In analogy to the rejection probability in Eq.~\ref{eq:abc_rejection_probability}, we can define a kernel density estimate for the likelihood function as
  \begin{equation}
    \hat{p}(x | \theta) = \frac 1 N \, \sum_e K_\epsilon (v(x), v(x_e)) \,.
    \label{eq:afc_likelihood_estimate}
  \end{equation}
The corresponding likelihood ratio estimator is
  \begin{equation}
    \hat{r} (x | \theta_0, \theta_1) = \frac {\hat{p}(x | \theta_0)} {\hat{p}(x | \theta_1)} \,.
  \end{equation}

  We use this approach point by point in $\theta_0$ with a fixed reference $\theta_1$. As summary statistics, we use subsets of kinematic variables, similar to the histogram approach. We give more details in Appendix~\ref{sec:appendix_models_afc}.

\item[Calibrated classifiers (\carl\footnotemark)]
\addtocounter{footnote}{-1}
\footnotetext{\textbf{Ca}librated \textbf{r}atios of \textbf{l}ikelihoods}
  As discussed in Sec.~\ref{sec:availability_likelihood_free}, the decision function $\hat{s}(x|\theta_0, \theta_1)$ of a classifier trained to discriminate between samples generated according to $\theta_0$ from $\theta_1$ can be turned into an estimator for the likelihood ratio
  \begin{equation}
    \hat{r}(x | \theta_0, \theta_1) = \frac {1 - \hat{s}(x| \theta_0, \theta_1)} {\hat{s}(x| \theta_0, \theta_1)} \,.
  \end{equation}
  This is illustrated in the left panel of Fig.~\ref{fig:illustration_carl_ratio_regression}.

  If the classifier does not learn the optimal decision function of Eq.~\eqref{eq:carl_optimal_decision_function}, but any monotonic function of the likelihood ratio, a calibration procedure can improve the performance significantly. We will discuss this in Sec.~\ref{sec:calibration} below.

  We implement this strategy point by point in $\theta_0$, as an agnostic parameterized classifier $\hat{r}(x | \theta_0, \theta_1)$ that learns the dependence on both $x$ and $\theta_0$, as well as a morphing-aware parameterized classifier. More details are given in Appendix~\ref{sec:appendix_models_carl}.
\item[Neural conditional density estimators (\nde)]
  Several other methods for conditional density estimation have been proposed, often based on neural networks~\cite{2012arXiv1212.1479F, NIPS2016_6084, 2016arXiv160206701P, 2016arXiv161110242D, gutmann2017likelihood, 2017arXiv170208896T, 2017arXiv170707113L}. One particularly interesting class of methods for density estimation is based on the idea of expressing the target density as a sequence of invertible transformations applied to a simple initial density, such as a Gaussian~\cite{2014arXiv1410.8516D, 2015arXiv150505770J, 2016arXiv160508803D, 2017arXiv170507057P, 2018arXiv180507226P}. The density in the target space is then given by the Jacobian determinant of the transformation and the base density. A closely related and successful alternative are neural autoregressive models~\cite{2016arXiv160502226U, 2016arXiv160903499V, 2016arXiv160605328V, 2016arXiv160106759V}, which factorize the target density as a sequence of simpler conditional densities.  Both classes of estimators are trained by maximizing the log likelihood.

  We leave a detailed discussion of these techniques for particle physics problems as well as an implementation in our example process for future work.
\end{description}

\subsubsection{Particle-physics structure}

As we argued in Sec.~\ref{sec:availability}, particle physics simulations let us extract the joint likelihood ratio $r(x_e, z_e | \theta_0, \theta_1)$ and the joint score $t(x_e, z_e | \theta_0, \theta_1)$, giving rise to strategies tailored to this class of problems:
\begin{description}
\item[Ratio regression (\rolr\footnotemark)]
\addtocounter{footnote}{-1}
\footnotetext{\textbf{R}egression \textbf{o}n \textbf{l}ikelihood \textbf{r}atio}
  We can directly regress the likelihood ratio $\hat{r}(x | \theta_0, \theta_1)$. As shown in the previous section, the squared error loss between a function $\hat{r}(x_e | \theta_0, \theta_1)$ and the available joint likelihood ratio $r(x_e, z_e | \theta_0, \theta_1)$ is minimized by the likelihood ratio $r(x | \theta_0, \theta_1)$, provided that the samples $(x_e, z_e)$ are drawn according to $\theta_1$. Conversely, the squared error of $1 / r(x_e, z_e | \theta_0, \theta_1)$ with $(x_e, z_e) \sim \intractablep(x, z | \theta_0)$ is also minimized by the likelihood ratio. We can combine these two terms into a combined loss function
  \begin{multline}
    L [ \hat{r}(x | \theta_0, \theta_1) ] = \frac 1 N \, \sum_{(x_e, z_e, y_e)} \Biggl( y_e \, |r(x_e,z_e | \theta_0, \theta_1) - \hat{r}(x| \theta_0, \theta_1)|^2 \\
    + (1-y_e)  \, \left |\frac 1 {r(x_e,z_e | \theta_0, \theta_1)} - \frac 1 {\hat{r}(x| \theta_0, \theta_1)} \right|^2 \Biggr)
    \label{eq:ratio_regression_loss}
  \end{multline}
  with $y_e=0$ for events generated according to $(x_e, z_e) \sim \intractablep(x, z | \theta_0)$ and $y_e=1$ for $(x_e, z_e) \sim \intractablep(x, z | \theta_1)$.  The factors of $y_e$ and $(1-y_e)$ ensure the correct sampling for each part of the loss functional. We illustrate this approach in the right panel of Fig.~\ref{fig:illustration_carl_ratio_regression}.

  This strategy is again implemented point by point in $\theta_0$, in an agnostic parameterized setup, as well as in a morphing-aware parameterized setup. We describe it in more detail in Appendix~\ref{sec:appendix_models_rolr}.

\item[\carl + score regression (\cascal\footnotemark)]
\addtocounter{footnote}{-1}
\footnotetext{\textbf{C}ARL \textbf{a}nd \textbf{sc}ore \textbf{a}pproximate \textbf{l}ikelihood ratio}
  The parameterized \carl strategy outlined above learns a classifier decision function $\hat{s}(x | \theta_0, \theta_1)$ as a function of $\theta_0$. If the classifier is realized with a differentiable architecture such as a neural network, we can calculate the gradient of this function and of the corresponding estimator for the likelihood ratio $\hat{r}(x | \theta_0, \theta_1)$ with respect to $\theta_0$ and derive the estimated score
  \begin{equation}
    \hat{t} (x | \theta_0) = \nabla_\theta \log \hat{r} (x | \theta, \theta_1 ) \Biggr |_{\theta_0}
                            = \nabla_\theta \log  \frac {1 - \hat{s}(x| \theta_0, \theta_1)} {\hat{s}(x| \theta_0, \theta_1)} \Biggr |_{\theta_0}\,.
                            \label{eq:parameterized_carl_score}
  \end{equation}
   If the estimator is perfect, we expect this estimated score to minimize the squared error with respect to the joint score data available from the simulator, following the arguments in Sec.~\ref{sec:availability}.

  We can turn this argument around and use the available score data during the training. Instead of training the classifier just by minimizing the cross-entropy, we can instead simultaneously minimize the squared error on this derived score with respect to the true joint score $t(x, z | \theta_0, \theta_1)$. The combined loss function is given by
  \begin{equation}
    L[\hat{s}] = \frac 1 N \, \sum_e \Biggl[
    y_e \, \log \hat{s}(x_e) + (1-y_e) \log (1- \hat{s}(x_e))
    + \alpha \, (1-y_e) \,  \left|t(x_e,z_e | \theta_0) - \hat{t} (x_e | \theta_0)\right|^2
    \Biggr]
    \label{eq:carl_plus_score_loss}
  \end{equation}
  with $\hat{t} (x | \theta_0)$ defined in Eq.~\eqref{eq:parameterized_carl_score} and a hyperparameter $\alpha$ that weights the two pieces of the loss function relative to each other. Again, $y_e=0$ for events generated according to $(x_e, z_e) \sim \intractablep(x, z | \theta_0)$ and $y_e=1$ for $(x_e, z_e) \sim \intractablep(x, z | \theta_1)$, and the factors of $y_e$ and $(1 - y_e)$ ensure the correct sampling for each part of the loss functional.

This strategy relies on the parameterized modeling of the likelihood ratio. We implement both an agnostic version as well as a morphing-aware model.  See Appendix~\ref{sec:appendix_models_cascal} for more details.

\item[Neural conditional density estimators + score (\scandal\footnotemark)]
\addtocounter{footnote}{-1}
\footnotetext{\textbf{Sc}ore \textbf{a}nd \textbf{n}eural \textbf{d}ensity \textbf{a}pproximate \textbf{l}ikelihood}
In the same spirit as the \cascal method, neural density estimators such as autoregressive flows can be augmented with score information. We have started to explore this class of algorithms in Ref.~\cite{companion_nips}, but leave a detailed study and the application to particle physics for future work.

\item[Ratio + score regression (\rascal\footnotemark)]
\addtocounter{footnote}{-1}
\footnotetext{\textbf{R}atio \textbf{a}nd \textbf{sc}ore \textbf{a}pproximate \textbf{l}ikelihood ratio}
The same trick works for the parameterized \rolr approach. If the regressor is implemented as a differentiable architecture such as a neural network, we can calculate the gradient of the parameterized estimator $\hat{r}(x | \theta_0, \theta_1)$ with respect to $\theta_0$ and calculate the score
  \begin{equation}
    \hat{t} (x | \theta_0) = \nabla_\theta \log \hat{r} (x | \theta, \theta_1 ) \Biggr |_{\theta_0} \,.
    \label{eq:parameterized_regression_score}
  \end{equation}
  Instead of training just on the squared likelihood ratio error, we can minimize the combined loss
  \begin{align}
    L [ \hat{r}(x | \theta_0, \theta_1) ] = \frac 1 N \, \sum_{(x_e, z_e, y_e)} \Biggl[
    &y_e \, |r(x_e,z_e | \theta_0, \theta_1) - \hat{r}(x_e| \theta_0, \theta_1)|^2 \notag\\
    &+ (1-y_e)  \, \left |\frac 1 {r(x_e,z_e | \theta_0, \theta_1)} - \frac 1 {\hat{r}(x_e| \theta_0, \theta_1)} \right|^2  \notag \\
    &+ \alpha \, (1-y_e) \,  \left|t(x_e,z_e | \theta_0) - \hat{t} (x_e | \theta_0)\right|^2
    \Biggr]
    \label{eq:ratio_plus_score_loss}
  \end{align}
  with $\hat{t} (x | \theta_0)$ defined in Eq.~\eqref{eq:parameterized_regression_score} and a hyperparameter $\alpha$. The likelihood ratios and scores again provide complementary information as shown in the Fig.~\ref{fig:illustration_3d}.

Once more we experiment with both an agnostic parameterized model as well as a morphing-aware version.

This technique uses all the available data from the simulator that we discussed in Sec.~\ref{sec:availability} to train an estimator of particularly high fidelity. It is essentially a machine-learning version of the Matrix Element Method. It replaces computationally expensive numerical integrals with an upfront regression phase, after which the likelihood ratio can be evaluated very efficiently. Instead of manually specifying simplified smearing functions, the effect of parton shower and detector is learned from full simulations. For more details on \rascal, see Appendix~\ref{sec:appendix_models_rascal}.

\item[Local score regression and density estimation (\sally\footnotemark)]
\addtocounter{footnote}{-1}
\footnotetext{\textbf{S}core \textbf{a}pproximates \textbf{l}ikelihood \textbf{l}ocall\textbf{y}}
  In the local model approximation discussed in Sec.~\ref{sec:local_model}, the score evaluated at some reference point $\theta_{\text{score}}$ is the sufficient statistics, carrying all the information on $\theta$. A precisely estimated score vector (with one component per parameter of interest) is therefore the ideal summary statistics, at least in the neighborhood of the Standard Model or any other reference parameter point.

In the last section we argued that we can extract the joint score $t(x_e, z_e | \theta_{\text{score}})$ from the simulator. We showed that the squared error between a function $\hat{t}(x | \theta_{\text{score}})$ and the joint score is minimized by the intractable score $\intractablet(x | \theta_{\text{score}})$, as long as the events are sampled as $(x_e, z_e) \sim \intractablep(x, z | \theta_{\text{score}})$. We can thus use the augmented data to train an estimator $\hat{t}(x | \theta_{\text{score}})$ for the score at the reference point.

In a second step, we can then estimate the likelihood $\hat{p}(\hat{t} (x | \theta_{\text{score}}) | \theta)$ with histograms, KDE, or any other density estimation technique, yielding the likelihood ratio estimator
  \begin{equation}
    \hat{r} (x | \theta_0, \theta_1) = \frac {\hat{p}\left(\hat{t}(x | \theta_{\text{score}}) \,\middle|\, \theta_0\right)}  {\hat{p}\left(\hat{t}(x | \theta_{\text{score}}) \,\middle|\, \theta_1\right)} \,.
    \label{eq:score_regression_density_estimation_score}
  \end{equation}

  This particularly straightforward strategy is a machine-learning analogue of Optimal Observables that learns the effect of parton shower and detector from data. After an upfront regression phase, the analysis of an event only requires the evaluation of one estimator to draw conclusions about all parameters.  See Appendix~\ref{sec:appendix_models_sally} for more details.

\item[Local score regression, compression to scalar, and density estimation (\sallino\footnotemark)]
\addtocounter{footnote}{-1}
The \sally technique compresses the information in a high-dimensional vector of observables $x$ into a lower-dimensional estimated score vector. But for measurements in high-dimensional parameter spaces, density estimation in the estimated score space might still be computationally expensive. Fortunately, the local model of Eq.~\eqref{eq:local_model} motivates an even more dramatic dimensionality reduction to one dimension, independent of the number of parameters: Disregarding the normalization constants, the ratio $r(x | \theta_0, \theta_1)$ only depends on the scalar product between the score and $\theta_0 - \theta_1$.

\footnotetext{\textbf{S}core \textbf{a}pproximates \textbf{l}ikelihood \textbf{l}ocally \textbf{in} \textbf{o}ne dimension}

Given the same score estimator $\hat{t}(x | \theta_{\text{score}})$ developed for the \sally method, we can define the scalar function
\begin{equation}
  \hat{h}(x | \theta_0, \theta_1) \equiv \hat{t}(x | \theta_{SM}) \cdot (\theta_0 - \theta_1) \,.
  \label{eq:sallino_h}
\end{equation}
Assuming a precisely trained score estimator, this scalar encapsulates all information on the likelihood ratio between $\theta_0$ and $\theta_1$, at least in the local model approximation. The likelihood ratio can then be estimated as
  \begin{equation}
    \hat{r} (x | \theta_0, \theta_1)
    = \frac {\hat{p}\left( \hat{h}(x| \theta_0, \theta_1) \, \middle|\, \theta_0\right)} {\hat{p}\left( \hat{h}(x| \theta_0, \theta_1) \, \middle|\, \theta_1\right)} \,,
    \label{eq:score_regression_density_estimation_score_times_theta}
  \end{equation}
where the $\hat{p} (\hat{h})$ are simple univariate density estimators.

This method allows us to condense any high-dimensional observation $x$ into a scalar function without losing sensitivity, at least in the local model approximation. It thus scales exceptionally well to problems with many theory parameters.  We describe \sallino in more detail in Appendix~\ref{sec:appendix_models_sallino}.

\end{description}

Several other variants are possible, including other combinations of the loss functionals discussed above. We leave this for future work.%
\footnote{Not all initially promising strategies work in practice. We experimented with an alternative strategy based on the morphing structure in Eq.~\eqref{eq:morphing_structure}. Consider the case in which the training sample consists of a number of sub-samples, each generated according to a morphing basis point $\theta_c$. Then the morphing basis sample $c$ used in the event generation is a latent variable that we can use instead of the parton-level momenta $z$ to define joint likelihood ratios and joint scores. Regressing on these quantities should converge to the true likelihood ratio and score, in complete analogy to the discussion in Sec.~\ref{sec:availability}. But the joint ratios and scores defined in this way span a huge range, and the densities of the different basis points are very similar. The convergence is therefore extremely slow and the results based on this method suffer from a huge variance.}

\begin{table}
  \small
  \begin{tabular}{l c ccc c cccc c c}
    \toprule
    \multirow{2}{*}{Strategy} && \multicolumn{3}{c}{Estimator versions} && \multicolumn{4}{c}{Loss function} && \multirow{2}{*}{Asymptotically exact} \\
    \cmidrule{3-5} \cmidrule{7-10}
    && PbP & Param & Aware && CE & ML & Ratio &Score && \\
    \midrule
    Histograms && \yep &  & (\yep) && & & & &&  \\
    \afc && \yep &  & (\yep) && & & & && \\
    \carl && \yep & \yep & \yep && \yep & & & && \yep \\
    \nde && (\yep) & (\yep) & (\yep) && & \yep & & &&  \yep \\
    \midrule
    \rolr && \yep & \yep & \yep  && & & \yep &  && \yep \\
    \cascal &&  & \yep & \yep && \yep & & & \yep && \yep \\
    \scandal &&  & (\yep) & (\yep) && & \yep & & \yep && \yep \\
    \rascal &&  & \yep & \yep &&  & & \yep & \yep && \yep \\
    \sally && \yep &  & (\yep) &&  & & & \yep &&  \\
    \sallino && \yep &  & (\yep) &&  & & & \yep &&  \\
    \bottomrule
  \end{tabular}
  \caption{Overview over the discussed measurement strategies. The first three techniques can be applied in the general likelihood-free setup, they only require sets of generated samples $\{x_e\}$. The remaining five methods are tailored to the particle physics structure and require the availability of $r(x_e,z_e | \theta_0, \theta_1)$ or $t(x_e,z_e | \theta_0)$ from the generator, as discussed in Sec.~\ref{sec:availability}. Brackets denote possible variations that we have not implemented for our example process. In the \sally and \sallino strategies, ``estimator versions'' refers to the density estimation step. In the loss function columns, ``CE'' stands for the cross-entropy, ``ML'' for maximum likelihood, ``ratio'' for losses of the type $|r(x,z) - \hat{r}(x) |^2$, and ``score'' for terms such as $|t(x,z) - \hat{t}(x) |^2$.}
  \label{tbl:strategies}
\end{table}

We summarize the different techniques in Table~\ref{tbl:strategies}. With this plethora of well-motivated analysis methods, the remaining key question for this paper is how well they do in practice, and which of them (if any) should be used. This will be the focus of Sec.~\ref{sec:results}. In the next sections, we will first discuss some important additional aspects of these methods.

\subsection{Calibration}
\label{sec:calibration}

While the likelihood ratio estimators described above work well in many cases, their performance can be further improved by an additional calibration step. Calibration takes place \emph{after} the ``raw'' or uncalibrated estimators $\hat{r}_{\text{raw}} (x | \theta_0, \theta_1)$ have been trained. In general, it defines a function $C$ with the aim that $\hat{r}_{\text{cal}} = C(\hat{r}_{\text{raw}})$ provides a better estimator of the true likelihood ratio than $\hat{r}_{\text{raw}}$. We consider two different approaches to defining this function $C$, which we call probability calibration and expectation calibration.

\subsubsection{Probability calibration}
\label{sec:probability_calibration}

Consider the \carl strategy, which trains a classifier with a decision function $s(x)$ as the basis for the likelihood ratio estimation. Even if this classifier can separate the two classes of events from $\theta_0$ and $\theta_1$ well, its decision function $\hat{s}(x | \theta_0, \theta_1)$ might not have a direct probabilistic interpretation: it might be any approximately monotonic function of the likelihood ratio rather than the ideal solution given in Eq.~\eqref{eq:carl_optimal_decision_function}. In this case, the \carl approach requires calibration, an additional transformation of the raw output $\hat{r}_{\text{raw}} = \hat{r}(x | \theta_0, \theta_1)$ into a calibrated decision function
\begin{equation}
  \hat{r}_{\text{cal}}  = C(\hat{r}_{\text{raw}}) = \frac {\hat{p} (\hat{r}_{\text{raw}} | \theta_0)} {\hat{p} (\hat{r}_{\text{raw}} | \theta_1)} \,,
\end{equation}
where the densities $\hat{p}(\hat{r}_{\text{raw}} | \theta)$ are estimated through a univariate density estimation technique such as histograms. This calibration procedure does not only apply to classifiers, but to any other likelihood ratio estimation strategy.

\subsubsection{Expectation calibration}
\label{sec:expectation_calibration}

Consider some likelihood ratio $r(x | \theta_0, \theta_1)$. The expectation value of this ratio assuming $\theta_1$ to be true is given by
\begin{equation}
  \expectation [ \intractabler(x | \theta_0, \theta_1) | \theta_1] =  \intx \intractablep(x | \theta_1) \, \frac {\intractablep(x | \theta_0)} {\intractablep(x | \theta_1)} = 1 \,.
\end{equation}
A good estimator for the likelihood ratio should reproduce this property. We can numerically approximate this expectation value by evaluating $\hat{r} (x | \theta, \theta_1)$ on a sample $\{x_e\}$ of $N$ events drawn according to $\theta_1$,
\begin{equation}
  \hat{R}(\theta) = \frac 1 N \, \sum_{x_e \sim \theta_1} \hat{r}(x_e | \theta, \theta_1) \approx 1 \,.
  \label{eq:ratio_expectation}
\end{equation}

If a likelihood ratio estimator $\hat{r}_{\text{raw}} (x | \theta, \theta_1)$ (which might be entirely uncalibrated or already probability calibrated) does not satisfy this condition, we can calibrate it by rescaling it as
\begin{equation}
  \hat{r}_{\text{cal}} (x | \theta, \theta_1) = \frac {\hat{r}_{\text{raw}} (x | \theta, \theta_1)} {\hat{R}_{\text{raw}}(\theta)} \,.
  \label{eq:expectation_calibration}
\end{equation}

For a perfect estimator with $\hat{r}(x | \theta_0, \theta_1) = \intractabler(x | \theta_0, \theta_1)$, we can even calculate the variance of the numeric calculation of the expectation value in Eq.~\eqref{eq:ratio_expectation}. We find
    \begin{equation}
      \var [ \hat{R}(\theta) ] = \frac 1 N \, \left[ \expectation \left[ \hat{r}(x | \theta, \theta_1) \middle| \theta \right] - 1 \right] \,,
      \label{eq:expectation_calibration_variance}
    \end{equation}
where $N$ is the number of events used to calculate the expectation value $R(\theta)$, and the expectation $\expectation \left[ \hat{r}(x | \theta, \theta_1) \middle| \theta \right]$ (under the \emph{numerator} hypothesis!) can be calculated numerically.

This calibration strategy can easily improve classifiers that are off by some $\theta$-dependent factor. However, a few rare events $x_e$ with large $\hat{r}(x_e | \theta, \theta_1)$ can dominate the expectation value. If these are mis-estimated, the expectation calibration \label{eq:eq:expectation_calibration} can actually degrade the performance of the estimator on the bulk of the distribution with smaller $\hat{r}(x | \theta, \theta_1)$.

\subsection{Implementation}
\label{sec:details}

\subsubsection{Neural networks}

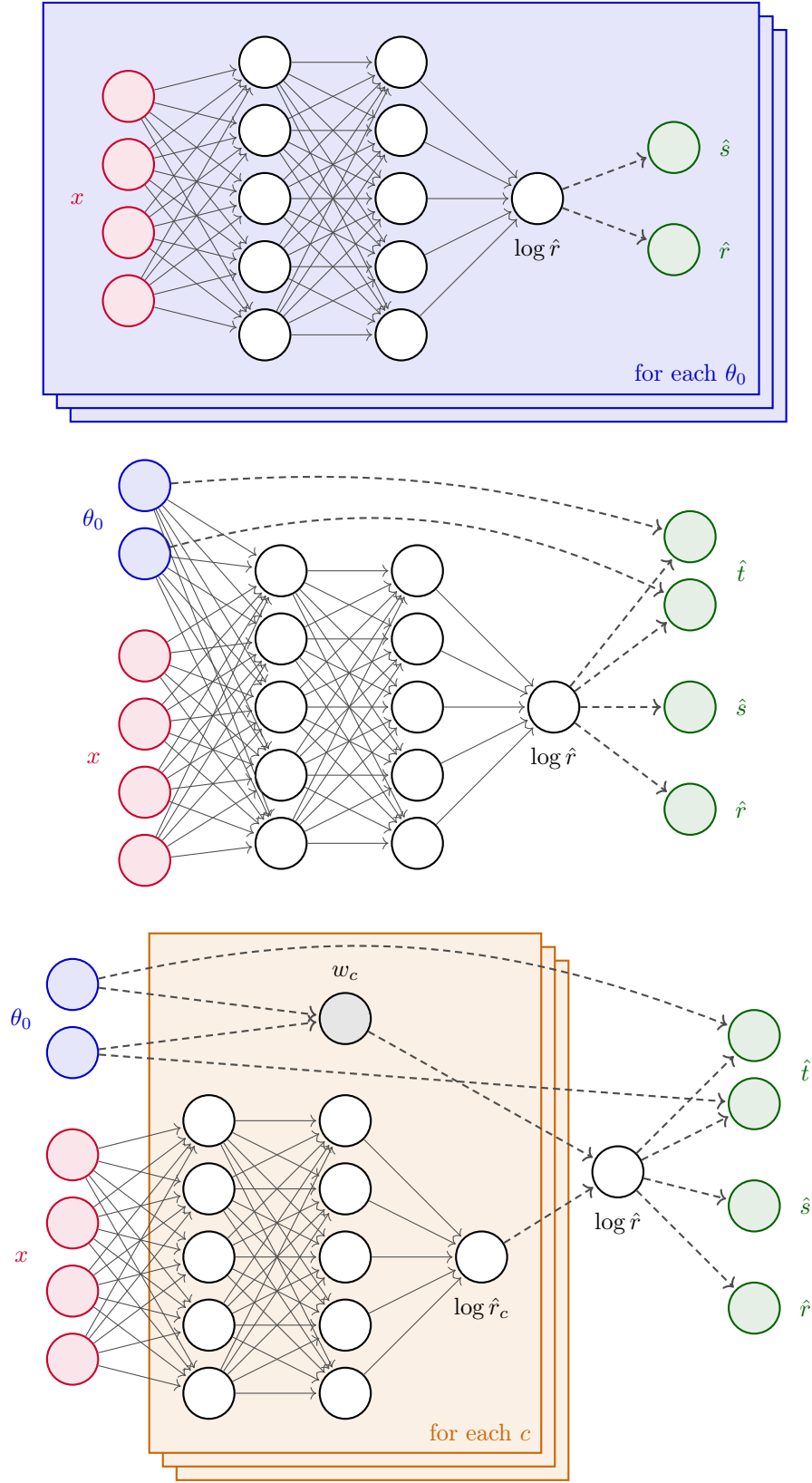
\begin{figure}
  \begin{tikzpicture}[]
  \tikzstyle{connection}=[shorten >=1pt,draw=black!70, -{>[scale=1]}, node distance=\layersep]
  \tikzstyle{function}=[shorten >=1pt,thick, dash pattern={on 3pt off 2pt}, draw=black!70, -{>[scale=1]}]
  \tikzstyle{neuron}=[draw,circle,thick,fill=white,minimum size=\nnnodesize]
  \tikzstyle{x neuron}=[neuron, draw=dark-red, fill=dark-red!10]
  \tikzstyle{hidden neuron}=[neuron, draw=black]
  \tikzstyle{final neuron}=[neuron, draw=dark-green, fill=dark-green!10]
  \tikzstyle{x label} = [text=dark-red]
  \tikzstyle{hidden label} = [text=black]
  \tikzstyle{final label} = [text=dark-green]
  \tikzstyle{theta label} = [text=dark-blue]
  \tikzstyle{box} = [draw=dark-blue,thick,fill=dark-blue!10]

  \foreach \shift / \name in {-2,...,0} {
    \draw[box, xshift= - \shift * \nnboxshift, yshift= \shift * \nnboxshift]
    (- \nnlabelsep - \nnbordersep , -2cm - \nnbordersep - 0.5*\nnnodesize)
    -- (4*\nnlayersep + \nnlabelsep + \nnbordersep, -2cm - \nnbordersep - 0.5*\nnnodesize)
    -- (4*\nnlayersep + \nnlabelsep + \nnbordersep, 2cm + \nnbordersep + 0.5*\nnnodesize)
    -- (- \nnlabelsep - \nnbordersep, 2cm + \nnbordersep + 0.5*\nnnodesize) --cycle;
  }

  \foreach \name / \y in {0,...,3}
  \node[x neuron] (x-\name) at (0,1.5cm -\y cm) {};

  \foreach \name / \y in {0,...,4}
  \node[hidden neuron] (h1-\name) at (\nnlayersep, 2 cm - \y cm) {}; 

  \foreach \name / \y in {0,...,4}
  \node[hidden neuron] (h2-\name) at (2*\nnlayersep,2 cm - \y cm) {};

  \node[hidden neuron] (logr) at (3*\nnlayersep, 0) {};

  \node[final neuron] (s) at (4*\nnlayersep, 0.75cm) {}; 
  \node[final neuron] (r) at (4*\nnlayersep, -0.75cm) {};

  \foreach \source in {0,...,3}
  \foreach \dest in {0,...,4}
  \path[connection] (x-\source) edge (h1-\dest); 

  \foreach \source in {0,...,4}
  \foreach \dest in {0,...,4}
  \path[connection] (h1-\source) edge (h2-\dest);

  \foreach \source in {0,...,4}
  \path[connection] (h2-\source) edge (logr); 

  \path[function] (logr) edge (s);
  \path[function] (logr) edge (r);

  \node[x label] at (-\nnlabelsep , 0) {$x$}; 
  \node[hidden label] at (3.*\nnlayersep, - \nnlabelsep) {$\log \hat{r}$}; 
  \node[final label] at (4*\nnlayersep + \nnlabelsep, 0.75 cm) {$\hat{s}$};
  \node[final label] at (4*\nnlayersep + \nnlabelsep, -0.75 cm) {$\hat{r}$};
  \node[theta label, xshift=-1cm, yshift=0.3cm] at  (4*\nnlayersep + \nnlabelsep + \nnbordersep, - 2cm - \nnbordersep - 0.5*\nnnodesize)
  {for each $\theta_0$};
\end{tikzpicture} \\[0.5cm]


\begin{tikzpicture}[]
  \tikzstyle{connection}=[shorten >=1pt,draw=black!70, -{>[scale=1]}, node distance=\nnlayersep]
  \tikzstyle{function}=[shorten >=1pt,thick, dash pattern={on 3pt off 2pt}, draw=black!70, -{>[scale=1]}]
  \tikzstyle{neuron}=[draw,circle,thick,fill=white,minimum size=\nnnodesize]
  \tikzstyle{x neuron}=[neuron, draw=dark-red, fill=dark-red!10]
  \tikzstyle{theta neuron}=[neuron, draw=dark-blue, fill=dark-blue!10]
  \tikzstyle{hidden neuron}=[neuron, draw=black]
  \tikzstyle{final neuron}=[neuron, draw=dark-green, fill=dark-green!10]
  \tikzstyle{x label} = [text=dark-red]
  \tikzstyle{hidden label} = [text=black]
  \tikzstyle{final label} = [text=dark-green]
  \tikzstyle{theta label} = [text=dark-blue]

  \foreach \name / \y in {0,...,1}
  \node[theta neuron] (theta-\name) at (0,2.75cm -\y cm) {};

  \foreach \name / \y in {0,...,3}
  \node[x neuron] (x-\name) at (0,0.25cm -\y cm) {};

  \foreach \name / \y in {0,...,4}
  \node[hidden neuron] (h1-\name) at (\nnlayersep, 1.5 cm - \y cm) {}; 

  \foreach \name / \y in {0,...,4}
  \node[hidden neuron] (h2-\name) at (2*\nnlayersep,1.5 cm - \y cm) {};

  \node[hidden neuron] (logr) at (3*\nnlayersep, -0.5cm) {};

  \node[final neuron] (t-0) at (4*\nnlayersep, 2cm) {}; 
  \node[final neuron] (t-1) at (4*\nnlayersep, 1cm) {}; 
  \node[final neuron] (s) at (4*\nnlayersep, -0.5cm) {}; 
  \node[final neuron] (r) at (4*\nnlayersep, -2) {}; 

  \foreach \source in {0,...,1}
  \foreach \dest in {0,...,4}
  \path[connection] (theta-\source) edge (h1-\dest); 

  \foreach \source in {0,...,3}
  \foreach \dest in {0,...,4}
  \path[connection] (x-\source) edge (h1-\dest); 

  \foreach \source in {0,...,4}
  \foreach \dest in {0,...,4}
  \path[connection] (h1-\source) edge (h2-\dest);

  \foreach \source in {0,...,4}
  \path[connection] (h2-\source) edge (logr); 

  \path[function] (logr) edge (s);
  \path[function] (logr) edge (r);
  \path[function] (logr) edge (t-0); 
  \path[function] (logr) edge (t-1);

  \path[function] (theta-0) edge[bend left=10] (t-0);
  \path[function] (theta-1) edge[bend left=20] (t-1);

  \node[theta label] at (-\nnlabelsep , 2.25 cm) {$\theta_0$}; 
  \node[x label] at (-\nnlabelsep , -1.25 cm) {$x$}; 
  \node[hidden label] at (3.*\nnlayersep, -0.5 cm - \nnlabelsep) {$\log \hat{r}$}; 
  \node[final label] at (4*\nnlayersep + \nnlabelsep, -0.5 cm) {$\hat{s}$};
  \node[final label] at (4*\nnlayersep + \nnlabelsep, -2 cm) {$\hat{r}$}; 
  \node[final label] at (4*\nnlayersep + \nnlabelsep,  1.5 cm) {$\hat{t}$}; 
\end{tikzpicture}\\[0.5cm]


\begin{tikzpicture}[]
  \tikzstyle{connection}=[shorten >=1pt,draw=black!70, -{>[scale=1]}, node distance=\layersep]
  \tikzstyle{function}=[shorten >=1pt,thick, dash pattern={on 3pt off 2pt}, draw=black!70, -{>[scale=1]}]
  \tikzstyle{neuron}=[draw,circle,thick,fill=white,minimum size=\nnnodesize]
  \tikzstyle{x neuron}=[neuron, draw=dark-red, fill=dark-red!10]
  \tikzstyle{weight neuron} = [neuron, draw=black, fill=black!10]
  \tikzstyle{theta neuron}=[neuron, draw=dark-blue, fill=dark-blue!10]
  \tikzstyle{hidden neuron}=[neuron, draw=black]
  \tikzstyle{final neuron}=[neuron, draw=dark-green, fill=dark-green!10]
  \tikzstyle{x label} = [text=dark-red]
  \tikzstyle{hidden label} = [text=black]
  \tikzstyle{final label} = [text=dark-green]
  \tikzstyle{theta label} = [text=dark-blue]
  \tikzstyle{morphing label} = [text=dark-orange]
  \tikzstyle{box} = [draw=dark-orange,thick,fill=dark-orange!10]

  \foreach \name / \y in {0,...,1}
  \node[theta neuron] (theta-\name) at (0,2.75cm -\y cm) {};

  \foreach \name / \y in {0,...,3}
  \node[x neuron] (x-\name) at (0,0.25cm -\y cm) {}; 

  \node[hidden neuron] (logr) at (4*\nnlayersep, 0) {};

  \node[final neuron] (t-0) at (5*\nnlayersep, 2cm) {}; 
  \node[final neuron] (t-1) at (5*\nnlayersep, 1cm) {}; 
  \node[final neuron] (s) at (5*\nnlayersep, -0.5cm) {}; 
  \node[final neuron] (r) at (5*\nnlayersep, -2) {}; 

  \foreach \shift in {-2,...,0} {
    \draw[box, xshift= - \shift * \nnboxshift, yshift= \shift * \nnboxshift]
    (\nnlayersep - 0.5*\nnnodesize - \nnbordersep, 2.25 cm + \nnbordersep + \nnlabelsep) 
    --(3*\nnlayersep + 0.5*\nnnodesize + \nnbordersep, 2.25 cm + \nnbordersep + \nnlabelsep)
    -- (3*\nnlayersep + 0.5*\nnnodesize + \nnbordersep, -3.25 cm - 0.5*\nnnodesize - \nnbordersep)
    -- (\nnlayersep - 0.5*\nnnodesize - \nnbordersep, -3.25 cm - 0.5*\nnnodesize - \nnbordersep)
    -- cycle;
  }

  \path[function] (theta-0) edge[bend left=17] (t-0);
  \path[function] (theta-1) edge[bend left=0] (t-1);

  \foreach \name / \y in {0,...,4}
  \node[hidden neuron] (h1-\name) at (\nnlayersep, 0.75 cm - \y cm) {}; 

  \foreach \name / \y in {0,...,4}
  \node[hidden neuron] (h2-\name) at (2*\nnlayersep, 0.75 cm - \y cm) {}; 

  \node[weight neuron] (wi) at (2*\nnlayersep, 2.25 cm) {}; 

  \node[hidden neuron] (logri) at (3*\nnlayersep, -1.25 cm) {}; 

  \foreach \source in {0,...,1}
  \path[function] (theta-\source) edge (wi); 

  \foreach \source in {0,...,3}
  \foreach \dest in {0,...,4}
  \path[connection] (x-\source) edge (h1-\dest); 

  \foreach \source in {0,...,4}
  \foreach \dest in {0,...,4}
  \path[connection] (h1-\source) edge (h2-\dest);

  \foreach \source in {0,...,4}
  \path[connection] (h2-\source) edge (logri); 

  \path[function] (logri) edge (logr); 
  \path[function] (wi) edge (logr); 

  \path[function] (logr) edge (s);
  \path[function] (logr) edge (r);
  \path[function] (logr) edge (t-0); 
  \path[function] (logr) edge (t-1);

  \node[x label] at (-\nnlabelsep , -1.25 cm) {$x$}; 
  \node[theta label] at (-\nnlabelsep , 2.25 cm) {$\theta_0$}; 
  \node[hidden label] at (2*\nnlayersep, 2.25cm + 0.9 * \nnlabelsep) {$w_c$}; 
  \node[hidden label] at (3*\nnlayersep, -1.25 cm - \nnlabelsep) {$\log \hat{r}_c$}; 
  \node[hidden label] at (4*\nnlayersep, - \nnlabelsep) {$\log \hat{r}$}; 
  \node[final label] at (5*\nnlayersep + \nnlabelsep, -2 cm) {$\hat{r}$};
  \node[final label] at (5*\nnlayersep + \nnlabelsep, -0.5 cm) {$\hat{s}$}; 
  \node[final label] at (5*\nnlayersep + \nnlabelsep,  1.5 cm) {$\hat{t}$}; 
  \node[morphing label, xshift=-0.9cm, yshift=0.3cm] at (3*\nnlayersep + 0.5*\nnnodesize + \nnbordersep, -3.25 cm - 0.5*\nnnodesize - \nnbordersep)
  {for each $c$};
\end{tikzpicture}
  \caption{Schematic neural network architectures for point-by-point (top), agnostic parameterized (middle), and morphing-aware parameterized (bottom) estimators. Solid lines denote dependencies with learnable weights, dashed lines show fixed functional dependencies.}
  \label{fig:neural_network_architecture}
\end{figure}

With the exception of the simple histogram and AFC methods, all strategies rely on a classifier $\hat{s}(x | \theta_0, \theta_1)$, score regressor $\hat{t}(x | \theta_0)$, or ratio regressor $\hat{r}(x | \theta_0, \theta_1)$ that is being learnt from training data. For our explicit example, we implement these functions as fully connected neural networks:
\begin{itemize}
\item In the point-by-point setup, the neural networks take the features $x$ as input and models $\log \hat{r}(x|\theta_0, \theta_1)$. For ratio regression, this is exponentiated to yield the final output $\hat{r}(x|\theta_0, \theta_1)$. In the \carl strategy, the network output is transformed to a decision function
  \begin{equation}
    \hat{s}(x|\theta_0, \theta_1) = \frac 1 {1 + \hat{r}(x|\theta_0, \theta_1)} \,.
    \label{eq:s_from_r}
  \end{equation}
\item In the agnostic parameterized setup, the neural networks take both the features $x$ as well as the numerator parameter $\theta_0$ as input and model $\log \hat{r}(x|\theta_0, \theta_1)$. In addition to the same subsequent transformations as in the point-by-point case, taking the gradient of the network output gives the estimator score.
\item In the morphing-aware setup, the estimator takes both $x$ and $\theta_0$ as input. The features $x$ are fed into a number of independent networks, one for each basis component $c$, that model the basis ratios $\log r_c(x)$. From $\theta_0$ the estimator calculates the component weights $w_c(\theta_0)$ analytically. The components are then combined with Eq.~\eqref{eq:morphing_aware}. For the \carl approaches, the output is again transformed with Eq.~\eqref{eq:s_from_r}, and for the score-based strategies the gradient of the output is calculated.
\end{itemize}
We visualize these architectures in Fig.~\ref{fig:neural_network_architecture}.

All networks are implemented in shallow, regular, and deep versions with 2, 3, and 5 hidden layers of 100 units each and $\tanh$ activation functions. They are trained with the \toolfont{Adam} optimizer~\cite{adam} over 50 epochs with early stopping and learning rate decay. We implement them in \toolfont{keras}~\cite{chollet2015keras} with a \toolfont{TensorFlow}~\cite{tensorflow} backend. Experiments modeling $s$ rather than $\log r$, with different activation functions, adding dropout layers, or using other optimizers and learning rate schedules have led to a worse performance.

\subsubsection{Training samples}
\label{sec:samples}

Starting from the weighted event sample described in Sec.~\ref{sec:event_generation}, we draw events $(x_e, z_e)$ randomly with probabilities given by the corresponding $\intractablep(x_e, z_e | \theta)$. Due to the form of the likelihood $\intractablep(x,z | \theta)$ and due to technical limitations of our simulator, individual data points can carry large probabilities  $\intractablep(x_e, z_e | \theta)$, leading to duplicate events in the training samples. However, we enforce that there is no duplication between training and evaluation samples, so this limitation can only degrade the performance.

For the point-by-point setup, we choose 100 values of $\theta_0$, 5 of which are fixed at the SM ($\theta_0 = (0,0)$) and at the corners of the considered parameter space, with the remaining 95 chosen randomly with a flat prior over $-1 \leq \theta_o \leq 1$. For each of these training points we sample $250\,000$ events according to $\theta_0$ and $250\,000$ events according to the reference hypothesis
\begin{equation}
  \theta_1 = (0.393, 0.492)^T \,.
  \label{eq:denom_benchmark}
\end{equation}
For the parameterized strategies, we compare three different training samples, each consisting of $10^7$ events:
\begin{description}
  \item[Baseline]
    For 1000 values of $\theta_0$ chosen randomly in $\theta$ space, we draw $5000$ events according to $\theta_0$ and $5000$ events according to $\theta_1$ given in Eq.~\eqref{eq:denom_benchmark}.
  \item[Random $\theta$]
    In this sample,  the value of $\theta_0$ is drawn randomly independently for each event. Again we use a flat prior over $\theta_0 \in [-1,1]^2$.
  \item[Morphing basis]
    For each of the 15 basis hypotheses $\theta_i$ from the morphing procedure, we generate $333\,000$ events according to $\theta_0 = \theta_i$ and  $333\,000$ according to $\theta_1$.
\end{description}

Finally, for the local score regression model we use a sample of $10^7$ events drawn according to the SM.

Our evaluation sample consists of $50\,000$ events drawn according to the SM. We evaluate the likelihood ratio for these events for a total of 1016 values of $\theta_0$, 1000 of which are the same as those used in the baseline training sample. Again we fix $\theta_1$ as in Equation~\eqref{eq:denom_benchmark}.

Each event is characterized by 42 features:
\begin{itemize}
  \item the energies, transverse momenta, azimuthal angles, and pseudo-rapidities of all six particles in the final state;
  \item the energies, transverse momenta, azimuthal angles, pseudo-rapidities, and invariant mass of the four-lepton system as well as the two-lepton systems that reconstruct the two $Z$ bosons; and
  \item the invariant mass, separation in pseudorapidity, and separation in azimuthal angle of the di-jet system.
\end{itemize}
The derived variables in the feature set help the neural networks pick up the relevant features faster, though we did not find that their choice affects the performance significantly.

\subsubsection{Calibration and density estimation}

We calibrate the classifiers for our example process with probability calibration as described in Sec.~\ref{sec:probability_calibration}. We determine the calibration function $C(r)$ with isotonic regression~\cite{Kruskal1964}, which constrains $C(r)$ to be monotonic. Experiments with other regression techniques based on histograms, kernel density estimation, and logistic regression did not lead to a better performance. We apply the calibration point by point in $\theta_0$. It is based on an additional event sample that is independent of the training and evaluation data. The same events are used to calibrate each value of $\theta$, with an appropriate reweighting. This strategy to minimize variance is based on the availability of the parton-level likelihood function $p(z | \theta)$.

The techniques based on local score regression require the choice of a reference point to evaluate the score. For the EFT problem, the natural choice is $\theta_{\text{score}} = \theta_{\text{SM}} = (0, 0)^T$.  In the \sally approach, we perform the density estimation based on two-dimensional histograms of the estimated score at the SM, point by point in $\theta_0$. For the \sallino technique, we use a one-dimensional histograms of $\hat{h}(x | \theta_{\text{SM}})$, point by point in $\theta_0$.

\subsection{Challenges and diagnostics}
\label{sec:challenges}

\subsubsection{Uncertainties}

Even the most evolved and robust estimators will have some deviations from the true likelihood ratio, which should be taken into account in an analysis as an additional modeling uncertainty. Most of the estimators developed above converge to the true likelihood ratio in the limit of infinite training and calibration samples. But with finite statistics, there are different sources of variance that affect some strategies more than others.

Consider the traditional histogram approach. In the point-by-point version, each separate estimator $\hat{r}(x | \theta_0, \theta_1)$ is trained on the small subset of the data generated from a specific value of $\theta_0$, so the variance from the finite size of the training data, \ie the statistical uncertainty from the Monte-Carlo simulation, is large. At $\theta_0$ values between the training points, there are additional sources of uncertainty from the interpolation. On the other hand, morphing-aware histograms use all of the training data to make predictions at all points, and since the dependence on $\theta_0$ is known, the interpolation is exact. But the large morphing weights $w_c(\theta_0)$ and the cancellations between them mean that even small fluctuations in the individual basis histograms can lead to huge errors on the combined estimator.

 \begin{figure}
  \includegraphics[width=0.5\textwidth]{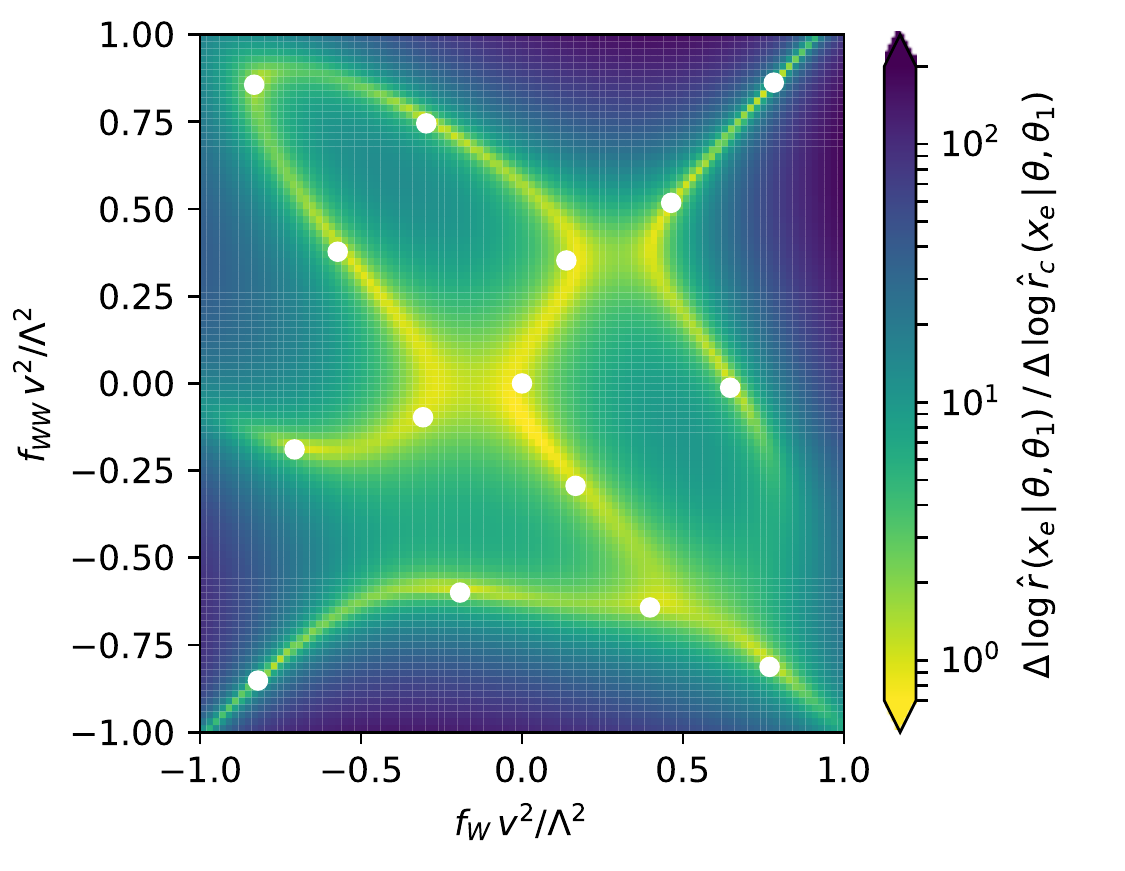}%
  \caption{Uncertainty $\Delta \log \hat{r}(x_e | \theta, \theta_1)$ of morphing-aware estimators due to uncertainties $\Delta \log \hat{r}_c(x_e | \theta, \theta_1)$ on the individual basis ratios as a function of $\theta$. We fix $\theta_1$ as in Eq.~\eqref{eq:denom_benchmark} and show one random event $x_e$, the results for other events are very similar. We assume iid Gaussian uncertainties on the $\log \hat{r}_c(x | \theta, \theta_1)$ and use Gaussian error propagation. The white dots show the position of the basis points $\theta_c$. Small uncertainties in the individual basis estimators $\hat{r}_c(x_e | \theta, \theta_1)$ are significantly increased due to the large morphing weights and can lead to large errors of the combined estimator $\hat{r}(x_e | \theta, \theta_1)$.}
  \label{fig:morphing_uncertainties}
\end{figure}

Similar patterns hold for the ML-based inference strategies. The point-by-point versions suffer from a larger variance due to small training samples at each point and interpolation uncertainties. The agnostic parameterized models have more statistics available, but have to learn the more complex full statistical model including the dependence on $\theta_0$. The morphing-aware versions make maximal use of the physics structure of the process and all the training data, but large morphing weights can dramatically increase the errors of the individual component estimators $\log \hat{r}_c(x)$. We demonstrate this for our example process in Fig.~\ref{fig:morphing_uncertainties}: in some regions of parameter space, in particular far away from the basis points, the errors on a morphing-aware estimator $\log \hat{r}(x | \theta, \theta_1)$ can easily be 100 times larger than the individual errors on the component estimators $\log \hat{r}_c(x)$. The $\theta_0$ dependence on this error depends on the choice of the basis points, this uncertainty can thus be somewhat mitigated by optimizing the basis points or by combining several different bases.

\subsubsection{Diagnostics}
\label{sec:diagnostics}

After discussing the sources of variance, let us now turn towards diagnostic tools that can help quantify the size of estimator errors and to assign a modeling uncertainty for the statistical analysis. These methods are generally closure tests: we can check the likelihood ratio estimators for some expected behaviour, and use deviations either to correct the results (as in the calibration procedures described in Sec.~\ref{sec:calibration}), to define uncertainty bands, or to discard estimators altogether. We suggest the following tests:
\begin{description}
  \item[Ensemble variance]
    Repeatedly generating new training data (or bootstrapping the same training sample) and training the estimators again gives us an ensemble of predictions $\{\hat{r}_r (x | \theta_0, \theta_1)\}$. We can use the ensemble variance as a measure of uncertainty of the prediction that is due to the variance in the training data and random seeds used during the training.

  \item[Reference hypothesis variation]
    Any estimated likelihood ratio between two hypotheses $\theta_A$, $\theta_B$
    \begin{equation}
      \hat{r}(x | \theta_A, \theta_B)  = \frac { \hat{r}(x | \theta_A, \theta_1) } { \hat{r}(x | \theta_B, \theta_1)}
    \end{equation}
    should be independent of the choice of the reference hypothesis $\theta_1$ used in the estimator $\hat{r}$. Training several independent estimators with different values of $\theta_1$ thus provides another check of the stability of the results~\cite{Cranmer:2015bka}.

Much like the renormalization and factorization scale variations that are ubiquitous in particle physics calculations, this technique does not have a proper statistical interpretation in terms of a likelihood function. We can still use it to qualitatively indicate the stability of the estimator under this hyperparameter change.

  \item[Ratio expectation]
    As discussed above, the expectation value of the estimated likelihood ratio assuming the denominator hypothesis should be very close to one. We can numerically calculate this expectation value $\hat{R}(\theta)$, see Eq.~\eqref{eq:ratio_expectation}. In Sec.~\ref{sec:expectation_calibration} we argued that this expectation value can be used to calibrate the estimators, but that this calibration can actually decrease the performance in certain situations.

    If expectation calibration is used, the calibration itself has a non-zero variance from the finite sample size used to calculate the expectation value $\hat{R}(\theta)$. As we pointed out in Sec.~\ref{sec:expectation_calibration}, we can calculate this source of statistical uncertainty, at least under the assumption of a perfect estimator with $\hat{r}(x | \theta_0, \theta_1) = \intractabler(x | \theta_0, \theta_1)$. The result given in Eq.~\eqref{eq:expectation_calibration_variance} provides us with a handle to calculate the statistical uncertainty of this calibration procedure from the finite size of the calibration sample. Note that for imperfect estimators, the variance of $R[\hat{R}]$ may be significantly larger.

    Independent of whether expectation calibration is part of the estimator, the deviation of the expectation $\hat{R} (\theta)$ from one can serve as a diagnostic tool to check for mismodelling of the estimator. We can take $\log \hat{R}(\theta)$ as a measure of the uncertainty of $\log \hat{r}(x | \theta, \theta_1)$. As in the case of the reference hypothesis variation, there is no consistent statistical interpretation of this uncertainty, but this does not mean that it is useless as a closure test.

  \item[Reweighting distributions]
    A  good estimator $\hat{r}(x | \theta_0, \theta_1)$ should satisfy
    \begin{equation}
      \intractablep(x | \theta_0) \approx \hat{r}(x | \theta_0, \theta_1) \, \intractablep(x | \theta_1) \,.
    \end{equation}
    We cannot evaluate the $\intractablep(x | \theta_0)$ to check this relation explicitly. However, we \emph{can} sample events $\{x_e \}$ from them. This provides another diagnostic tool~\cite{Cranmer:2015bka}: we can draw a first sample as $x_e \sim \intractablep(x_e | \theta_0)$, and draw a second sample as $x_e \sim \intractablep(x_e | \theta_1)$ and reweight it with $\hat{r}(x_e | \theta_0, \theta_1)$. For a good likelihood ratio estimator, the two samples should have similar distributions. This can easily be tested by training a discriminative classifier between the samples. If a classifier can distinguish between the sample from the first hypothesis and the sample drawn from the second hypothesis and reweighted with the estimated likelihood ratio, then $\hat{r} (x | \theta_0, \theta_1)$ is not a good approximation of the true likelihood ratio $r (x | \theta_0, \theta_1)$. Conversely, if the classifier cannot separate the two classes, the classifier is either not efficient, or the likelihood ratio is estimated well.
\end{description}

Note that passing these closure tests is not a guarantee for a good estimator of the likelihood ratio. In Sec.~\ref{sec:neyman} we will discuss how we can nevertheless derive exclusion limits that are guaranteed to be statistically correct, \ie that might not be optimal, but are never wrong.

In our example process, we will use a combination of the first two ideas of this list: we will create copies of estimators with independent training samples and random seeds during training, as well as with different choices of the reference hypothesis $\theta_1$, and analyse the median and envelope of the predictions.

\section{Limit setting}
\label{sec:inference}

The final objective of any EFT analysis are exclusion limits on the parameters of interest at a given confidence level. These can be derived in one of two ways. The Neyman construction based on toy experiments provides a generic and fail-safe method, we will discuss it in Sec.~\ref{sec:neyman}. But since the techniques developed in the previous section directly provide an estimate for the likelihood ratio, we can alternatively apply existing statistical methods for likelihood ratios as test statistics. This much more efficient approach will be the topic of the following section.

\subsection{Asymptotics}
\label{sec:asymptotics}

Consider the test statistics
\begin{equation}
  q(\theta) = -2 \, \sum_e \log \intractabler(x_e | \theta, \hat{\theta})
  = -2 \, \sum_e \left( \log \intractabler(x_e | \theta, \theta_1) - \log \intractabler(x_e | \hat{\theta}, \theta_1) \right)
  \label{eq:log_likelihood_ratio_test_statistics}
\end{equation}
for a fixed number $N$ of observed events $\{x_e\}$ with the maximum-likelihood estimator
\begin{equation}
  \hat{\theta} = \argmax_\theta \sum_e \log \intractabler(x_e | \theta, \theta_1)\,.
\end{equation}

In the asymptotic limit, the distribution according to the null hypothesis, $p(q(\theta) | \theta)$, is given by a chi-squared distribution. The number of degrees of freedom $k$ is equal to the number of parameters $\theta$. This result by Wilks~\cite{Wilks:1938dza} allows us to translate an observed value $q_{\text{obs}}(\theta)$ directly to a $p$-value that measures the confidence with which $\theta$ can be excluded:
\begin{equation}
  p_\theta \equiv \int_{q_{\text{obs}}(\theta)}^\infty \! \diff q \; p(q | \theta)
  = 1 - F_{\chi^2} \left(  q_{\text{obs}}(\theta) \middle| k \right)
  \label{eq:wilks}
\end{equation}
where $F_{\chi^2}(x | k)$ is the cumulative distribution function of the chi-squared distribution with $k$ degrees of freedom. In our example process $k=2$, for which this simplifies to
\begin{equation}
  p_\theta = \exp \left( - \frac {q_{\text{obs}}(\theta)} 2 \right) \,.
\end{equation}

In particle physics it is common practice to calculate ``expected exclusion contours'' by calculating the expected value of $q_{\text{obs}}(\theta)$ based on a large ``Asimov'' data set generated according to some $\theta'$~\cite{Cowan:2010js}. With Eq.~\eqref{eq:wilks} this value is then translated into an expected $p$-value.\footnote{For $k=1$, this standard procedure reproduces the median expected $p$-value. Note however that this is not true anymore for more than one parameter of interest. In this case, the median expected $p$-value can be calculated based on a different, but not commonly used, procedure. It is based on the fact that the distribution of $q$ according to an alternate hypothesis, $p(q(\theta) | \theta')$, is given by a non-central chi-squared distribution~\cite{Wald}. In the asymptotic limit, the non-centrality parameter is equal to the expectation value $\expectation [q(\theta) | \theta']$~\cite{Cowan:2010js}. This allows us to calculate for instance the median expected $q(\theta)$ assuming some value $\theta'$ based on the Asimov data. Combining all the pieces, the median expected $p$-value $p_\theta$ with which $\theta$ can be excluded under the assumption that $\theta'$ is true is then given by $p_\theta^{\text{expected from } \theta'} = 1 - F_{\chi^2} (  F_{\chi^2_{nc}}^{-1} ( \frac 1 2  | k, \expectation [q(\theta) | \theta'] )  | k )$, where $F_{\chi^2}(x | k)$ is the cumulative distribution function for the chi-squared distribution with $k$ degrees of freedom and $F_{\chi^2_{nc}}^{-1}(p | k, \Lambda)$ is the inverse cumulative distribution function for the non-central chi-squared distribution with $k$ degrees of freedom and non-centrality parameter $\Lambda$.}

In practice we cannot access the true likelihood ratio defined on the full observable space and thus also not $q(\theta)$. But if the error of an estimator $\hat{r}(x|\theta_0, \theta_1)$ compared to the true likelihood ratio is negligible, we can simply calculate
\begin{equation}
  \hat{q}(\theta) = -2 \, \sum_e \log \hat{r}(x_e | \theta, \hat{\theta})
  \label{eq:q_hat}
\end{equation}
with maximum likelihood estimator $\hat{\theta}$ also based on the estimated likelihood ratio. The $p$-value can then be read off directly from the estimator output, substituting $\hat{q}$ for $q$ in Eq.~\eqref{eq:wilks}.

Under this assumption and in the asymptotic limit, constructing confidence intervals is thus remarkably simple and computationally cheap: after training an estimator $\hat{r}(x | \theta, \theta_1)$ as discussed in the previous section, the observed events $\{x_e \}$ are fed into the estimator for each value of $\theta$ on some parameter grid. From the results we can read off the maximum likelihood estimator $\hat{q}(\theta)$ and calculate the observed value of the test statistics $\hat{q}(\theta)$ for each $\theta$. Equation~\eqref{eq:wilks} then translates these values to $p$-values, which can then be interpolated between the tested $\theta$ points to yield the final contours.

To check whether these asymptotic properties apply to a likelihood ratio estimator, we can use the diagnostic tools discussed in Sec.~\ref{sec:diagnostics}. In addition, we can explicitly check whether the distribution of $q(\theta)$ actually follows a chi-squared distribution by generating toy experiments for a few $\theta$ points. If it does, the asymptotic results are likely to apply at other points in parameter space as well. If the variance of the toy experiments is larger than expected from the chi-squared distribution, the residual variance may be taken as an error estimate on the estimator prediction.

\subsection{Neyman construction}
\label{sec:neyman}

Rather than relying on the asymptotic properties of the likelihood ratio test, we can construct the distribution of a test statistic with toy experiments. This is computationally more expensive, but useful if the number of events is not in the asymptotic regime or if the uncertainty of the estimators cannot be reliably quantified. Constraints derived in this way are conservative: even if the likelihood ratio is estimated poorly, the resulting contours might not be optimal, but they are never wrong (at a specified confidence level).

A good choice for the test statistics is the estimated profile log likelihood ratio $\hat{q}(\theta)$ given in Eq.~\eqref{eq:q_hat}, which allows us to compare the distribution of the toy experiments directly to the asymptotic properties discussed in the previous section. However, its construction requires finding the maximum likelihood estimator for every toy experiment. This increases the necessary computation time substantially, especially in high-dimensional parameter spaces. An alternative test statistics is the estimated log likelihood ratio with respect to some fixed hypothesis, which need not be identical to the reference denominator $\theta_1$ used in the likelihood ratio estimators. In the EFT approach, the natural choice is the estimated likelihood ratio with respect to the SM,
\begin{equation}
  \hat{q}'(\theta) \equiv -2 \, \sum_e \log \hat{r}(x_e | \theta, \theta_{SM}) \,.
\end{equation}
Using this test statistic rather than the profile likelihood ratio defined in Eq.~\eqref{eq:log_likelihood_ratio_test_statistics} is expected to lead to stronger constraints if the true value of $\theta$ is close to the SM point, as expected in the EFT approach, and less powerful bounds if the true value is substantially different from the SM.

In practice we can efficiently calculate the distribution of $\hat{q}^{(\prime)}(\theta)$ after $n$ events by first calculating the distribution of $\hat{q}^{(\prime)} (\theta)$ for one event and convolving the result with itself $(n-1)$ times.

\subsection{Nuisance parameters}
\label{sec:nuisance}

The tools developed above also support nuisance parameters, for instance to model systematic uncertainties in the theory calculation or the detector model. One strategy is to train parameterized estimators on samples generated with different values of the nuisance parameters $\nu$ and let them learn the likelihood ratio
\begin{equation}
  \hat{r}(x| \theta_0, \theta_1; \nu_0, \nu_1) \equiv \frac {\hat{p}(x | \theta_0; \nu_0)} {\hat{p}(x | \theta_1; \nu_1)}
\end{equation}
with its dependence on the nuisance parameters. As test statistics we can then use the estimator version of the usual profile log likelihood ratio,
\begin{equation}
  \hat{q}(\theta)
  = -2 \, \sum_e \log \left[ r\!\left(x_e \middle| \theta, \hat{\theta}; \hhat{\nu}, \hat{\nu}\right)
  \frac {q(\hhat{\nu})} {q(\hat{\nu}) \vphantom{\left(\hat{\theta}\right)}} \right]
\end{equation}
with constraint terms $q(\nu)$,
\begin{align}
  \hhat{\nu} &= \argmax_\nu \sum_e \log \left[ \hat{r}(x_e | \theta, \theta_1; \nu, \nu_1) \frac {q(\nu)} {q(\nu_1)} \right] \,, \quad \text{and} \\
  \left(\hat{\theta}, \hat{\nu}\right) &= \argmax_{(\theta,\nu)} \sum_e \log \left[ \hat{r}(x_e | \theta, \theta_1; \nu, \nu_1) \frac {q(\nu)} {q(\nu_1)} \right] \,.
\end{align}
The profile log likelihood ratio has two advantages: it is pivotal, \ie its value and its distribution do not depend on the value of the nuisance parameter, and it has the asymptotic properties discussed in Sec.~\ref{sec:asymptotics}.

Similarly, we can train the score including nuisance parameters,
\begin{equation}
  \hat{t}(x | \theta_0; \nu_0) = \nabla_{(\theta, \nu)} \log \left[ \hat{p}(x|\theta; \nu) q(\nu) \right] \Biggr|_{\theta_0, \nu_0} \,.
\end{equation}
If the constraints $q(\nu)$ limit the nuisance parameters to a relatively small region around some $\nu_0$, \ie a range in which the shape of the likelihood function does not change significantly, the \sally and \sallino methods seem particularly appropriate.

Finally, an adversarial component in the training procedure lets us directly train pivotal estimators $\hat{r}(x | \theta_0, \theta_1)$, \ie that do not depend on the value of the nuisance parameters~\cite{Louppe:2016ylz}. Compared to learning the explicit dependence on $\nu$, this can dramatically reduce the dimensionality of the parameter space as early as possible, and does not require manual profiling. However, the estimators will generally not converge to the profile likelihood ratio, so its asymptotic properties do not apply and limit setting requires the Neyman construction.

\section{Results}
\label{sec:results}

We now apply the analysis techniques to our example process of WBF Higgs production in the $4\ell$ decay mode. We first study the idealized setup discussed in Sec.~\ref{sec:event_generation}, in which we can assess the techniques by comparing their predictions to the true likelihood ratio. In Sec.~\ref{sec:results_smearing} we then calculate limits in a more realistic setup.

\subsection{Idealized setup}
\label{sec:results_idealized}

\subsubsection{Quality of likelihood ratio estimators}

\begin{table}
  \begin{tabular}{ll rr c}
    \toprule
    \multirow{2}{*}{Strategy} & \multirow{2}{*}{Setup} & \multicolumn{2}{c}{Expected MSE} & \multirow{2}{*}{Figures} \\
    \cmidrule{3-4}
    && All & Trimmed \\
    \midrule
   Histogram & $p_{T,j1}, \Delta \phi_{jj}$ & $\mathbf{0.056}$ & $0.0106$ & $\yep$ \\
    & $p_{T,j1}$ & $0.088$ & $0.0230$ \\
    & $\Delta \phi_{jj}$ & $0.160$ & $0.0433$ \\
   \afc & $p_{T,j1}, \Delta \phi_{jj}$ & $0.059$ & $\mathbf{0.0091}$ \\
    & $p_{T,j1}, m_{Z2}, m_{jj}, \Delta \eta_{jj}, \Delta \phi_{jj}$ & $0.078$ & $0.0101$\\
   \midrule
   \carl (PbP) & PbP & $0.030$ & $0.0111$ & Fig.~\ref{fig:pbp_parameterized_aware}\\
   \carl (parameterized) & Baseline & $0.012$ & $\mathbf{0.0026}$ & $\yep$ \\
    & Random $\theta$ & $\mathbf{0.012}$ & $0.0028$ \\
   \carl (morphing-aware) & Baseline & $0.076$ & $0.0200$ & Fig.~\ref{fig:pbp_parameterized_aware}\\
    & Random $\theta$ & $0.086$ & $0.0226$ \\
    & Morphing basis & $0.156$ & $0.0618$ \\
   \midrule
   \rolr (PbP) & PbP &$0.005$ & $0.0022$ \\
   \rolr (parameterized) & Baseline & $0.003$ & $0.0017$ & $\yep$ \\
    & Random $\theta$ & $\mathbf{0.003}$ & $\mathbf{0.0014}$\\
   \rolr (morphing-aware) & Baseline & $0.024$ & $0.0063$ \\
    & Random $\theta$ & $0.022$ & $0.0052$ \\
    & Morphing basis & $0.130$ & $0.0485$ \\
   \midrule
   \sally && $\mathbf{0.013}$ & $\mathbf{0.0002}$ & $\yep$\\
   \sallino && $0.021$ & $0.0006$ \\
   \midrule
   \cascal (parameterized) & Baseline & $\mathbf{0.001}$ & $\mathbf{0.0002}$ & $\yep$ \\
    & Random $\theta$ & $0.001$ & $0.0002$\\
   \cascal (morphing-aware) & Baseline & $0.136$ & $0.0427$ \\
    & Random $\theta$ & $0.092$ & $0.0268$ \\
    & Morphing basis & $0.040$ & $0.0081$ \\
   \midrule
   \rascal (parameterized) & Baseline & $0.001$ & $0.0004$ & $\yep$ \\
    & Random $\theta$ & $\mathbf{0.001}$ & $\mathbf{0.0004}$\\
   \rascal (morphing-aware) & Baseline & $0.125$ & $0.0514$\\
    & Random $\theta$ & $0.132$ & $0.0539$ \\
    & Morphing basis & $0.031$ & $0.0072$\\
    \bottomrule
  \end{tabular}
  \caption{Performance of the different likelihood ratio estimation techniques in our example process. The metrics shown are the expected mean squared error on the log likelihood ratio with and without trimming, as defined in the text. Checkmarks in the last column denotes estimators shown in the following figures. Here we only give results based on default settings, which are defined in Appendix~\ref{sec:appendix_models}. An extended list of results that covers more estimators is given in Appendix~\ref{sec:appendix_results}.}
  \label{tbl:results}
\end{table}

Table~\ref{tbl:results} summarizes the performance of the different likelihood ratio estimators in the idealized setup. For 50\,000 events $\{x_e\}$ drawn according to the SM, we evaluate the true likelihood ratio $r(x_e| \theta_0, \theta_1)$ as well as the estimated likelihood ratios $\hat{r}(x_e| \theta_0, \theta_1)$ for 1000 values of $\theta_0$ sampled randomly in $[-1,1]^2$. As a metric we use the expected mean squared error on the log likelihood ratio
\begin{equation}
  \varepsilon[\hat{r}(x)] = \sum_{\theta_0} \pi(\theta_0) \; \frac 1 N \, \sum_e \left[ \Bigl(\log \hat{r}(x_e| \theta_0, \theta_1) - \log \intractabler(x_e| \theta_0, \theta_1)\Bigr)^2 \right] \,.
  \label{eq:expected_trimmed_mse}
\end{equation}
The 1000 tested values of $\theta_0$ are weighted with a Gaussian prior
\begin{equation}
  \pi(\theta) = \frac 1 Z \, \normal { ||\theta|| } {0, 2 \cdot 0.2^2}
  \label{eq:prior}
\end{equation}
with normalization factor $Z$ such that $\sum_{\theta_0} \pi(\theta) = 1$. In addition we show the expected trimmed mean squared error, which truncates the top 5\% and bottom 5\% of events for each $\theta$. This allows us to analyse the quality of the estimators for the bulk of the phase space without being dominated by a few outliers. In Table~\ref{tbl:results} and in the figures of this section, we only show results for a default set of hyperparameters for each likelihood ratio estimators. These default setups are defined in Appendix~\ref{sec:appendix_models}. Results for other hyperparameter choices are given in Appendix~\ref{sec:appendix_results}.

\begin{figure}
  \includegraphics[width=\textwidth]{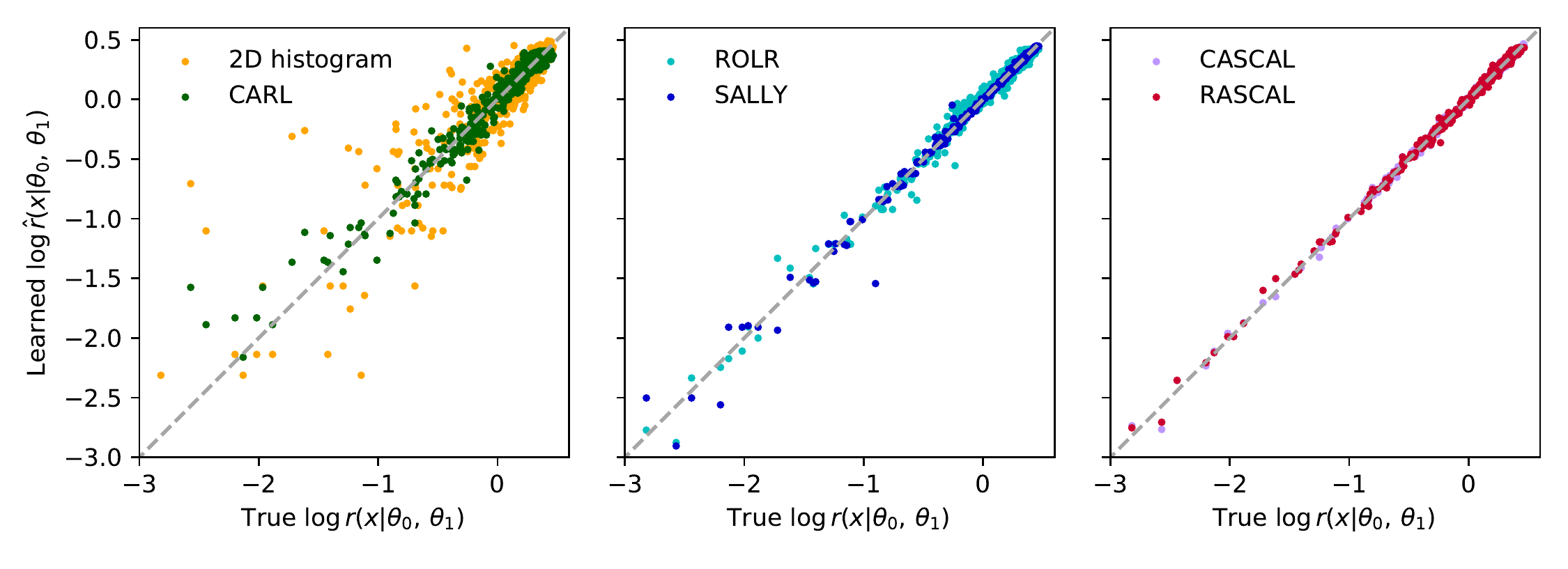}%
  \caption{True vs.\ estimated likelihood ratios for a benchmark hypothesis $\theta_0 = (-0.5, -0.5)^T$. Each dot corresponds to one event $x_e$. The \cascal (right, red), \rascal (right, orange), and \sally (middle, blue) techniques can predict the likelihood ratio extremely accurately over the whole phase space. All new techniques clearly lead to more precise estimates than the traditional histogram approach (left, orange).}
  \label{fig:r_scatter}
\end{figure}

\begin{figure}
  \includegraphics[width=\textwidth]{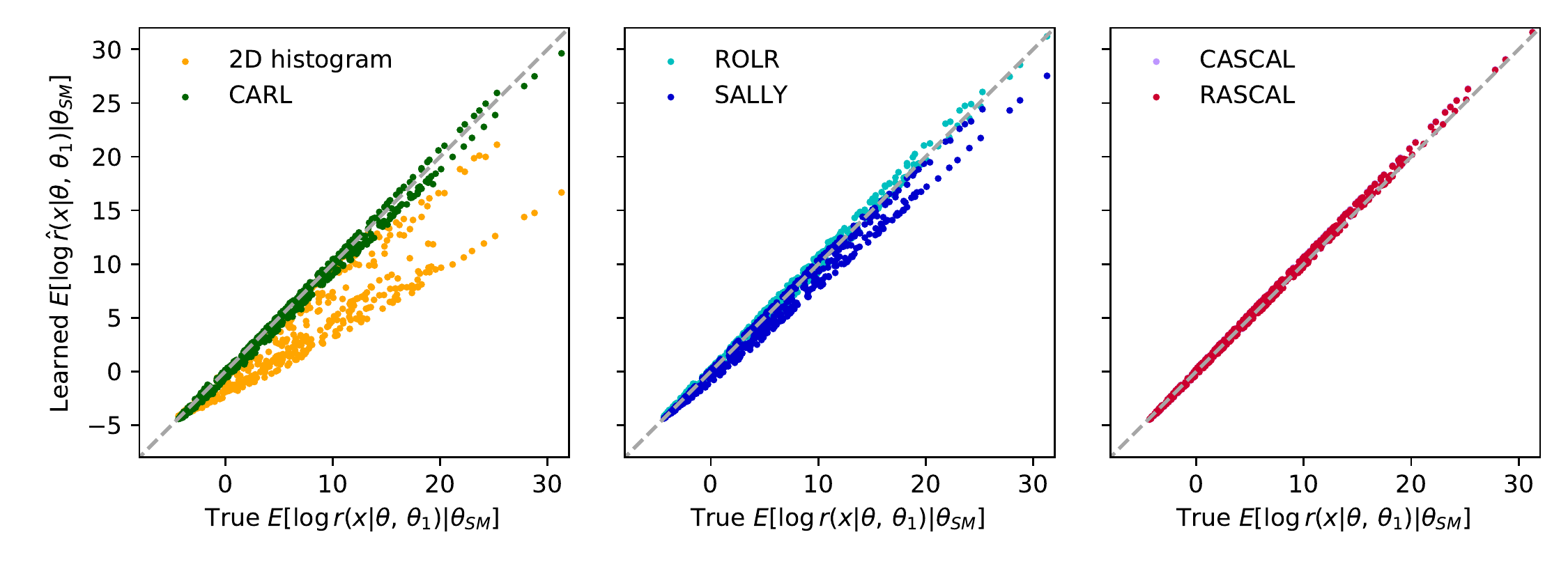}%
  \caption{True vs.\ estimated expected log likelihood ratio. Each dot corresponds to one value of $\theta_0$, where we take the expectation over $x \sim \intractablep(x | \theta_{\text{SM}})$. The new techniques are less biased than the histogram approach (left, orange).}
  \label{fig:expected_log_r_scatter}
\end{figure}

The best results come from parameterized estimators that combine either a classifier decision function or ratio regression with regression on the score: the \cascal and \rascal strategies provide very accurate estimates of the log likelihood ratio. \sally, parameterized \rolr, and parameterized \carl perform somewhat worse. For \carl and \rolr, parameterized estimators consistently perform better then the corresponding point-by-point versions. All these ML-based strategies significantly outperform the traditional one- or two-dimensional histograms and the Approximate Frequentist Computation.

The morphing-aware versions of the parameterized estimators lead to a poor performance, comparable or worse than the two-dimensional histogram approach. As anticipated in Sec.~\ref{sec:challenges}, the large weight factors and the sizable cancellations between them blow up small errors on the estimation of the individual basis estimators $\hat{r}_i(x)$ to large errors on the combined estimator.

We find that probability calibration as discussed in Sec.~\ref{sec:probability_calibration} improves the results in almost all cases, in particular for the \carl method. An additional step of expectation calibration (see Sec.~\ref{sec:expectation_calibration}) after the probability calibration does not lead to a further improvement, and in fact often increases the variance of the estimator predictions. We therefore only use probability calibration for the results presented here.

The choice of the training sample is less critical, with nearly identical results between the baseline and random $\theta$ samples. For the \carl approach, shallow networks with two hidden layers perform better, while \rolr works best for three hidden layers and the score-based strategies benefit from a deeper network with five hidden layers.

In Fig.~\ref{fig:r_scatter} we show scatter plots between the true and estimated likelihood ratios for a fixed hypothesis $\theta_0$. The likelihood ratio estimate from histograms of observables is widely spread around the true likelihood ratio, reflecting the loss of information from ignoring most directions in the observable space. \carl performs clearly better. \rolr and \sally offer a further improvement. Again, the best results come from the \cascal and \rascal strategies, both giving predictions that are virtually one-to-one with the true likelihood ratio.

We go beyond a single benchmark point $\theta_0$ in Fig.~\ref{fig:expected_log_r_scatter}. This scatter plot compares true and estimated likelihood ratios for different values of $\theta_0$, taking the expectation value over $x$. We find that the \carl, \rolr, \cascal, and \rascal approaches converge to the correct likelihood ratio in this expectation value. For the \sally and \sallino techniques we find larger deviations, pointing towards the breakdown of the local model approximation. Much more obvious is the loss of information in the traditional histogram approach, which is clearly not asymptotically exact.

\begin{figure}
  \includegraphics[width=0.89\textwidth]{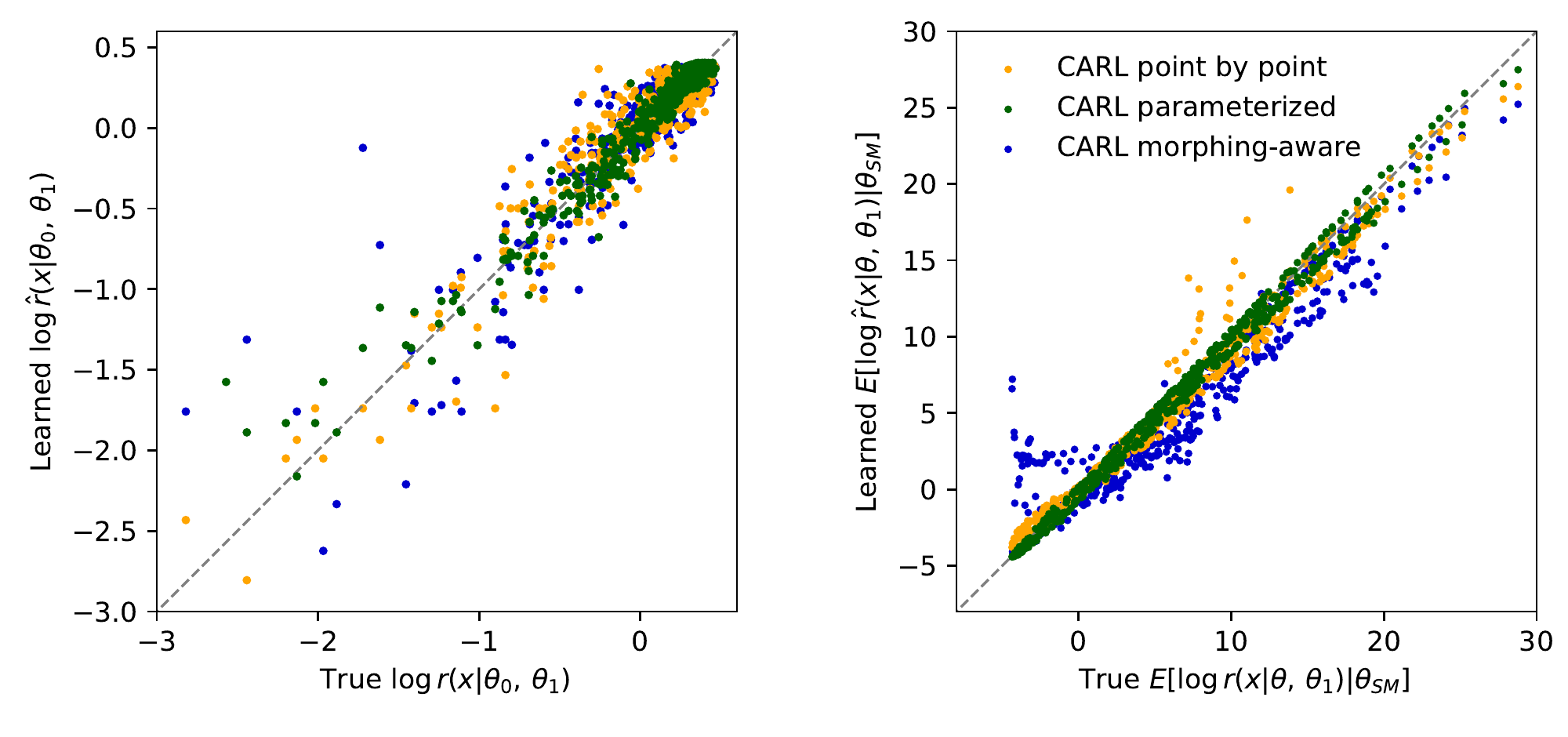}%
  \caption{Comparison of the point-by-point, parameterized, and morphing-aware versions of \carl. Top left: True vs.\ estimated likelihood ratios for a benchmark hypothesis $\theta_0 = (-0.5, -0.5)^T$, as in Fig.~\ref{fig:r_scatter}. Each dot corresponds to one event $x_e$. Top right: True vs.\ estimated expected log likelihood ratio, as in Fig.~\ref{fig:expected_log_r_scatter}. Each dot corresponds to one value of $\theta_0$, where we take the expectation over $x \sim p(x | \theta_{\text{SM}})$. The parameterized estimator outperforms the point-by-point one and particularly the morphing-aware version.}
  \label{fig:pbp_parameterized_aware}
\end{figure}

The point-by-point, agnostic parameterized, and morphing-aware versions of the \carl strategy are compared in Fig.~\ref{fig:pbp_parameterized_aware}. As expected from Table~\ref{tbl:results}, the parameterized strategy performs better than the point-by-point version, and both are clearly superior to the morphing-aware estimator.

\subsubsection{Efficiency and speed}

\begin{figure}
  \includegraphics[width=0.89\textwidth]{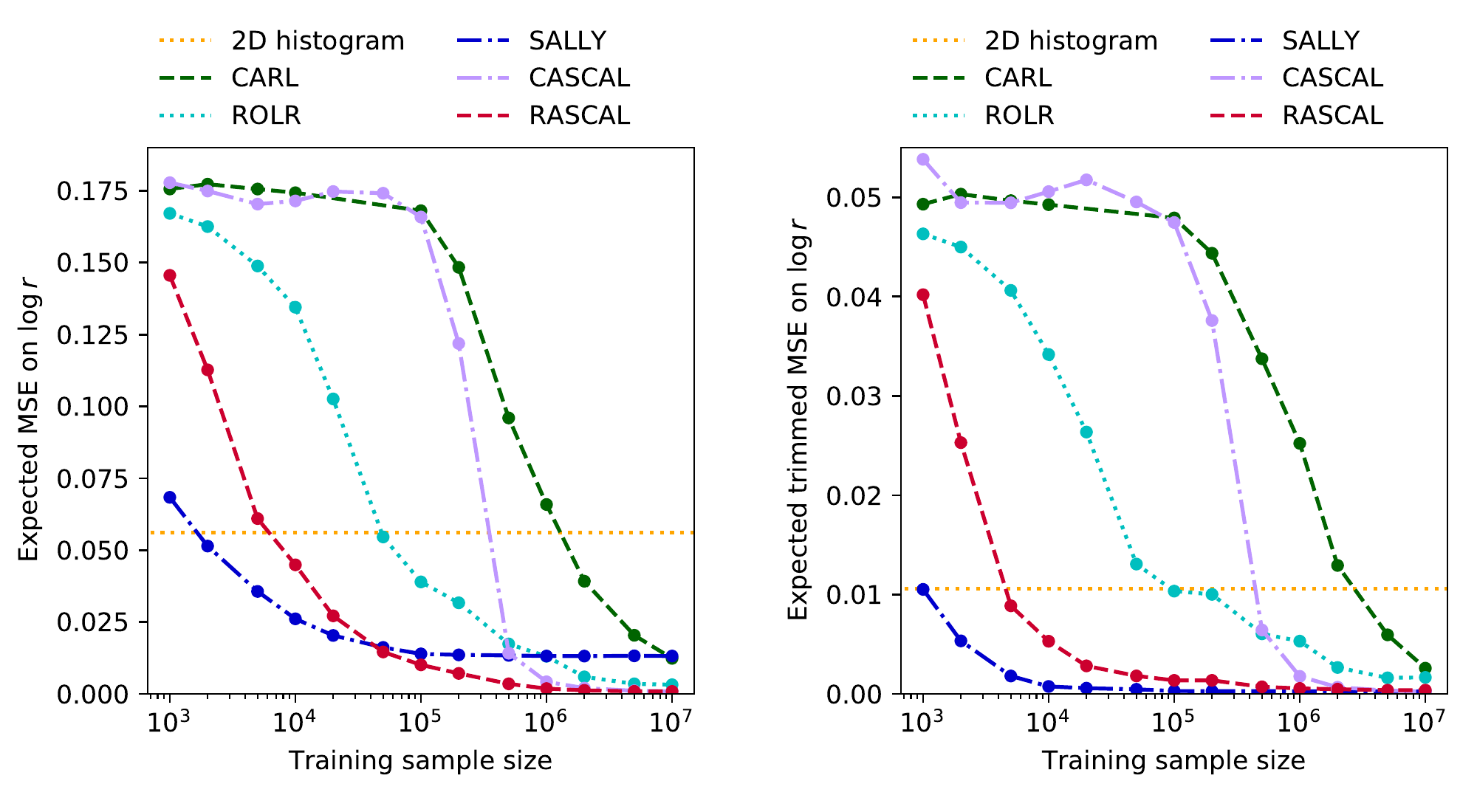}%
  \caption{Performance of the techniques as a function of the training sample size. As a metric, we show the mean squared error (left) and trimmed mean squared error on $\log \intractabler(r | \theta_0, \theta_1)$ weighted with a Gaussian prior, as discussed in the text. Note that we do not vary the size of the calibration data samples. The number of epochs are increased such that the number of epochs times the training sample size is constant, all other hyperparameters are kept constant. The \sally method works well even with very little data, but plateaus eventually due to the limitations of the local model approximation. The other algorithms learn faster the more information from the simulator is used.}
  \label{fig:sample_size}
\end{figure}

With infinite training data, many of the algorithms should converge to the true likelihood ratio. But generating training samples can be expensive, especially when a full detector simulation is used. An important question is therefore how much training data the different techniques require to perform well. In Fig.~\ref{fig:sample_size} we show the performance as a function of the training sample size.

The \sally approach performs very well even with very little data. Its precision stops improving eventually, showing the limitations of the local model approximation. For the other methods we find that the more information a technique uses, the less training data points it requires. The \rascal technique utilizes the most information from the simulator, leading to an exceptional performance with training samples of approximately 100\,000 events. This is in contrast to the most general \carl method, which does not use any of the extra information from the simulator and requires a two orders of magnitude larger training sample for a comparable performance.

\begin{figure}[p]
  \includegraphics[width=0.89\textwidth]{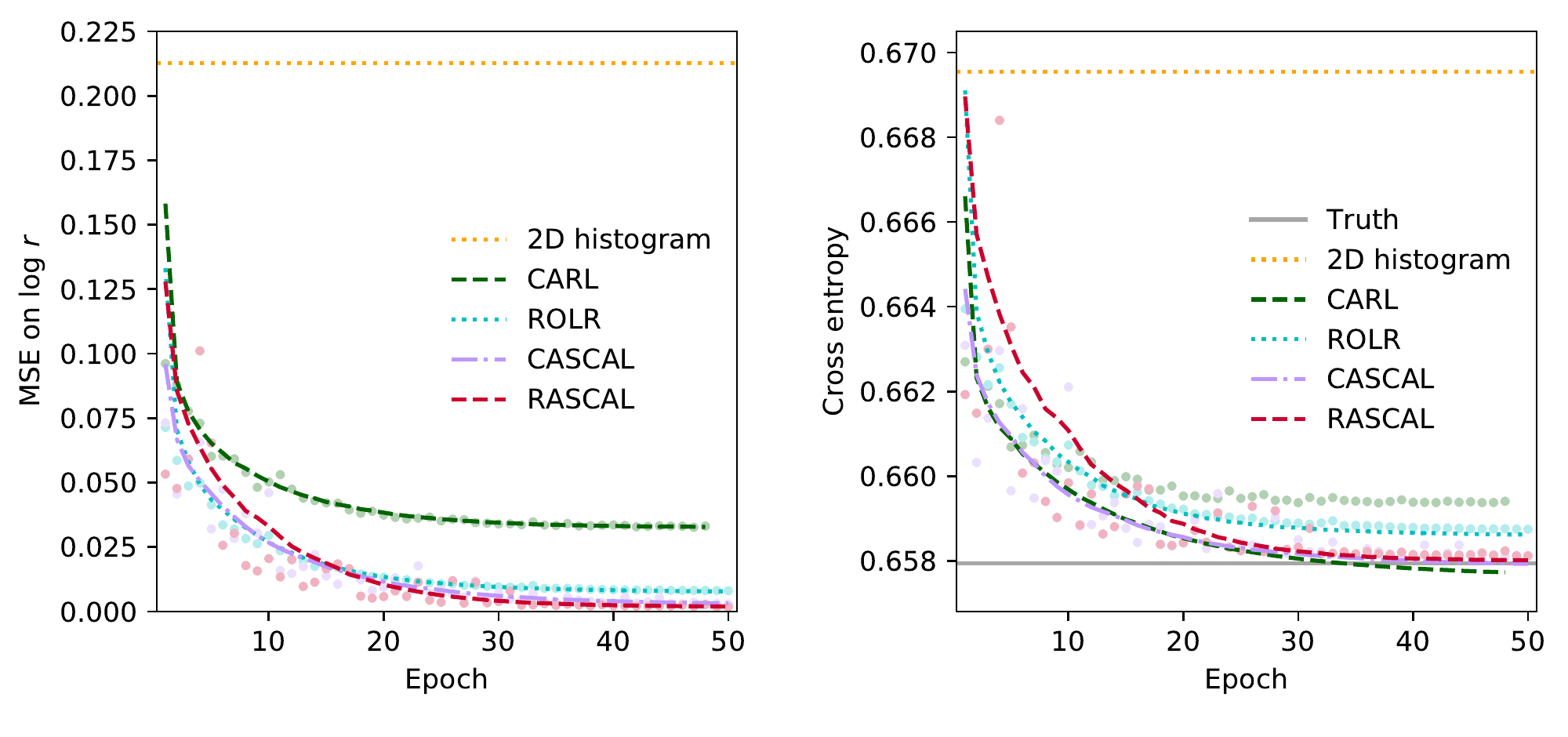}%
  \caption{Learning curve of the parameterized models. The solid lines show the metrics evaluated on the training sample, the dots indicate the performance on the validation sample. Note that these numbers are not comparable to the metrics in Table~\ref{tbl:results} and Fig.~\ref{fig:sample_size}, which are weighted with the prior in Eq.~\eqref{eq:prior}. These results also do not include calibration. Left: Mean squared error of $\log \hat{r}(x | \theta_0, \theta_1)$. Right: binary cross-entropy of the classification based on $\hat{s}(x | \theta_0, \theta_1)$ between the numerator and denominator samples. The solid grey line shows the ``optimal'' performance based on the true likelihood ratio. The \cascal and \rascal techniques converge to a performance close to the theoretic optimum. The \carl approach (green), based on minimizing the cross entropy, shows signs of overfitting. All machine-learning-based methods outperform traditional histograms (dashed orange).}
  \label{fig:learning_curve}
\end{figure}

\begin{table}[p]
  \vspace*{1.5cm}
  \begin{tabular}{lrr}
    \toprule
    \multirow{2}{*}{Algorithm} & \multicolumn{2}{c}{Evaluation time [$\mu$s]} \\
    \cmidrule{2-3}
    & per $x_e$ & per $x_e$ and $\theta_0$ \\
    \midrule
    Histogram & & $0.2$ \\
    \carl & & $19.7$ \\
    \sally & $25.4$ & $0.1$ \\
    \rolr & & $19.7$ \\
    \cascal & & $25.1$\\
    \rascal & & $21.7$ \\
    \bottomrule
  \end{tabular}
  \caption{Computation times of evaluating $\hat{r}(x | \theta_0, \theta_1)$ in the different algorithms. We distinguish between steps that have to be calculated once per $x$ and and those which have to be repeated for every evaluated value of $\theta_0$. These numbers are from one run of our algorithms with default settings on the NYU HPC cluster on machines equipped with Intel Xeon E5-2690v4 2.6GHz CPUs and NVIDIA P40 GPUs with 24 GB RAM, using a batch of $50\,000$ events $\{x_e\}$, and taking the mean over $1\,017$ values of $\theta_0$. The local score regression method and the traditional histogram method are particularly fast. But all techniques are many orders of magnitude faster to evaluate than the matrix element method or optimal observables.}
  \label{tbl:evaluation_times}
  \vspace*{0.8cm}
\end{table}

In Fig.~\ref{fig:learning_curve} we show the evolution of the likelihood estimation error and the cross-entropy of the classification problem during the training of the parameterized estimators. For comparison, we also show the optimal metrics based on the true likelihood ratio, and the results of the two-dimensional histogram approach. Once again we see that either \cascal or \rascal leads to the best results. This result also holds true for the cross entropy, hinting that the techniques we use to measure continuous parameters might also improve the power of estimators in discrete classification problems. Note that the \carl approach is more prone to overfitting than the others, visible as a significant difference between the metrics evaluated on the training and validation samples.

Equally important to the training efficiency is the computation time taken up by evaluating the likelihood ratio estimators $\hat{r}(x_e | \theta_0, \theta_1)$. We compare example evaluation times in Table~\ref{tbl:evaluation_times}. The traditional histogram approach takes the shortest time. But all tested algorithms are very fast: the likelihood ratio for fixed hypotheses $(\theta_0, \theta_1)$ for $50\,000$ events $\{x_e\}$ can always be estimated in around one second or less. The local score regression method is particularly efficient, since the estimator $\hat{t}(x | \theta_{\text{score}}, \theta_1)$ has to be evaluated only once to estimate the likelihood ratio for any value of $\theta_0$. Only the comparably fast step of density estimation has to be repeated for each tested value of $\theta_0$.

So after investing some training time upfront, all the measurement strategies developed here can be evaluated on any events with very little computational cost and amortize quickly. While this is not the focus of our paper, note that this distinguishes our approaches from the Matrix Element Method and Optimal Observable techniques. These well-established methods require the computationally expensive evaluation of complicated numerical integrals for every evaluation of the likelihood ratio estimator.

\subsubsection{Physics results}

\begin{figure}
  \includegraphics[width=\textwidth]{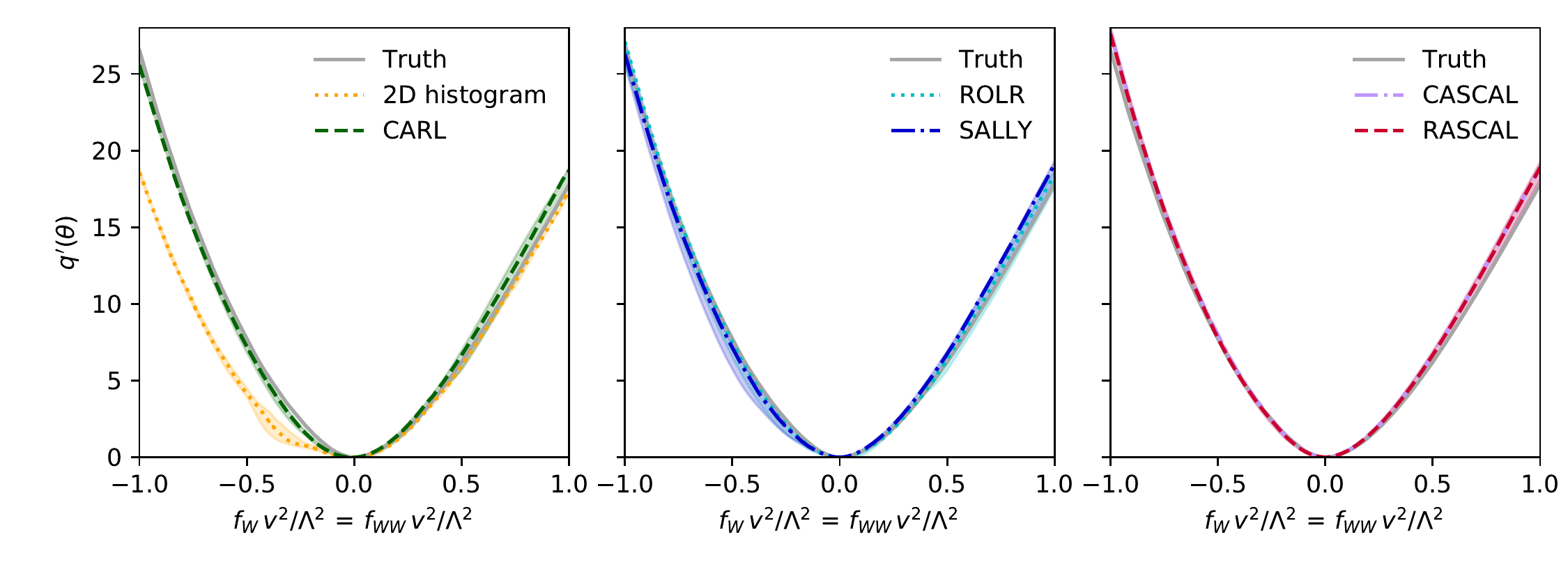}%
  \caption{Expected likelihood ratio with respect to the Standard Model along a one-dimensional slice of the parameter space. We assume 36 observed events and the SM to be true. For each estimator, we generate five sets of predictions with different reference hypotheses, independent data samples, and different random seeds. The lines show the median of this ensemble, the shaded error bands the envelope. All machine-learning-based methods reproduce the true likelihood function well, while the doubly differential histogram method underestimates the likelihood ratio in the region of negative Wilson coefficients.}
  \label{fig:expected_likelihood_ratio_slice}
\end{figure}

\begin{figure}
  \includegraphics[width=\textwidth]{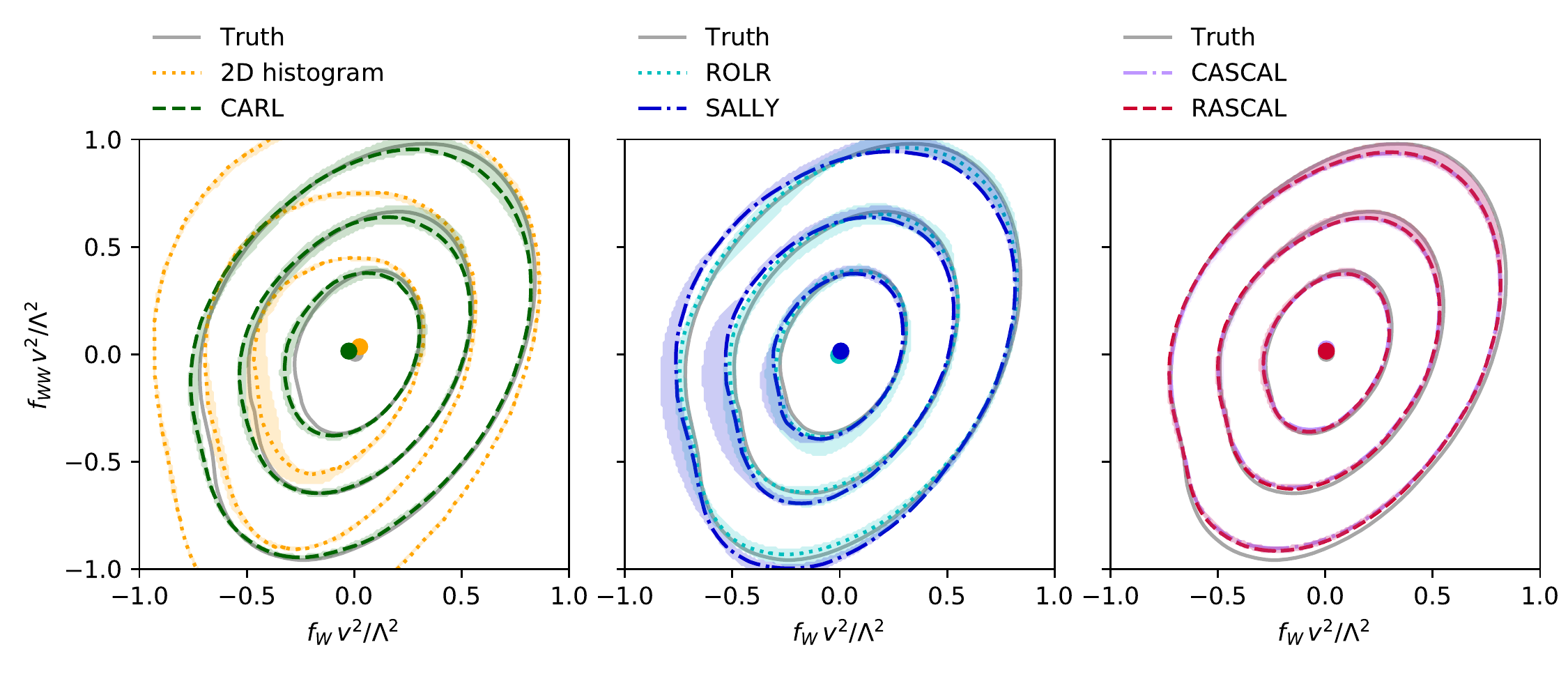}%
  \caption{Expected exclusion contours based on asymptotics at $68\%$~CL (innermost lines), $95\%$~CL, and $99.7\%$~CL (outermost lines). We assume 36 observed events and the SM to be true. As test statistics, we use the profile likelihood ratio with respect to the maximum-likelihood estimator. For each estimator, we generate five sets of predictions with different reference hypotheses, independent data samples, and different random seeds. The lines show the median of this ensemble, the shaded error bands the envelope. The new techniques based on machine learning, in particular the \cascal and \rascal techniques, lead to expected exclusion contours very close to those based on the true likelihood ratio. An analysis of a doubly differential histogram leads to much weaker bounds.}
  \label{fig:limits_asymptotics}
\end{figure}

The most important result of an EFT measurement are observed and expected exclusion contours, either based on asymptotics or toy experiments. In the asymptotic approach, the expected contours are determined just by the likelihood ratio evaluated on a large ``Asimov'' data set, as described in Sec.~\ref{sec:asymptotics}. Figure~\ref{fig:expected_likelihood_ratio_slice} shows this expected log likelihood ratio in the SM after 36 events over a one-dimensional slice of the parameter space. In Fig.~\ref{fig:limits_asymptotics} we show the corresponding expected exclusion limits on the two Wilson coefficients. To estimate the robustness of the likelihood ratio estimators, each algorithm is run five times with different choices of the reference hypothesis; independent training, calibration, and evaluation samples; and independent random seeds during training. The lines show the median of the five replicas, while the shaded bands show the envelope. While this error band does not have a clear statistic interpretation, it does provide a diagnostic tool for the variance of the estimators.

A traditional histogram-based analysis of jet $p_T$ and $\Delta \phi_{jj}$ leads to overly conservative results. It is interesting to note that this simple analysis works reasonably well in the region of parameter space with $f_W > 0$ and $f_{WW} > 0$, which is exactly the part of parameter space where informative high-energy events interfere mostly constructively with the SM amplitude. In the $f_W < 0$ region of parameter space, destructive interference dominates in the important regions of phase space with large momentum transfer. An extreme example is the ``amplitude-through-zero'' effect shown in the left panel of Fig.~\ref{fig:features}. Simple histograms with a rough binning generally lead to a poor estimation of the likelihood ratio in such complicated kinematic signatures.

We find that the new ML-based strategies allow us to place visibly tighter constraints on the Wilson coefficients than the doubly differential histogram. In particular the \carl + score and regression + score estimators lead to exclusion contours that are close to the contours based on the true likelihood ratio. In this analysis based on asymptotics, however, it is possible for the estimated contours to be slightly too tight, wrongly marking parameter regions as excluded at a given confidence level. This problem can be mitigated by profiling over systematic uncertainties assigned to the likelihood ratio estimates. 

\begin{figure}
  \includegraphics[width=\textwidth]{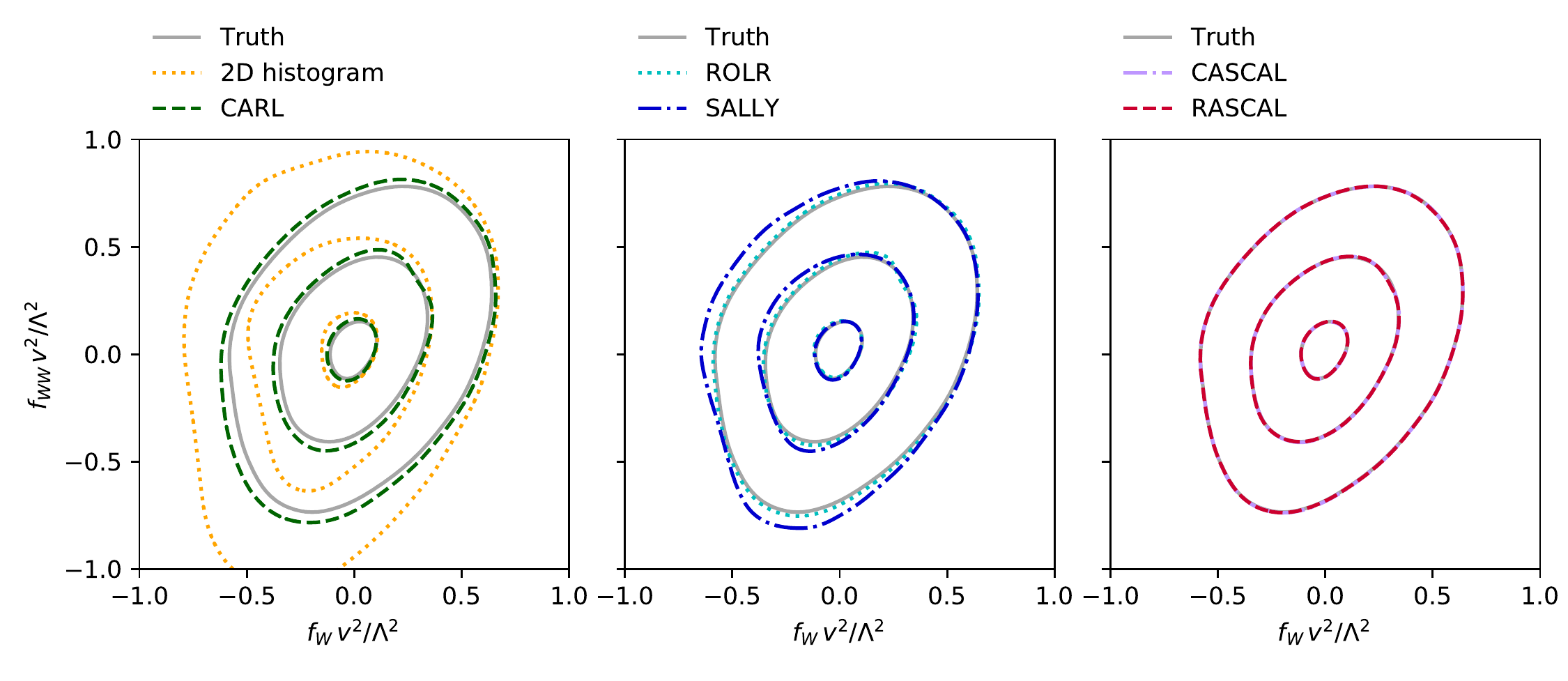}%
  \caption{Expected exclusion contours based on the Neyman construction with toy experiments at $68\%$~CL (innermost lines), $95\%$~CL, and $99.7\%$~CL (outermost lines). We assume 36 observed events and the SM to be true. As test statistics, we use the likelihood ratio with respect to the SM. All machine-learning-based methods let us impose much tighter bounds on the Wilson coefficients than the traditional histogram approach (left, dotted orange). The Neyman construction guarantees statistically correct results: no contour based on estimators excludes parameter points that should not be excluded. The expected limits based on the \cascal (right, lavender) and \rascal (right, red) techniques are virtually indistinguishable from the true likelihood contours.}
  \label{fig:limits_nc}
\end{figure}

Exclusion limits based on the Neyman construction do not suffer from this issue: contours derived in this way might be not optimal, but they are never wrong. We generate toy experiments to estimate the distribution of the likelihood ratio with respect to the SM for individual events. Repeatedly convolving this single-event distribution with itself, we find the distribution of the likelihood ratio after 36 observed events. 

The expected corresponding expected exclusion limits are shown in Fig.~\ref{fig:limits_nc}. Indeed, errors in the likelihood ratio estimation never lead to undercoverage, \ie the exclusion of points that should not be excluded based on the true likelihood ratio. Again we find that histograms of kinematic observables only allow us to place rather weak bounds on the Wilson coefficients. \sally performs clearly better, with excellent performance close to the SM. Deviations from the optimal bounds become visible at the $2\sigma$ level, hinting at the breakdown of the local model approximation there. The best results come once more from the \cascal and \rascal methods. Both of these strategies yield exclusion bounds that are virtually indistinguishable from those based on the true likelihood ratio.

As a side note, a comparison of the expected contours based on asymptotics to those based on the Neyman construction shows the Neyman results to be tighter. This reflects the different test statistics used in the two figures: in the asymptotics case, we use the profile likelihood ratio with respect to the maximum likelihood estimator, which itself fluctuates around the true value of $\theta$ (which in our case is assumed to be the SM). In the Neyman construction we use the likelihood ratio with respect to the SM, leading to tighter contours if the true value is in fact close to the SM, and weaker constraints if it is very different.

\subsection{Detector effects}
\label{sec:results_smearing}

\begin{figure}
  \includegraphics[width=\textwidth]{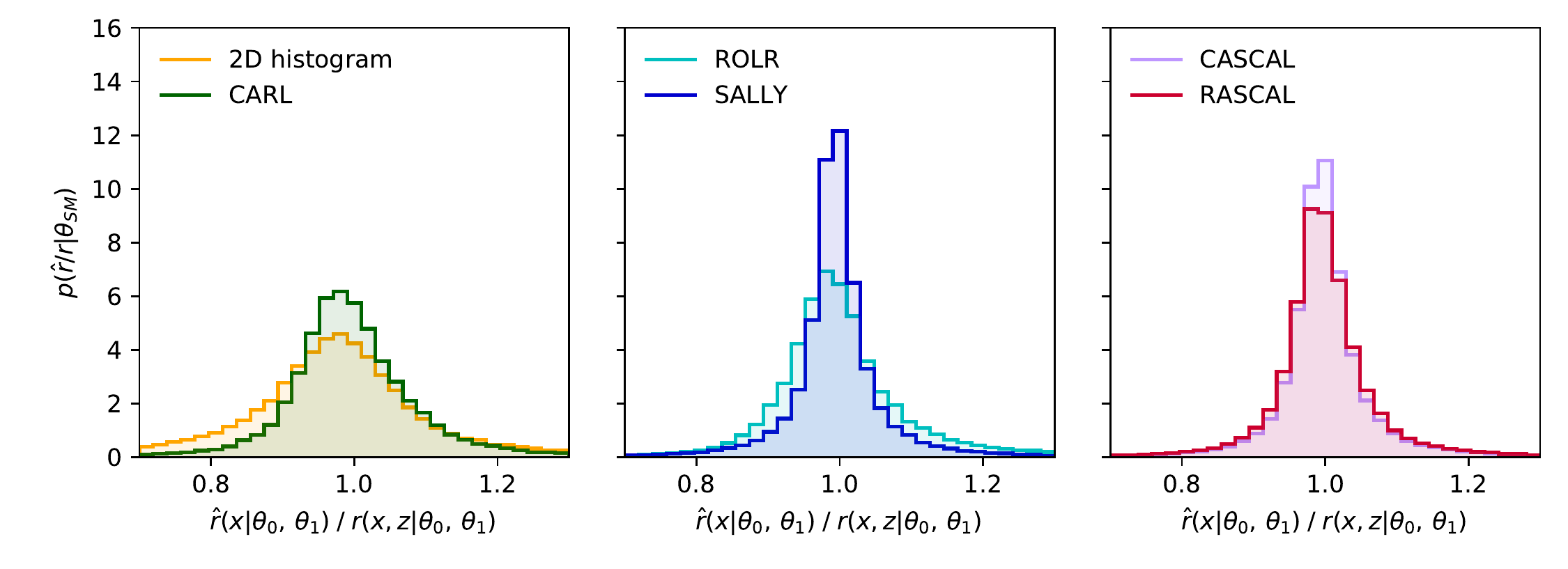}%
  \caption{Ratio of the estimated likelihood ratio $\hat{r} (x | \theta_0, \theta_1)$ to the \emph{joint} likelihood ratio $r(x, z | \theta_0, \theta_1)$, which is conditional on the parton-level momenta and other latent variables. As a benchmark hypothesis we use $\theta_0 = (-0.5, -0.5)^T$, the events are drawn according to the SM. The spread common to all methods shows the effect of the smearing on the likelihood ratio. The additional spread in the histogram, \carl, and \rolr methods is due to a poorer performance of these techniques.}
  \label{fig:r_smearing}
\end{figure}

We have now established that the measurement strategy work very well in an idealized setup, where we can compare them to the true likelihood ratio. In a next step, we turn towards a setup with a rudimentary smearing function that models the effect of the parton shower and the detector response on the observables. In this setting, the true likelihood is intractable, so we cannot use it as a baseline to validate the predictions any more. But we can still discuss the relative ordering of the exclusion contours predicted by the different estimators.

Figure~\ref{fig:r_smearing} shows the relation between the true \emph{joint} likelihood ratio $r(x, z | \theta_0, \theta_1)$, which is conditional on the parton-level momenta $z$ and other latent variables, to the estimated likelihood ratio $\hat{r}(x | \theta_0, \theta_1)$, which only depends on the observables $x$. We see that this relation is stochastically smeared out around 1. Recall that in the idealized scenario the best estimators described the true likelihood ratio perfectly, as shown in Fig.~\ref{fig:r_scatter}. This strongly suggests that the spread visible here is not due to errors of the likelihood ratio estimators, but rather shows the difference between the joint and true likelihood ratios, as illustrated in Fig.~\ref{fig:illustration_carl_ratio_regression}. 

\begin{figure}
  \vspace*{0.8cm}
  \includegraphics[width=\textwidth]{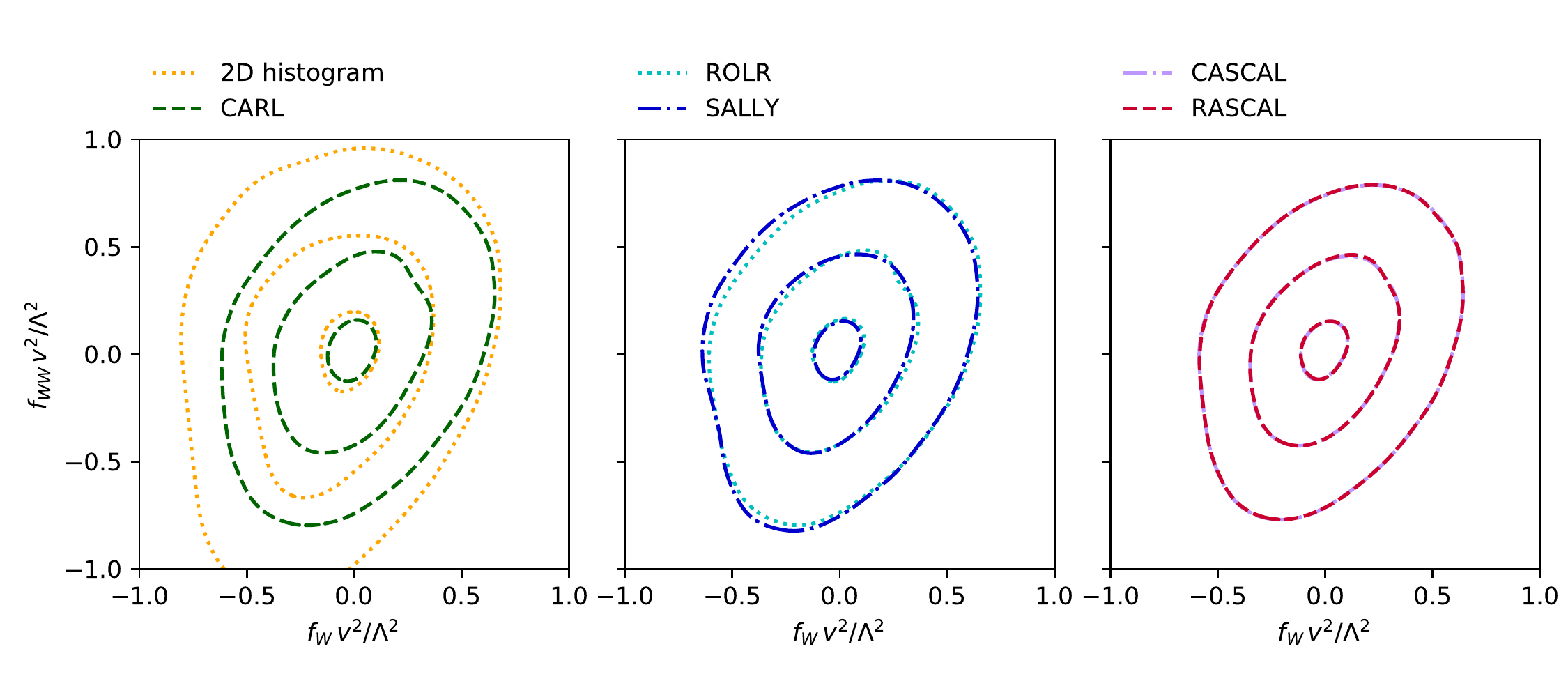}%
  \caption{Expected exclusion contours based on the Neyman construction with toy experiments  at $68\%$~CL, $95\%$~CL, and $99.7\%$~CL with smearing. We assume 36 observed events and the SM to be true. As test statistics, we use the likelihood ratio with respect to the SM. In the setup with smearing we cannot these results to the true likelihood contours. But since the Neyman construction is guaranteed to cover, these expected limits are correct. The new techniques, in particular \cascal (right, dashed red) and \rascal (right, dash-dotted orange), allow us to set much tighter bounds on the Wilson coefficients than a traditional histogram analysis (left, dotted orange). }
  \label{fig:limits_nc_smearing}
  \vspace*{0.8cm}
\end{figure}

In Fig.~\ref{fig:limits_nc_smearing} we show the expected exclusion contours based on the Neyman construction, which guarantees statistically correct results. The conclusions from the idealized setting are confirmed: a measurement based on the likelihood ratio estimators leads to robust bounds that are clearly more powerful than those based on a histogram. Once again, the \cascal and \rascal algorithms lead to the strongest limits.

\section{Conclusions}
\label{sec:conclusions}

We have developed and analysed a suite of new analysis techniques for measurements of continuous parameters in LHC experiments based on simulations and machine learning. Exploiting the structure of particle physics processes, they extract additional information from the event generators, and use this information to train precise estimators for likelihood ratios.

Our approach is designed for problems with large numbers of observables, where the likelihood function is not tractable and traditional methods based on individual kinematic variables often perform poorly. It scales well to high-dimensional parameter spaces such as that of effective field theories. The new methods do not require any approximations on the hard process, parton shower, or detector effects, and the likelihood ratio for any event and hypothesis pair can be evaluated in microseconds. These two properties set it apart from the Matrix Element Method or Optimal Observables, which rely on crude approximations for the shower and detector and require the evaluation of typically very expensive integrals.

Using Higgs production in weak boson fusion in the four-lepton mode as a specific example process, we have evaluated the performance of the different methods and compared them to a classical analysis of the jet momentum and azimuthal angle between the tagging jets. We find that the new algorithms provide very precise estimates of arbitrary likelihood ratios. Using them as a test statistics allows us to impose significantly tighter constraints on the EFT coefficients than the traditional kinematic histograms.

Out of the several methods introduced and discussed in this paper, two stand out. The first, which we call \sally, is designed for parameter regions close to the Standard Model:
\begin{enumerate}
\item As training data, the algorithm requires a sample of fully simulated events, each accompanied by the corresponding \emph{joint score} at the SM: the relative change of the parton-level likelihood function of the parton-level momenta associated with this event under small changes of the theory parameters away from the SM. This can be calculated by evaluating the squared matrix element at the same phase-space points for different theory parameters. We can thus extract this quantity from Monte-Carlo generators such as \toolfont{MadGraph}.
\item Regressing on this data, we train an estimator (for instance realized as a neural network) that takes as input an observation and returns the score at the SM. This function compresses the high-dimensional observable space into a vector with as many components as parameters of interest. If the parameter space is high-dimensional, this can be even further compressed into the scalar product between the score vector and the difference between two parameter points.
\item The estimated score (or the scalar product between score and parameter difference) can then be treated like any set of observables in a traditional analysis. We can fill histograms of this quantity for different hypotheses, and calculate likelihood ratios from them.
\end{enumerate}

There are two key ideas that underlie this strategy. First, note that the training data only consists of the \emph{joint score}, which depends on the parton-level four-momenta of an event. But during the training the estimator converges to the actual score of the distribution of the observables, \ie the relative change of the actual likelihood function under infinitesimal changes of the parameters. We have proven this powerful, yet surprisingly simple relation in this paper.

Second, close to the Standard Model (or any other reference parameter point), the score provides the sufficient statistics: it encapsulates all information on the local approximation of the statistical model. In other words, if the score is estimated well, the dimensionality reduction from high-dimensional observables into a low-dimensional vector does not lose any information on the parameters. The estimated score is a machine-learning version of the Optimal Observable idea, but requires neither approximations of the parton shower or detector treatment nor numerically expensive integrals.

As a matter of fact, the dimensionality reduction can be taken one step further. We have introduced the \sallino technique that compresses the estimated score vector to a single scalar function, again without loss of power in the local approximation, and independent of the number of theory parameters.

In our example process, these simple and robust analysis strategies work remarkably well, especially close to the Standard Model. Deviations appear at the $2\sigma$ level, but even there it allows for much stronger constraints than a traditional analysis of kinematic variables. It requires significantly less data to train than the other discussed methods. Since the \sallino method can compress any observation into a single number without losing much sensitivity, even for hundreds of theory parameters, this approach scales exceptionally well to high-dimensional parameter spaces, as in the case of the SMEFT.

The second algorithm we want to highlight here is the \rascal technique. Using even more information available from the simulator, it learns a parameterized likelihood ratio estimator: one function that takes both the observation and a theory parameter point as input and returns an estimate for the likelihood ratio between this point and a reference hypothesis given the observation. This estimator is constructed as follows:
\begin{enumerate}
  \item Training this parameterized estimator requires data for many different values of the tested parameter point (the numerator in the likelihood ratio). For simplicity, the reference hypothesis (the denominator in the likelihood ratio) can be kept fixed. For each of these hypothesis pairs, event samples are generated according to the numerator and denominator hypothesis. In addition, we extract the joint likelihood ratio from the simulator: essentially the squared matrix element according to the numerator theory parameters divided by the squared matrix element according to the denominator hypothesis, evaluated at the generated parton-level momenta. Again, we also need the joint score, \ie the relative change of the parton-level likelihood function under infinitesimal changes of the theory parameters. Both quantities can be extracted from matrix element codes.
\item A neural network models the estimated likelihood ratio as a function of both the observables and the value of the theory parameters (of the numerator in the likelihood ratio). We can calculate the gradient of the network output with respect to the theory parameter and thus also the estimated score. The network is trained by minimizing the squared error of the likelihood ratio plus the squared error of the score, in both cases with respect to the joint quantities extracted from the simulator.
  \item After the training phase, the likelihood ratio can optionally be calibrated, for instance through isotonic regression.
\end{enumerate}

This technique relies on a similar trick as the local score regression method: the likelihood ratio learned during the training converges to the true likelihood ratio, even though the joint ratio information in the training data is conditional on the parton-level momenta. The \rascal method is among the best-performing methods of all analysed techniques. It requires significantly smaller training samples than all other approaches, with the exception of \sally and \sallino. Expected exclusion limits derived in this way are virtually indistinguishable from those based on the true likelihood ratio.

On top of these two approaches, we have developed, analysed, and compared several other methods. We refer the reader to the main part and the appendices of this document, where all these algorithms are discussed in depth.

All tools developed here are suitable for large-scale LHC analyses. On the software side, only few modifications of existing tools are necessary. Most importantly, matrix-element generators should provide a user-friendly interface to calculate the squared matrix element for a given configuration of four-momenta and a given set of physics parameters. With such an interface, one could easily calculate the joint score and joint likelihood ratio data that is needed for the new algorithms. The training of the estimators is then straightforward, in particular for the \sally and \sallino methods. The limit setting follows established procedures, either based on the Neyman construction with toy experiments, or (since the tools provide direct estimates for the likelihood ratio) using asymptotic formulae.

While we have focussed on the example of effective field theory measurements, these techniques equally apply to other measurements of continuous parameter in collider experiments as well as to a large class of problems outside of particle physics~\cite{companion_nips}. Some of the techniques can also be applied to improve the training of machine-learning-based classifiers. Finally, while we restricted our analysis to frequentist confidence intervals, as is common in particle physics, the same ideas can be used in a Bayesian setting.

All in all, we have presented a range of new inference techniques based on machine learning, which exploit the structure of particle physics processes to augment training data. They scale well to large-scale LHC analyses with many observables and high-dimensional parameter spaces. They do not require any approximations of the hard process, parton shower, or detector effects, and the likelihood ratio can be evaluated in microseconds. In an example analysis, these new techniques have demonstrated the potential to substantially improve the precision and new physics reach of the LHC legacy results.

\subsubsection*{Acknowledgments}

We would like to thank Cyril Becot and Lukas Heinrich, who contributed to this project at an early stage. We are grateful to Felix Kling, Tilman Plehn, and Peter Schichtel for providing the \toolfont{MadMax} code and helping us use it. KC wants to thank CP3 at UC Louvain for their hospitality. Finally, we would like to thank At{\i}l{\i}m G\"{u}ne\c{s} Baydin, Lydia Brenner, Joan Bruna, Kyunghyun Cho, Michael Gill, Ian Goodfellow, Daniela Huppenkothen, Hugo Larochelle, Yann LeCun, Fabio Maltoni, Jean-Michel Marin, Iain Murray, George Papamakarios, Duccio Pappadopulo, Dennis Prangle, Rajesh Ranganath, Dustin Tran, Rost Verkerke, Wouter Verkerke, Max Welling, and Richard Wilkinson for interesting discussions.

JB, KC, and GL are grateful for the support of the Moore-Sloan data science environment at NYU. KC and GL were supported through the NSF grants ACI-1450310 and PHY-1505463. JP was partially supported by the Scientific and Technological Center of Valpara\'{i}so (CCTVal) under Fondecyt grant BASAL FB0821. This work was supported in part through the NYU IT High Performance Computing resources, services, and staff expertise.

\appendix
\section{Appendix}
\label{sec:appendix}

\subsection{Simplified detector description}
\label{sec:appendix_smearing}

While most of our results are based on an idealized perfect measurement of parton-level momenta, we also consider a toy smearing representing the effect of parton shower and the detector. The total smearing function is given by
\begin{equation}
  \intractablep(x | z) = \prod_{\ell \in \text{leptons}} \!\!\!\!\!  \intractablep_\ell(x_\ell| z_\ell) \; \prod_{j \in \text{jets}} \!\! \intractablep_j(x_j | z_j) \,.
\end{equation}
Lepton momenta $x_\ell = (\hat{E}, \hat{p}_T, \hat{\eta}, \hat{\phi})$ are smeared by
\begin{multline}
  \intractablep_\ell ( \hat{E}, \hat{p}_T, \hat{\eta}, \hat{\phi} | E, p_T, \eta, \phi)
  = \normal{\hat{p}_T }{ p_T, (3\cdot 10^{-4} \gev^{-1} p_T^2)^2}\\
\cdot \delta(\hat{E} - E_0(\hat{p}_T, \hat{\eta}; m_\ell))
\ \delta(\hat{\eta} - \eta)
\ \delta(\hat{\phi} - \phi) \,,
\end{multline}
while the distribution of the jet properties depending on the quark momenta is given by
\begin{multline}
  \intractablep_j ( \hat{E}, \hat{p}_T, \hat{\eta}, \hat{\phi} | E, p_T, \eta, \phi)
  = \Biggl( \normal{\hat{E}}{a_0 + a_1 \sqrt{E} + a_2 E ,  (b_0 + b_1 \sqrt{E} + b_2 E)^2}\\
              \qqqquad + c \normal{\hat{E}}{d_0 + d_1 \sqrt{E} + d_2 E ,  (e_0 + e_1 \sqrt{E} + e_2 E)^2} \Biggr) \\
\cdot \delta(\hat{p}_T - p_{T0}(\hat{E}, \hat{\eta}; m_\ell))
\ \normal {\hat{\eta} }{ \eta, 0.1^2 }
\ \normal {\hat{\phi} }{ \phi, 0.1^2 } \,.
\end{multline}
Here $\normal{x}{ \mu, \sigma^2}$ is the Gaussian distribution with mean $\mu$ and variance $\sigma^2$. The jet energy resolution parameters $a_i$, $b_i$, $c$, $d_i$, and $e_i$ are based on the default settings of the jet transfer function in \toolfont{MadWeight}~\cite{Mertens:2014iya}. The functions $E_0(p_T, \eta, m)$ and $p_{T0} (E, \eta, m)$ refer to the energy and transverse momentum corresponding to an on-shell particle with mass $m$.

\subsection{Model almanac}
\label{sec:appendix_models}

\begin{figure}
  \includegraphics[width=0.89\textwidth]{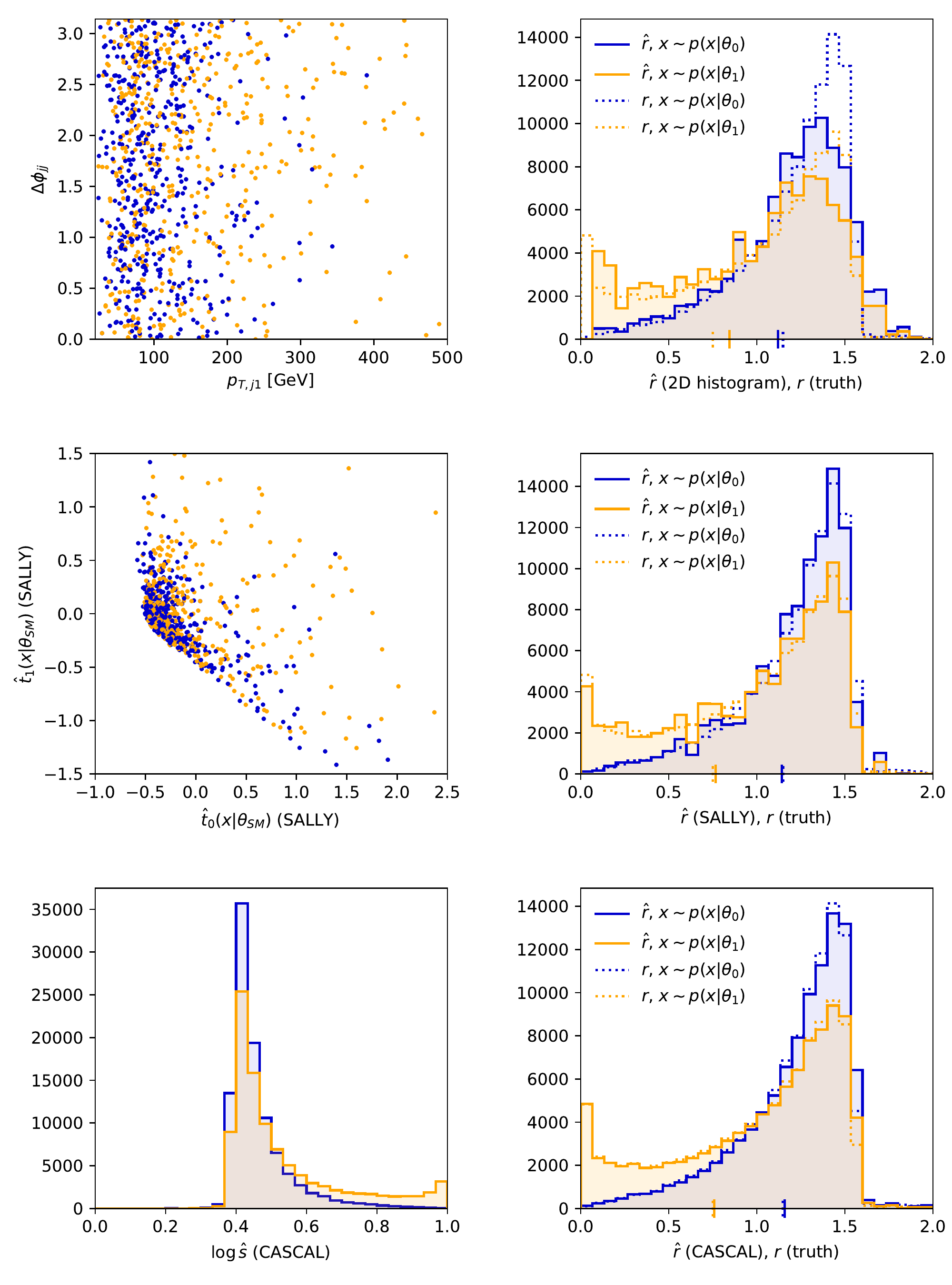}%
  \caption{Example distributions to illustrate the doubly differential histogram analysis (top), the \sally technique (middle), and the \cascal method (bottom). The left panels show the different spaces in which the densities and their ratios are estimated. On the right we show the corresponding distributions of the estimated ratio $\hat{r}$ (solid) and compare them to the true likelihood ratio distributions (dotted). We use the benchmark $\theta_0 = (-0.5,-0.5)^T$ (blue) and $\theta_1$ as in Eq.~\eqref{eq:denom_benchmark} (orange).}
  \label{fig:inference_examples}
\end{figure}

In Sec.~\ref{sec:eft_modeling} we developed different estimators for the likelihood ratio, focussing on the key ideas over technical details. Here we fill in the gaps, explain all strategies in a self-contained way, and document the settings we use for our example process. To facilitate their comparison, we describe all models in terms of a ``training'' and an ``evaluation'' part, even if this language is not typically used \eg for histograms.

\begin{table}
  \small
  \begin{tabular}{l r r c }
    \toprule
    Strategy & NN layers & $\alpha$ & Calibration\,/\,density estimation \\
    \midrule
    Histogram & & & Histogram \\
    AFC & & & Gaussian KDE \\
    \midrule
    \carl (PbP) & $3$ && Isotonic probability calibration \\
    \carl (parameterized) & $2$ && Isotonic probability calibration  \\
    \carl (morphing-aware) & $2$ && Isotonic probability calibration  \\
    \midrule
    \sally & $5$ && Histogram \\
    \sallino & $5$ && Histogram \\
    \midrule
    \rolr (PbP) & $3$ && Isotonic probability calibration \\
    \rolr (parameterized) & $3$ && Isotonic probability calibration \\
    \rolr (morphing-aware) & $2$ && Isotonic probability calibration \\
    \midrule
    \cascal (parameterized) & $5$ & $5$ & Isotonic probability calibration  \\
    \cascal (morphing-aware) & $2$ & $5$ & Isotonic probability calibration  \\
    \midrule
    \rascal (parameterized) & $5$ & $100$ & Isotonic probability calibration  \\
    \rascal (morphing-aware) & $2$ & $100$ & Isotonic probability calibration  \\
    \bottomrule
  \end{tabular}
  \caption{Default settings for the analysis techniques. The neural network (NN) layers each have 100 units with $\tanh$ activation functions. The hyperparameter $\alpha$ multiplies the score squared error in the combined loss functions of Eq.~\eqref{eq:carl_plus_score_loss} and \eqref{eq:ratio_plus_score_loss}.}
  \label{tbl:default_settings}
\end{table}

\subsubsection{Histograms of observables}
\label{sec:appendix_models_histos}

\begin{description}
  \item[Idea]
    Most collider measurements are based on the number of events or the cross section of a process in a given phase-space region or on the differential cross section or distribution of one or at most a few kinematic observables $v$. Typical choices are the reconstructed energies, momenta, angles, or invariant masses of particles. Choosing the right set of observables for a given measurement problem is all but trivial, but many processes have been studied extensively in the literature. Once this choice is made, this strategy is simple, fast, and intuitive. We illustrate the information used by this approach in the top panels of Fig.~\ref{fig:inference_examples}.

 \item[Requirements]
   The histogram approach can be used in the general likelihood-free setting: it only requires a simulator that can generate samples $\{x\} \sim \intractablep(x | \theta)$.
 \item[Structure]
   Histograms are most commonly used point by point in $\theta$. If the problem has the morphing structure discussed in Sec.~\ref{sec:morphing}, they can also be applied in a morphing-aware parameterized way (we have not implemented this for our example process).
  \item[Training]
   After generating samples for both the numerator and denominator hypotheses, the values of the chosen kinematic variables $v(x)$ are extracted, and binned into separate histograms for the two hypotheses.
  \item[Calibration]
   With sufficient training data, histograms should be well calibrated, so we do not experiment with an additional calibration stage.
  \item[Evaluation]
    To estimate the likelihood ratio between two hypotheses $\theta_0$ and $\theta_1$ for a given set of observables $x$, one has to extract the kinematic variables $v(x)$ and look up the corresponding bin contents in the histograms for $\theta_0$ and $\theta_1$. Assuming equal binning for both histograms, the likelihood ratio is simply estimated as the ratio of bin contents.
  \item[Parameters]
    The only parameters of this approach are the choices of kinematic observables and the histogram binning.

    In our example process, we consider six different variants:
    \begin{itemize}
      \item A one-dimensional histogram of the transverse momentum $p_{T,j1}$ of the leading (higher-$p_T$) jet with 80 bins.
      \item A one-dimensional histogram of the absolute value of the azimuthal angle $\Delta \phi_{jj}$  between the two jets with 20 bins.
      \item A ``coarse'' two-dimensional histogram of these two variables with 10 bins in the $p_{T,j1}$ direction and 5 bins along $\Delta \phi_{jj}$.
      \item A ``medium'' two-dimensional histogram of these two variables with 20 bins in the $p_{T,j1}$ direction and 10 bins along $\Delta \phi_{jj}$.
      \item A ``fine'' two-dimensional histogram of these two variables with 30 bins in the $p_{T,j1}$ direction and 15 bins along $\Delta \phi_{jj}$.
      \item A ``very fine'' two-dimensional histogram of these two variables with 50 bins in the $p_{T,j1}$ direction and 20 bins along $\Delta \phi_{jj}$.
      \item An ``asymmetric'' two-dimensional histogram of these two variables with 50 bins in the $p_{T,j1}$ direction and 5 bins along $\Delta \phi_{jj}$.
      \end{itemize}
      For each pair $(\theta_0, \theta_1)$ and each observable, the bin edges are chosen such that the same expected number of events according to $\theta_0$ plus the expected number of events according to $\theta_1$ is the same in each bin.
\end{description}

\subsubsection{Approximate Frequentist Computation (\afc)}
\label{sec:appendix_models_afc}

\begin{description}
  \item[Idea]
    Approximate Bayesian Computation is a very common technique for likelihood-free inference in a Bayesian setup. In its simplest form it keeps samples according to the rejection probability of Eq.~\eqref{eq:abc_rejection_probability}. This amounts to an approximation of the likelihood function through kernel density estimation, which we can isolate from the \abc sampling mechanism and use in a frequentist setting. We call it Approximate Frequentist Computation (AFC) to stress the relation to \abc. Just as \abc or the histogram approach, it requires the choice of a summary statistics $v(x)$, which in our example process we take to be a two-dimensional or five-dimensional subset of the kinematic variables.

 \item[Requirements]
   AFC can be used in the general likelihood-free setting: it only requires a simulator that can generate samples $\{x\} \sim \intractablep(x | \theta)$.
 \item[Structure]
   We use AFC point by point in $\theta$. If the problem has the morphing structure discussed in Sec.~\ref{sec:morphing}, it can also be applied in a morphing-aware parameterized way.
  \item[Training]
    For each event in the numerator and denominator training samples, the summary statistics $v(x)$ are calculated and saved.
  \item[Calibration]
   AFC can be calibrated as any other technique on this list, but we left this for future work.
  \item[Evaluation]
    The summary statistics $v(x)$ is extracted from the observation. For numerator and denominator hypothesis separately, the likelihood at this point is estimated with Eq.~\eqref{eq:afc_likelihood_estimate}. The likelihood ratio estimate is then simply given by the ratio between the estimated numerator and denominator densities.
  \item[Parameters]
    Just as for histograms, the choice of the summary statistics is the most important parameter. The performance of AFC also crucially depends on the kernel and bandwidth $\varepsilon$. Too small values for the bandwidth make large training samples necessary, too large values lead to an oversmoothening and loss of information.

    In our example process, we consider two different variants:
    \begin{itemize}
      \item A two-dimensional summary statistics space of the leading jet $p_T$ and $\Delta \phi_{jj}$ (see above). Both variables are rescaled to zero mean and unit variance.
      \item A five-dimensional summary statistics space of the leading jet $p_T$, $\Delta \phi_{jj}$, the dijet invariant mass $m_{jj}$, the separation in pseudorapidity between the jets $\Delta \eta_{jj}$, and the invariant mass of the lighter (off-shell) reconstructed $Z$ boson $m_{Z2}$. All variables are rescaled to zero mean and unit variance.
      \end{itemize}
      We use Gaussian kernels with bandwidths between $0.01$ and $0.5$.
\end{description}

\subsubsection{Calibrated classifiers (\carl)}
\label{sec:appendix_models_carl}

\begin{figure}
  \includegraphics[width=\textwidth]{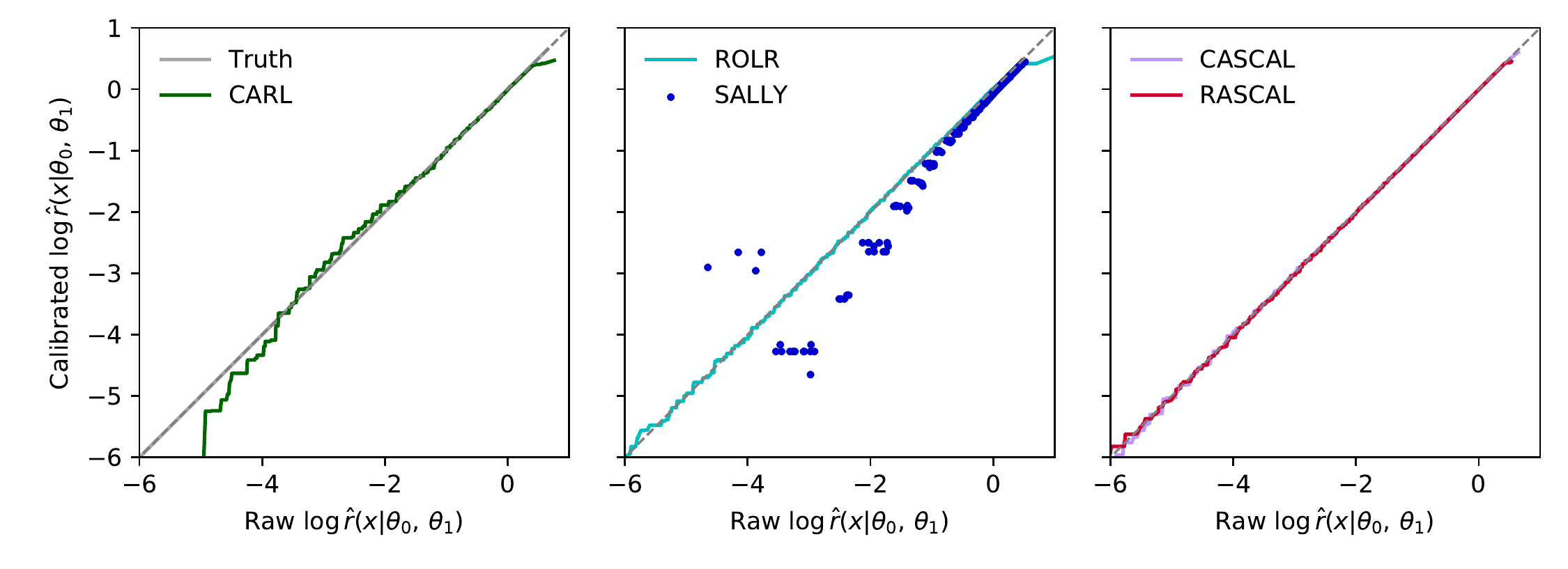}%
  \caption{Calibration curves for different estimators, comparing the uncalibrated (``raw'') estimator to the estimator after probability calibration. The calibration curve for the truth prediction is a cross-check for consistency, we do not actually use calibration for the truth predictions. For the local score regression technique, we show the value of $\hat{h}(x|\theta_0, \theta_1)$ (essentially the log likelihood ratio in the local model) versus the estimated likelihood ratio after density estimation.}
  \label{fig:calibration}
\end{figure}

\begin{description}
  \item[Idea]
    \carl was developed in Ref.~\cite{Cranmer:2015bka}. The authors showed that the likelihood ratio is invariant under any transformation that is monotonic with the likelihood ratio. In practice, this means that we can train a classifier between two samples generated from the numerator and denominator hypotheses and turn the classifier decision function $\hat{s}(x)$ into an estimator for the likelihood ratio $\hat{r}(x)$. This relation between $\hat{s}(x)$ and $\hat{r}(x)$ can follow the ideal relation in Eqs.~\eqref{eq:r_from_s} and \eqref{eq:s_from_r}. But even if this relation does not hold, we can still extract a likelihood ratio estimator from the classifier output through probability calibration.
 \item[Requirements]
   \carl can be used in the general likelihood-free setting: it only requires a simulator that can generate samples $\{x\} \sim \intractablep(x | \theta)$.
 \item[Structure]
   \carl can be used either point by point, in an agnostic parameterized version, or (if the morphing condition in Eq.~\eqref{eq:morphing_structure} holds) in a morphing-aware version. Figure~\ref{fig:neural_network_architecture} illustrates the structure of the estimator in these three cases.
  \item[Training]
A classifier with decision function $\hat{s}(x | \theta_0, \theta_1)$ is trained to discriminate between numerator (label $y = 0$) and denominator (label $y = 1$) samples by minimizing the binary cross-entropy given in Eq.~\eqref{eq:cross-entropy} (other loss functions are possible, but we have not experimented with them).

In the point-by-point version, the inputs to the classifiers are just the observables $x$, and the events in the numerator sample are generated according to one specific value $\theta_0$. In the parameterized versions of the estimator, the numerator training samples do not come from a single parameter $\theta_0$, but rather a combination of many different subsamples. In the agnostic parameterized setup, the value of $\theta_0$ used in each event is then one of the inputs to the neural network. In the morphing-aware versions, it is used to calculate the weights $w_c(\theta_0)$ that multiply the different component networks $\hat{r}_c(x)$, as visualized in the bottom panel of Fig.~\ref{fig:neural_network_architecture}.
  \item[Calibration]
In a next step, the classifier output is optionally calibrated as discussed in Sec.~\ref{sec:probability_calibration} using isotonic regression. The calibration curve is shown in the left panel of Fig.~\ref{fig:calibration}. We have also experimented with an additional step of expectation calibration, see Sec.~\ref{sec:expectation_calibration}.
  \item[Evaluation]
 For a given $x$ (and in the parameterized versions $\theta_0$), the classifier decision function $\hat{s}(x | \theta_0, \theta_1)$ is evaluated. This is turned into a likelihood ratio estimator with the relation given in Eq.~\eqref{eq:r_from_s}, and optionally calibrated.
  \item[Parameters]
The parameters of this approach are the architecture of the neural network, \ie the number of layers and elements, the activation function, the optimizer used for training, its parameters, and optionally regularization terms.

For our example process we consider fully connected neural networks with two (``shallow''), three, or five (``deep'') layers of 100 neurons each and $\tanh$ activation functions. They are trained with the \toolfont{Adam} optimizer~\cite{adam} over 50 epochs with early stopping and learning rate decay.
Our default settings are given in Table~\ref{tbl:default_settings}. Experiments with different architectures, other activation functions, additional dropout layers, other optimizers, and different learning rate schedules yielded a worse performance.
\end{description}

\subsubsection{Ratio regression (\rolr)}
\label{sec:appendix_models_rolr}

\begin{description}
  \item[Idea]
    Particle physics generators do not only provide sets of observables $\{x_e\}$, but also the corresponding parton-level momenta $\{z_e\}$. From matrix element codes such as \toolfont{MadGraph} we can extract the squared matrix element $| \mathcal{M} |^2 (z | \theta)$ given parton-level momenta $z$ and theory parameter points $\theta$. This allows us to calculate the joint likelihood ratio
\begin{equation}
  r(x_e, z_e | \theta_0 , \theta_1)
  = \frac {p(z_{e} | \theta_0)} {p(z_{e} | \theta_1)}
  = \frac {| \mathcal{M} |^2 (z_e | \theta_0)} {| \mathcal{M} |^2 (z_e | \theta_1)} \, \frac {\sigma(\theta_1)} {\sigma(\theta_0)}
\end{equation}
for any of the generated events.

    In Sec.~\ref{sec:availability} we have shown that regressing a function $\hat{r}(x)$ on the generated events $\{x_e\}$ and the corresponding joint likelihood ratios $r(x_e, z_e | \theta_0 , \theta_1)$ will converge to
\begin{equation}
  \hat{r}(x) \to \intractabler(x) = \frac {\intractablep(x | \theta_0)} {\intractablep (x | \theta_1)} \,,
\end{equation}
provided that the events are sampled according to $x_e \sim \intractablep(x | \theta_1)$.
 \item[Requirements]
   The \rolr technique requires a generator with access to the joint likelihood ratios $r(x_e, z_e | \theta_0 , \theta_1)$. In the particle physics case, this means we have to be able to evaluate the squared matrix elements for given phase-space points and theory parameters.
 \item[Structure]
   \rolr can be used either point by point, in an agnostic parameterized version, or (if the morphing condition in Eq.~\eqref{eq:morphing_structure} holds) in a morphing-aware version. Figure~\ref{fig:neural_network_architecture} illustrates the structure of the estimator in these three cases.
  \item[Training]
   The training phase is straightforward regression. It consists of minimizing the squared error loss between a flexible function $\hat{r}(x|\theta_0, \theta_1)$ (for instance a neural network) and the training data $\{x_e, r(x_e, z_e | \theta_0 , \theta_1)\}$, which was generated according to $\theta_1$.

In the point-by-point version, the input to the regressor are just the observables $x$, and the ratio is between two fixed hypotheses $\theta_0$ and $\theta_1$. In the parameterized versions of the estimator, the ratios are based on various values $\theta_0$, while $\theta_1$ is still kept fixed. In the agnostic parameterized setup, the value of $\theta_0$ used in each event is then one of the inputs to the neural network. In the morphing-aware versions, it us used to calculate the weights $w_c(\theta_0)$ that multiply the different component networks $\hat{r}_c(x)$, as visualized in the bottom panel of Fig.~\ref{fig:neural_network_architecture}.

In all cases we can slightly improve the structure by adding samples generated according to $\theta_0$ to the training samples, regressing on $1/r$ instead of $r$ on these events. The full loss functional is given in Eq.~\eqref{eq:ratio_regression_loss}.
  \item[Calibration]
In a next step, the classifier output is optionally calibrated as discussed in Sec.~\ref{sec:probability_calibration} using isotonic regression. The calibration curve is shown in the middle panel of Fig.~\ref{fig:calibration}. We have also experimented with an additional step of expectation calibration, see Sec.~\ref{sec:expectation_calibration}.
  \item[Evaluation]
 For a given $x$ (and in the parameterized versions $\theta_0$), the regressor $\hat{r}(x | \theta_0, \theta_1)$ is evaluated. The result is optionally calibrated.
  \item[Parameters]
The parameters of this approach are the architecture of the neural network, \ie the number of layers and elements, the activation function, the optimizer used for training, its parameters, and optionally regularization terms.

For our example process we consider fully connected neural networks with two (``shallow''), three, or five (``deep'') layers of 100 neurons each and $\tanh$ activation functions. They are trained with the \toolfont{Adam} optimizer~\cite{adam} over 50 epochs with early stopping and learning rate decay.
Our default settings are given in Table~\ref{tbl:default_settings}. Experiments with different architectures, other activation functions, additional dropout layers, other optimizers, and different learning rate schedules yielded a worse performance.
\end{description}

\subsubsection{\carl + score regression (\cascal)}
\label{sec:appendix_models_cascal}

\begin{description}
  \item[Idea]
    The parameterized \carl technique learns the full statistical model $\hat{r}(x | \theta_0, \theta_1)$, including the dependency on $\theta_0$. If it is realized as a differentiable classifier (such as a neural network), we can calculate the gradient of $\hat{r}(x | \theta_0, \theta_1)$ with respect to $\theta_0$, and thus the estimated score of this model. If the estimator is perfect, we expect this estimated score to minimize the squared error with respect to the joint score data available for the training data. This is based on the same argument as the local score regression technique, see Sec.~\ref{sec:availability} for the proof.

   We can turn this argument around and use the available score information during the training. To this end, we combine two terms in a combined loss function: the \carl{}-style cross-entropy and the squared error between the estimated score and the joint score of the training data. These two pieces contain complementary information: the \carl part contains the information of the likelihood ratio for a fixed hypothesis comparison $(\theta_0, \theta_1)$, while the score part describes the relative change of the likelihood ratio under changes in $\theta_0$.
 \item[Requirements]
   The \cascal technique requires a generator with access to the joint score $t(x_e, z_e | \theta_0)$. In the particle physics case, this means we have to be able to evaluate the squared matrix elements for given phase-space points and theory parameters.
 \item[Structure]
   Since the \cascal method relies on the extraction of the estimated score from the estimator, it can only be used for parameterized estimators, either in an agnostic or morphing-aware version. The middle and bottom panels of Fig.~\ref{fig:neural_network_architecture} illustrate the structure of the estimator in these two cases.
  \item[Training]
A differentiable classifier with decision function $\hat{s}(x | \theta_0, \theta_1)$ is trained to discriminate between numerator (label $y = 0$) and denominator (label $y = 1$) samples, while the derived estimated score $\hat{t}(x | \theta_0)$ is compared to the joint score on the training samples generated from $y = 0$. The loss function that is minimized is thus a combination of the \carl{}-style cross-entropy and the squared error on the score, weighted by a hyperparameter $\alpha$. It is given in Eq.~\eqref{eq:carl_plus_score_loss}.

The numerator ($y = 0$) training samples do not come from a single parameter $\theta_0$, but rather a combination of many different subsamples. In the agnostic parameterized setup, the value of $\theta_0$ used in each event is then one of the inputs to the neural network. In the morphing-aware versions, it is used to calculate the weights $w_c(\theta_0)$ that multiply the different component networks $\hat{r}_c(x)$, as visualized in the bottom panel of Fig.~\ref{fig:neural_network_architecture}.
  \item[Calibration]
In a next step, the classifier output is optionally calibrated as discussed in Sec.~\ref{sec:probability_calibration} using isotonic regression. The calibration curve is shown in the right panel of Fig.~\ref{fig:calibration}. We have also experimented with an additional step of expectation calibration, see Sec.~\ref{sec:expectation_calibration}.
  \item[Evaluation]
 For a given $x$ and $\theta_0$, the classifier decision function $\hat{s}(x | \theta_0, \theta_1)$ is evaluated. This is turned into a likelihood ratio estimator with the relation given in Eq.~\eqref{eq:r_from_s}, and optionally calibrated.
  \item[Parameters]
The key hyperparameter of this technique is the factor $\alpha$ that weights the two terms in the loss function. Additional parameters set the architecture of the neural network, \ie the number of layers and elements, the activation function, the optimizer used for training, its parameters, and optionally regularization terms.

For our example process we consider fully connected neural networks with two (``shallow''), three, or five (``deep'') layers of 100 neurons each and $\tanh$ activation functions. They are trained with the \toolfont{Adam} optimizer~\cite{adam} over 50 epochs with early stopping and learning rate decay.
Our default settings are given in Table~\ref{tbl:default_settings}. Experiments with different architectures, other activation functions, additional dropout layers, other optimizers, and different learning rate schedules yielded a worse performance.
\end{description}

\subsubsection{Ratio + score regression (\rascal)}
\label{sec:appendix_models_rascal}

\begin{description}
  \item[Idea]
    The parameterized \rolr technique learns the full statistical model $\hat{r}(x | \theta_0, \theta_1)$, including the dependency on $\theta_0$. If it is realized as a differentiable regressor, we can calculate the gradient of $\hat{r}(x | \theta_0, \theta_1)$ with respect to $\theta_0$, and thus the score of this model. If the estimator is perfect, we expect this estimated score to minimize the squared error with respect to the joint score data available for the training data.

   We can turn this argument around and use the available likelihood ratio and score information during the training. To this end, we combine two terms in a combined loss function: the squared errors on the ratio and the score. These two pieces contain complementary information: the ratio regression part contains the information of the likelihood ratio for a fixed hypothesis comparison $(\theta_0, \theta_1)$, while the score part describes the relative change of the likelihood ratio under changes in $\theta_0$.
 \item[Requirements]
   The \rascal technique requires a generator with access to the joint likelihood ratio $r(x_e, z_e | \theta_0 , \theta_1)$  and score $t(x_e, z_e | \theta_0)$. In the particle physics case, this means we have to be able to evaluate the squared matrix elements for given phase-space points and theory parameters.
 \item[Structure]
   Since the \rascal method relies on the extraction of the estimated score from the estimator, it can only be used for parameterized estimators, either in an agnostic or morphing-aware version. The middle and bottom panels of Fig.~\ref{fig:neural_network_architecture} illustrate the structure of the estimator in these two cases.
  \item[Training]
An estimator $\hat{r}(x | \theta_0, \theta_1)$ is trained through regression on the joint likelihood ratio, while the derived estimated score $\hat{t}(x | \theta_0)$ is compared to the joint score on the training samples generated from $y = 0$. The loss function that is minimized is thus a combination of the squared error on the ratio and the squared error on the score, weighted by a hyperparameter $\alpha$. It is given in Eq.~\eqref{eq:carl_plus_score_loss}.

The numerator ($y = 0$) training samples do not come from a single parameter $\theta_0$, but rather a combination of many different subsamples. In the agnostic parameterized setup, the value of $\theta_0$ used in each event is then one of the inputs to the neural network. In the morphing-aware versions, it is used to calculate the weights $w_c(\theta_0)$ that multiply the different component networks $\hat{r}_c(x)$, as visualized in the bottom panel of Fig.~\ref{fig:neural_network_architecture}.
  \item[Calibration]
In a next step, the classifier output is optionally calibrated as discussed in Sec.~\ref{sec:probability_calibration} using isotonic regression. The calibration curve is shown in the right panel of Fig.~\ref{fig:calibration}. We have also experimented with an additional step of expectation calibration, see Sec.~\ref{sec:expectation_calibration}.
  \item[Evaluation]
 For a given $x$ and $\theta_0$, the estimator $\hat{r}(x | \theta_0, \theta_1)$ is evaluated and optionally calibrated.
  \item[Parameters]
The key hyperparameter of this technique is the factor $\alpha$ that weighs the two terms in the loss function. Additional parameters set the architecture of the neural network, \ie the number of layers and elements, the activation function, the optimizer used for training, its parameters, and optionally regularization terms.

For our example process we consider fully connected neural networks with two (``shallow''), three, or five (``deep'') layers of 100 neurons each and $\tanh$ activation functions. They are trained with the \toolfont{Adam} optimizer~\cite{adam} over 50 epochs with early stopping and learning rate decay.
Our default settings are given in Table~\ref{tbl:default_settings}. Experiments with different architectures, other activation functions, additional dropout layers, other optimizers, and different learning rate schedules yielded a worse performance.
\end{description}

\subsubsection{Local score regression and density estimation (\sally)}
\label{sec:appendix_models_sally}

\begin{description}
  \item[Idea]
    In Sec.~\ref{sec:local_model} we introduced the score, the relative gradient of the likelihood with respect to the theory parameters. The score evaluated at some reference parameter point is the sufficient statistics of the local approximation of the likelihood given in Eq.~\eqref{eq:local_model}. In other words, we expect the score vector to be a set of ``optimal observables'' that includes all the information on the theory parameters, at least in the vicinity of the reference parameter point. If we can estimate the score from an observation, we can use it like any other set of observables. In particular, we can fill histograms of the score for any parameter point and thus estimate the likelihood ratio in score space.

To estimate the score, we again make use of the particle physics structure. Particle physics generators do not only provide sets of observables $\{x_e\}$, but also the corresponding parton-level momenta $\{z_e\}$. From matrix element codes such as \toolfont{MadGraph} we can extract the squared matrix element $| \mathcal{M} |^2 (z | \theta)$ given parton-level momenta $z$ and theory parameter points $\theta$. This allows us to calculate the joint score
\begin{equation}
  t(x_e, z_e | \theta_0)
  = \nabla_\theta \log  p(z_{e} | \theta) \Biggr |_{\theta_0}
  = \frac {\nabla_\theta | \mathcal{M} |^2 (z_e | \theta_0)} {| \mathcal{M} |^2 (z_e | \theta_0)} - \frac {\nabla_\theta \sigma(\theta_0)} {\sigma(\theta_0)}
  \label{eq:joint_score_from_me}
\end{equation}
for any of the generated events. The derivatives in Eq.~\eqref{eq:joint_score_from_me} can always be evaluated numerically. If the process has the morphing structure of Eq.~\eqref{eq:morphing_structure}, one can alternatively calculate it from the morphing weights.

In Sec.~\ref{sec:availability} we have shown that regressing a function $\hat{t}(x)$ on the generated events $\{x_e\}$ and the corresponding joint scores $t(x_e, z_e | \theta)$ will converge to
\begin{equation}
  \hat{t}(x) \to \intractablet(x) =  \nabla_\theta \log \intractablep (x | \theta)  \Biggr |_{\theta_0}\,,
\end{equation}
provided that the events are sampled according to $x_e \sim \intractablep(x | \theta_0)$.

This technique is illustrated in the middle panels of Fig.~\ref{fig:inference_examples}. The middle panel of Fig.~\ref{fig:calibration} shows the relation between the scalar product of estimated score and $\theta_0 - \theta_1$ and the estimated likelihood ratio.
 \item[Requirements]
   The \sally technique requires a generator with access to the joint score $t(x_e, z_e | \theta_0)$. In the particle physics case, this means we have to be able to evaluate the squared matrix elements for given phase-space points and theory parameters.
 \item[Structure]
   The technique consists of two separate steps: the score regression and the density estimation in the estimated score space. The score regression step is independent of the tested hypothesis and realized as a simple fully connected neural network. The subsequent density estimation is realized through multi-dimensional histograms, point by point in parameter space (if the morphing condition in Eq.~\eqref{eq:morphing_structure} holds, a morphing-aware version is also possible).
  \item[Training]
   The first part of the training is regression on the score (evaluated at some reference hypothesis). It consists of minimizing the squared error loss between a flexible vector-valued function $\hat{t}(x|\theta_{\text{score}})$ (implemented for instance as a neural network) and the training data $\{x_e, t(x_e, z_e | \theta_{\text{score}})\}$, which was sampled according to $\theta_{\text{score}}$.

   The second step is density estimation in the estimated score space. We only consider histograms, but other density estimation techniques are also possible. For each value of $\theta_0$ or $\theta_1$ that is tested, we generate samples of events, estimate the corresponding score vectors, and fill a multidimensional histogram of the estimated score.
  \item[Calibration]
The density estimation step already calibrates the results, so we do not experiment with an additional calibration step.
  \item[Evaluation]
 For a given observation $x$, the score regressor $\hat{t}(x)$ is evaluated. For each tested $(\theta_0, \theta_1)$ pair, we then extract the corresponding bin contents from the numerator and denominator histograms, and calculate the estimated likelihood ratio with Eq.~\eqref{eq:score_regression_density_estimation_score}.
  \item[Parameters]
Both the score regression part and the subsequent density estimation have parameters. The first and most important choice is the reference hypothesis $\theta_{\text{score}}$, at which the score is evaluated. For effective field theories the Standard Model is the natural choice, and we use it in our example process.

The score regression also depends on the hyperparameters of the neural network, \ie the number of layers and elements, the activation function, the optimizer used for training, its parameters, and optionally regularization terms. For our example process we consider fully connected neural networks with two (``shallow''), three, or five (``deep'') layers of 100 neurons each and $\tanh$ activation functions. They are trained with the \toolfont{Adam} optimizer~\cite{adam} over 50 epochs with early stopping and learning rate decay. Our default settings are given in Table~\ref{tbl:default_settings}. Experiments with different architectures, other activation functions, additional dropout layers, other optimizers, and different learning rate schedules yielded a worse performance.

The only parameter of the density estimation stage is the histogram binning. For our example process we consider two different variations:
\begin{itemize}
\item Density estimation with a ``fixed'' binning, where the bin axes are aligned with the score components. We use 40 bins for each of the two score components.
\item Density estimation with a ``dynamic'' binning, in which the bin axes are aligned with the  $\theta_0 - \theta_1$ direction and the orthogonal one. We use 80 bins along the $\Delta \theta$ direction, which carries the relevant information in the local model approximation, and 10 along the orthogonal vector.
\end{itemize}
For each pair $(\theta_0, \theta_1)$ and each dimension, the bin edges are chosen such that the expected number of events according to $\theta_0$ plus the expected number of events according to $\theta_1$ is the same in each bin.
\end{description}

\subsubsection{Local score regression, compression to scalar, and density estimation (\sallino)}
\label{sec:appendix_models_sallino}

\begin{description}
  \item[Idea]
    In the proximity of the Standard Model (or any other reference parameter point), likelihood ratios only depend on the scalar product between the score and the difference between the numerator and denominator parameter points. If we can estimate the score from an observation, we can calculate this scalar product $\hat{h}(x | \theta_0, \theta_1)$, defined in Eq.~\eqref{eq:sallino_h}, and use it like any other observable. In particular, we can fill histograms of $\hat{h}$ for any parameter point and thus estimate the likelihood ratio in $\hat{h}$ space.

To estimate the score, we once again exploit particle physics structure. Particle physics generators do not only provide sets of observables $\{x_e\}$, but also the corresponding parton-level momenta $\{z_e\}$. From matrix element codes such as \toolfont{MadGraph} we can extract the squared matrix element $| \mathcal{M} |^2 (z | \theta)$ given parton-level momenta $z$ and theory parameter points $\theta$. This allows us to calculate the joint score with Eq.~\eqref{eq:joint_score_from_me} for any of the generated events.

In Sec.~\ref{sec:availability} we have shown that regressing a function $\hat{t}(x)$ on the generated events $\{x_e\}$ and the corresponding joint scores $t(x_e, z_e | \theta)$ will converge to $\intractablet(x)$, provided that the events are sampled according to $x_e \sim \intractablep(x | \theta_0)$.

 \item[Requirements]
   The \sallino technique requires a generator with access to the joint score $t(x_e, z_e | \theta_0)$. In the particle physics case, this means we have to be able to evaluate the squared matrix elements for given phase-space points and theory parameters.
 \item[Structure]
   The technique consists of two separate steps: the score regression, and the density estimation in $\hat{h}$ space. The score regression step is independent of the tested hypothesis and realized as a simple fully connected neural network. The subsequent density estimation in $\hat{h}$ space is realized through one-dimensional histograms, point by point in parameter space (if the morphing condition in Eq.~\eqref{eq:morphing_structure} holds, a morphing-aware version is also possible).
  \item[Training]
   The first part of the training is regression on the score (evaluated at some reference hypothesis). It consists of minimizing the squared error loss between a flexible vector-valued function $\hat{t}(x|\theta_{\text{score}})$ (implemented for instance as a neural network) and the training data $\{x_e, t(x_e, z_e | \theta_{\text{score}})\}$, which was sampled according to $\theta_{\text{score}}$.

   The second step is density estimation in $\hat{h}$ space. We only consider histograms, but other density estimation techniques are also possible. For each value of $\theta_0$ or $\theta_1$ that is tested, we generate samples of events, estimate the corresponding score vectors, calculate the scalar product in Eq.~\eqref{eq:sallino_h} to get $\hat{h}(x | \theta_0, \theta_1)$, and fill a one-dimensional histogram of this quantity.
  \item[Calibration]
The density estimation step already calibrates the results, so we do not experiment with an additional calibration step.
  \item[Evaluation]
 For a given observation $x$, the score regressor $\hat{t}(x)$ is evaluated. For each tested $(\theta_0, \theta_1)$ pair, we multiply it with $\theta_0 - \theta_1$ to get $\hat{h}(x | \theta_0, \theta_1)$, extract the corresponding bin contents from the numerator and denominator histograms, and calculate the estimated likelihood ratio with Eq.~\eqref{eq:score_regression_density_estimation_score_times_theta}.
  \item[Parameters]
Both the score regression part and the subsequent density estimation have parameters. The first and most important choice is the reference hypothesis $\theta_{\text{score}}$, at which the score is evaluated. For effective field theories the Standard Model is the natural choice, and we use it in our example process.

The score regression also depends on the hyperparameters of the neural network, \ie the number of layers and elements, the activation function, the optimizer used for training, its parameters, and optionally regularization terms. For our example process we consider fully connected neural networks with two (``shallow''), three, or five (``deep'') layers of 100 neurons each and $\tanh$ activation functions. They are trained with the \toolfont{Adam} optimizer~\cite{adam} over 50 epochs with early stopping and learning rate decay. Our default settings are given in Table~\ref{tbl:default_settings}. Experiments with different architectures, other activation functions, additional dropout layers, other optimizers, and different learning rate schedules yielded a worse performance.

The only parameter of the density estimation stage is the histogram binning.  We use 100 bins. For each pair $(\theta_0, \theta_1)$ and each dimension, the bin edges are chosen such that the same expected number of events according to $\theta_0$ plus the expected number of events according to $\theta_1$ is the same in each bin.
\end{description}

\subsection{Additional results}
\label{sec:appendix_results}

In Tbls.~\ref{tbl:detailed_results1} to \ref{tbl:detailed_results6} we compare the performance of different versions of the likelihood ratio estimators. As metric we use the expected mean squared error on $\log \intractabler (x | \theta_0, \theta_1)$ as well as a trimmed version, as defined in Sec.~\ref{sec:results_idealized}. The estimators are an extended list of those given in Table~\ref{tbl:results}, adding variations with different hyperparameter choices and the results for uncalibrated (``raw'') estimators. By default, we use neural networks with 3 hidden layers, the labels ``shallow'' and ``deep'' refer to 2 and 5 hidden layers, respectively. We highlight the versions of the estimators that were shown in the main part of this paper.

\begin{figure}[ht]
  \includegraphics[width=\textwidth]{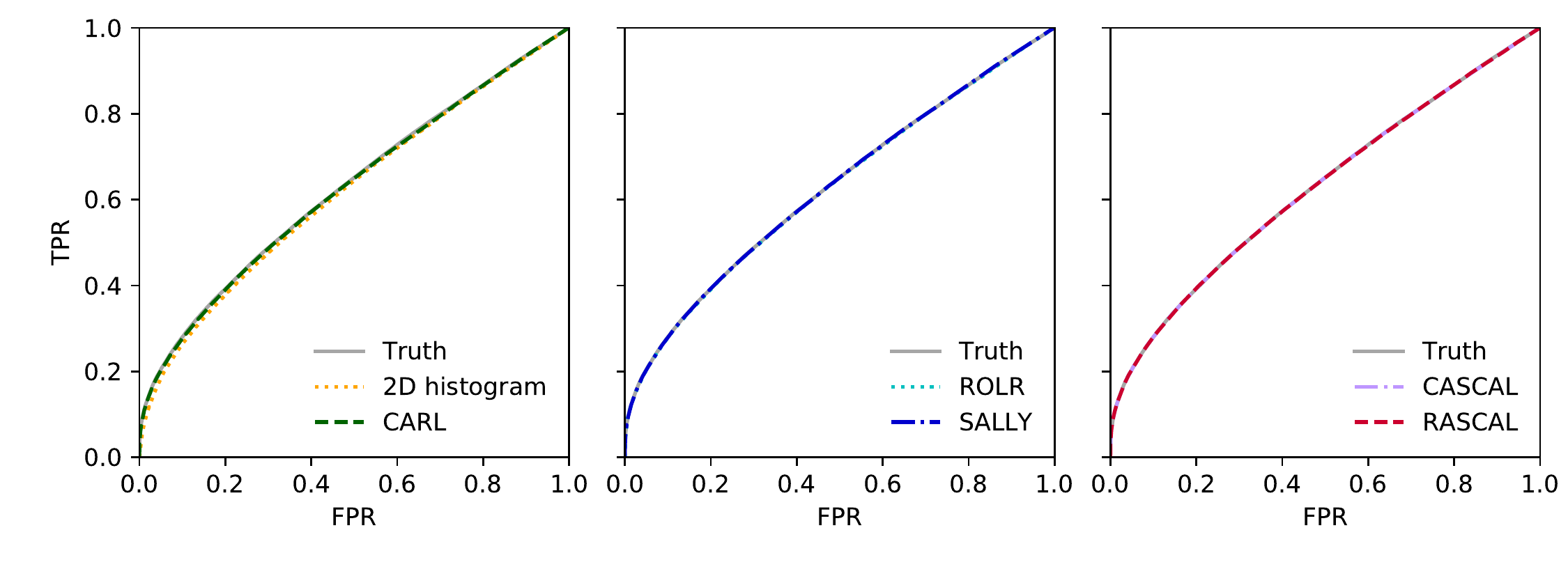}%
  \caption{Receiver operating characteristic (ROC) curves of true positive rates (TPR) vs.\ false positive rates (FPR) for the classification between the benchmark scenarios $\theta_0 = (-0.5,-0.5)^T$ and $\theta_1$ as in Eq.~\eqref{eq:denom_benchmark}. The ROC AUC based on the true likelihood is 0.6276. The results show how much the probability distributions for these two hypotheses overlap.}
  \label{fig:roc}
\end{figure}

Because of the duality between density estimation and probabilistic classification (see Eqs.~\eqref{eq:r_from_s} and \eqref{eq:s_from_r}), we can use all techniques to define classifiers. In  Fig.~\ref{fig:roc} we show the ROC curves for two benchmark parameter points. Note how badly the two scenarios can be separated. This is not a shortcoming of the discrimination power of the classifiers, but due to the genuine overlap of the probability distributions, as can be seen from the identical ROC curve based on the true likelihood ratio.

\begin{table}
  \small
  \begin{tabular}{ll rr cc}
    \toprule
    \multirow{2}{*}{Strategy} & \multirow{2}{*}{Setup} & \multicolumn{2}{c}{Expected MSE} & \multirow{2}{*}{Table~\ref{tbl:results}} & \multirow{2}{*}{Figures}\\
    \cmidrule{3-4}
    && All & Trimmed \\
    \midrule
   Histogram & $p_{T,j1}$ & $0.0879$ & $0.0230$ & \yep\\
    & $\Delta \phi_{jj}$ & $0.1595$ & $0.0433$ & \yep\\
    & 2d (coarse binning) & $0.0764$ & $0.0117$\\
    & 2d (medium binning) & $0.0630$ & $\mathbf{0.0101}$\\
    & 2d (fine binning) & $0.0597$ & $0.0115$\\
    & 2d (very fine binning) & $0.0603$ & $0.0153$\\
    & 2d (asymmetric binning) & $\mathbf{0.0561}$ & $0.0106$ & \yep  & \yep\\
   \midrule
   \afc & 2d, $\epsilon = 1$ & $0.1243$ & $0.0257$\\
    & 2d, $\epsilon = 0.5$ & $0.0797$ & $0.0144$\\
    & 2d, $\epsilon = 0.2$ & $\mathbf{0.0586}$ & $\mathbf{0.0091}$ & \yep \\
    & 2d, $\epsilon = 0.1$ & $0.0732$ & $0.0103$\\
    & 2d, $\epsilon = 0.05$ & $0.3961$ & $0.0160$\\
    & 2d, $\epsilon = 0.02$ & $13.6816$ & $0.0550$\\
    & 2d, $\epsilon = 0.01$ & $241.3264$ & $0.2143$\\
    & 5d, $\epsilon = 1$ & $0.1252$ & $0.0226$\\
    & 5d, $\epsilon = 0.5$ & $0.0779$ & $0.0101$ & \yep \\
    & 5d, $\epsilon = 0.2$ & $0.0734$ & $0.0128$\\
    & 5d, $\epsilon = 0.1$ & $0.9560$ & $0.1833$\\
    & 5d, $\epsilon = 0.05$ & $38.1854$ & $3.6658$\\
    & 5d, $\epsilon = 0.02$ & $2050.5289$ & $57.0410$\\
    & 5d, $\epsilon = 0.01$ & $50024.7997$ & $1668.8988$\\
    \bottomrule
  \end{tabular}
  \caption{Comparison of techniques based on manually selected kinematic observables. The metrics shown are the expected mean squared error on the log likelihood ratio with and without trimming, as defined in the text. Checkmarks in the last two columns denote estimators included in Table~\ref{tbl:results} and the figures in the main part of this paper, respectively.}
  \label{tbl:detailed_results1}
\end{table}

\begin{table}
  \small
  \begin{tabular}{ll rr cc}
    \toprule
    \multirow{2}{*}{Strategy} & \multirow{2}{*}{Setup} & \multicolumn{2}{c}{Expected MSE} & \multirow{2}{*}{Table~\ref{tbl:results}} & \multirow{2}{*}{Figures}\\
    \cmidrule{3-4}
    && All & Trimmed \\
   \midrule
   \carl (PbP, raw) & PbP & $\mathbf{0.0409}$ & $\mathbf{0.0213}$\\
   \midrule
   \carl (PbP, cal.) & PbP & $\mathbf{0.0301}$ & $\mathbf{0.0111}$ & Fig.~\ref{fig:pbp_parameterized_aware} \\
   \midrule
   \carl (parameterized, raw) & Baseline & $0.0157$ & $0.0040$\\
    & Baseline, shallow & $0.0134$ & $\mathbf{0.0035}$\\
    & Baseline, deep & $0.0161$ & $0.0038$\\
    & Random $\boldtheta$ & $0.0148$ & $0.0037$\\
    & Random $\boldtheta$, shallow & $\mathbf{0.0130}$ & $0.0037$\\
    & Random $\boldtheta$, deep & $0.0164$ & $0.0038$\\
   \midrule
   \carl (parameterized, cal.) & Baseline & $0.0156$ & $0.0032$\\
    & Baseline, shallow & $0.0124$ & $\mathbf{0.0026}$ & \yep  & \yep \\
    & Baseline, deep & $0.0160$ & $0.0029$\\
    & Random $\boldtheta$ & $0.0147$ & $0.0029$\\
    & Random $\boldtheta$, shallow & $\mathbf{0.0122}$ & $0.0028$ & \yep \\
    & Random $\boldtheta$, deep & $0.0155$ & $0.0029$\\
   \midrule
   \carl (morphing-aware, raw) & Baseline & $0.1598$ & $0.0350$\\
    & Baseline, shallow & $\mathbf{0.1483}$ & $\mathbf{0.0331}$\\
    & Random $\boldtheta$ & $0.1743$ & $0.0429$\\
    & Random $\boldtheta$, shallow & $0.1520$ & $0.0369$\\
    & Morphing basis, shallow & $10.1231$ & $7.9314$\\
   \midrule
   \carl (morphing-aware, cal.) & Baseline & $0.1036$ & $0.0282$\\
    & Baseline, shallow & $\mathbf{0.0762}$ & $\mathbf{0.0200}$ & \yep & Fig.~\ref{fig:pbp_parameterized_aware} \\
    & Random $\boldtheta$ & $0.1076$ & $0.0289$\\
    & Random $\boldtheta$, shallow & $0.0858$ & $0.0226$ & \yep \\
    & Morphing basis, shallow & $0.1564$ & $0.0618$ & \yep \\
    \bottomrule
  \end{tabular}
  \caption{Comparison of different versions of the \carl technique. The metrics shown are the expected mean squared error on the log likelihood ratio with and without trimming, as defined in the text. Checkmarks in the last two columns denote estimators included in Table~\ref{tbl:results} and the figures in the main part of this paper, respectively.}
  \label{tbl:detailed_results2}
\end{table}

\begin{table}
  \small
  \begin{tabular}{ll rr cc}
    \toprule
    \multirow{2}{*}{Strategy} & \multirow{2}{*}{Setup} & \multicolumn{2}{c}{Expected MSE} & \multirow{2}{*}{Table~\ref{tbl:results}} & \multirow{2}{*}{Figures}\\
    \cmidrule{3-4}
    && All & Trimmed \\
    \midrule
   \rolr (PbP, raw) & PbP & $\mathbf{0.0052}$ & $\mathbf{0.0023}$\\
   \midrule
   \rolr (PbP, cal.) & PbP & $\mathbf{0.0049}$ & $\mathbf{0.0022}$ & \yep \\
   \midrule
   \rolr (param., raw) & Baseline & $0.0034$ & $0.0019$\\
    & Baseline, shallow & $0.0069$ & $0.0037$\\
    & Baseline, deep & $0.0041$ & $0.0022$\\
    & Random $\boldtheta$ & $\mathbf{0.0034}$ & $0.0017$\\
    & Random $\boldtheta$, shallow & $0.0070$ & $0.0036$\\
    & Random $\boldtheta$, deep & $0.0036$ & $\mathbf{0.0017}$\\
   \midrule
   \rolr (param., cal.) & Baseline & $0.0032$ & $0.0017$ & \yep  & \yep \\
    & Baseline, shallow & $0.0059$ & $0.0030$\\
    & Baseline, deep & $0.0038$ & $0.0019$\\
    & Random $\boldtheta$ & $\mathbf{0.0030}$ & $\mathbf{0.0014}$ & \yep \\
    & Random $\boldtheta$, shallow & $0.0060$ & $0.0030$\\
    & Random $\boldtheta$, deep & $0.0034$ & $0.0015$\\
   \midrule
   \rolr (morph.-aware, raw) & Baseline & $0.2029$ & $0.1449$\\
    & Baseline, shallow & $0.1672$ & $0.1305$\\
    & Random $\boldtheta$ & $0.1908$ & $0.1353$\\
    & Random $\boldtheta$, shallow & $\mathbf{0.1160}$ & $\mathbf{0.0755}$\\
    & Morphing basis, shallow & $5.6668$ & $3.8335$\\
   \midrule
   \rolr (morph.-aware, cal.) & Baseline & $0.0328$ & $0.0088$\\
    & Baseline, shallow & $0.0243$ & $0.0063$ & \yep \\
    & Random $\boldtheta$ & $0.0321$ & $0.0089$\\
    & Random $\boldtheta$, shallow & $\mathbf{0.0224}$ & $\mathbf{0.0052}$ & \yep \\
    & Morphing basis, shallow & $0.1300$ & $0.0485$ & \yep \\
    \bottomrule
  \end{tabular}
  \caption{Comparison of different versions of the \rolr technique. The metrics shown are the expected mean squared error on the log likelihood ratio with and without trimming, as defined in the text. Checkmarks in the last two columns denote estimators included in Table~\ref{tbl:results} and the figures in the main part of this paper, respectively.}
  \label{tbl:detailed_results3}
\end{table}

\begin{table}
  \small
  \begin{tabular}{ll rr cc}
    \toprule
    \multirow{2}{*}{Strategy} & \multirow{2}{*}{Setup} & \multicolumn{2}{c}{Expected MSE} & \multirow{2}{*}{Table~\ref{tbl:results}} & \multirow{2}{*}{Figures}\\
    \cmidrule{3-4}
    && All & Trimmed \\
    \midrule
    \sally & Fixed 2D histogram & $0.0174$ & $0.0005$\\
    & Fixed 2D histogram, shallow & $0.0170$ & $0.0005$\\
    & Fixed 2D histogram, deep & $0.0171$ & $0.0005$\\
    & Dynamic 2D histogram & $0.0132$ & $0.0003$\\
    & Dynamic 2D histogram, shallow & $0.0133$ & $0.0003$\\
    & Dynamic 2D histogram, deep & $\mathbf{0.0132}$ & $\mathbf{0.0002}$ & \yep  & \yep \\
    \midrule
    \sallino & 1D histogram & $0.0213$ & $0.0006$\\
    & 1D histogram, shallow & $0.0215$ & $0.0007$\\
    & 1D histogram, deep & $0.0213$ & $0.0006$ & \yep \\
    \bottomrule
  \end{tabular}
  \caption{Comparison of different versions of the \sally and \sallino techniques. The metrics shown are the expected mean squared error on the log likelihood ratio with and without trimming, as defined in the text. Checkmarks in the last two columns denote estimators included in Table~\ref{tbl:results} and the figures in the main part of this paper, respectively.}
  \label{tbl:detailed_results4}
\end{table}

\begin{table}[p]
  \small
  \begin{tabular}{ll rr cc}
    \toprule
    \multirow{2}{*}{Strategy} & \multirow{2}{*}{Setup} & \multicolumn{2}{c}{Expected MSE} & \multirow{2}{*}{Table~\ref{tbl:results}} & \multirow{2}{*}{Figures}\\
    \cmidrule{3-4}
    && All & Trimmed \\
    \midrule
   \cascal (param., raw) & Baseline, $\alpha = 5$  & $0.0019$ & $0.0004$\\
    & Baseline, $\alpha = 5$ , shallow & $0.0037$ & $0.0004$\\
    & Baseline, $\alpha = 5$ , deep & $0.0010$ & $0.0003$\\
    & Baseline, $\alpha = 0.5$, deep & $0.0017$ & $0.0006$\\
    & Baseline, $\alpha = 1$, deep & $0.0014$ & $0.0005$\\
    & Baseline, $\alpha = 2$, deep & $0.0017$ & $0.0008$\\
    & Baseline, $\alpha = 10$, deep & $0.0013$ & $0.0004$\\
    & Baseline, $\alpha = 20$, deep & $0.0016$ & $0.0004$\\
    & Baseline, $\alpha = 50$, deep & $0.0024$ & $0.0007$\\
    & Random $\boldtheta$ & $0.0022$ & $0.0006$\\
    & Random $\boldtheta$, shallow & $0.0038$ & $0.0005$\\
    & Random $\boldtheta$, deep & $\mathbf{0.0010}$ & $\mathbf{0.0003}$\\
   \midrule
   \cascal (param., cal.) & Baseline, $\alpha = 5$  & $0.0012$ & $0.0002$\\
    & Baseline, $\alpha = 5$ , shallow & $0.0025$ & $0.0003$\\
    & Baseline, $\alpha = 5$ , deep & $\mathbf{0.0008}$ & $0.0002$ & \yep  & \yep \\
    & Baseline, $\alpha = 0.5$, deep & $0.0013$ & $0.0003$\\
    & Baseline, $\alpha = 1$, deep & $0.0011$ & $0.0003$\\
    & Baseline, $\alpha = 2$, deep & $0.0010$ & $\mathbf{0.0002}$\\
    & Baseline, $\alpha = 10$, deep & $0.0010$ & $0.0003$\\
    & Baseline, $\alpha = 20$, deep & $0.0011$ & $0.0003$\\
    & Baseline, $\alpha = 50$, deep & $0.0016$ & $0.0005$\\
    & Random $\boldtheta$, $\alpha = 5$  & $0.0013$ & $0.0003$\\
    & Random $\boldtheta$, $\alpha = 5$ , shallow & $0.0027$ & $0.0004$\\
    & Random $\boldtheta$, $\alpha = 5$ , deep & $0.0009$ & $0.0002$ & \yep \\
   \midrule
   \cascal (morph.-aw., raw) & Baseline, $\alpha = 5$  & $0.1935$ & $0.0810$\\
    & Baseline, $\alpha = 5$ , shallow & $0.1870$ & $0.0732$\\
    & Random $\boldtheta$, $\alpha = 5$ , shallow & $0.1624$ & $0.0643$\\
    & Morph.\ basis, $\alpha = 5$ , shallow & $\mathbf{0.0707}$ & $\mathbf{0.0109}$\\
   \midrule
   \cascal (morph.-aw., cal.) & Baseline, $\alpha = 5$  & $0.1408$ & $0.0508$\\
    & Baseline, $\alpha = 5$ , shallow & $0.1359$ & $0.0427$ & \yep \\
    & Random $\boldtheta$, $\alpha = 5$ , shallow & $0.0922$ & $0.0268$ & \yep \\
    & Morph.\ basis, $\alpha = 5$ , shallow & $\mathbf{0.0403}$ & $\mathbf{0.0081}$ & \yep \\
    \bottomrule
  \end{tabular}
  \caption{Comparison of different versions of the \cascal technique. The metrics shown are the expected mean squared error on the log likelihood ratio with and without trimming, as defined in the text. Checkmarks in the last two columns denote estimators included in Table~\ref{tbl:results} and the figures in the main part of this paper, respectively.}
  \label{tbl:detailed_results5}
\end{table}

\begin{table}[p]
  \small
  \begin{tabular}{ll rr cc}
    \toprule
    \multirow{2}{*}{Strategy} & \multirow{2}{*}{Setup} & \multicolumn{2}{c}{Expected MSE} & \multirow{2}{*}{Table~\ref{tbl:results}} & \multirow{2}{*}{Figures}\\
    \cmidrule{3-4}
    && All & Trimmed \\
    \midrule
   \rascal (param., raw) & Baseline, $\alpha = 100$ & $0.0010$ & $\mathbf{0.0003}$\\
    & Baseline, $\alpha = 100$, shallow & $0.0025$ & $0.0006$\\
    & Baseline, $\alpha = 100$, deep & $0.0009$ & $0.0004$\\
    & Baseline, $\alpha = 10$, deep & $0.0011$ & $0.0005$\\
    & Baseline, $\alpha = 20$, deep & $0.0009$ & $0.0004$\\
    & Baseline, $\alpha = 50$, deep & $0.0009$ & $0.0004$\\
    & Baseline, $\alpha = 200$, deep & $0.0009$ & $0.0004$\\
    & Baseline, $\alpha = 500$, deep & $0.0011$ & $0.0006$\\
    & Baseline, $\alpha = 1000$, deep & $0.0012$ & $0.0007$\\
    & Random $\boldtheta$ & $0.0011$ & $0.0004$\\
    & Random $\boldtheta$, shallow & $0.0030$ & $0.0010$\\
    & Random $\boldtheta$, deep & $\mathbf{0.0008}$ & $0.0004$\\
   \midrule
   \rascal (param., cal.) & Baseline, $\alpha = 100$ & $0.0010$ & $\mathbf{0.0003}$\\
    & Baseline, $\alpha = 100$, shallow & $0.0021$ & $0.0005$\\
    & Baseline, $\alpha = 100$, deep & $0.0009$ & $0.0004$ & \yep  & \yep \\
    & Baseline, $\alpha = 10$, deep & $0.0010$ & $0.0004$\\
    & Baseline, $\alpha = 20$, deep & $0.0009$ & $0.0004$\\
    & Baseline, $\alpha = 50$, deep & $0.0009$ & $0.0004$\\
    & Baseline, $\alpha = 200$, deep & $0.0008$ & $0.0004$\\
    & Baseline, $\alpha = 500$, deep & $0.0009$ & $0.0005$\\
    & Baseline, $\alpha = 1000$, deep & $0.0012$ & $0.0006$\\
    & Random $\boldtheta$, $\alpha = 100$ & $0.0010$ & $0.0004$\\
    & Random $\boldtheta$, $\alpha = 100$, shallow & $0.0025$ & $0.0008$\\
    & Random $\boldtheta$, $\alpha = 100$, deep & $\mathbf{0.0008}$ & $0.0004$ & \yep \\
   \midrule
   \rascal (morph.-aw., raw) & Baseline, $\alpha = 100$ & $0.2880$ & $0.2024$\\
    & Baseline, $\alpha = 100$, shallow & $0.3569$ & $0.2861$\\
    & Random $\boldtheta$, $\alpha = 100$ & $0.2705$ & $0.1825$\\
    & Random $\boldtheta$, $\alpha = 100$, shallow & $0.3243$ & $0.2488$\\
    & Morph.\ basis, $\alpha = 100$, shallow & $\mathbf{0.1909}$ & $\mathbf{0.1673}$\\
   \midrule
   \rascal (morph.-aw., cal.) & Baseline, $\alpha = 100$ & $0.1530$ & $0.0673$\\
    & Baseline, $\alpha = 100$, shallow & $0.1250$ & $0.0514$ & \yep \\
    & Random $\boldtheta$, $\alpha = 100$ & $0.1358$ & $0.0627$\\
    & Random $\boldtheta$, $\alpha = 100$, shallow & $0.1316$ & $0.0539$ & \yep \\
    & Morph.\ basis, shallow, $\alpha = 100$ & $\mathbf{0.0307}$ & $\mathbf{0.0072}$ & \yep \\
    \bottomrule
  \end{tabular}
  \caption{Comparison of different versions of the \rascal technique. The metrics shown are the expected mean squared error on the log likelihood ratio with and without trimming, as defined in the text. Checkmarks in the last two columns denote estimators included in Table~\ref{tbl:results} and the figures in the main part of this paper, respectively.}
  \label{tbl:detailed_results6}
\end{table}


\end{fmffile}

\clearpage

\bibliography{references}

\begin{thebibliography}{93}%
\makeatletter
\providecommand \@ifxundefined [1]{%
 \@ifx{#1\undefined}
}%
\providecommand \@ifnum [1]{%
 \ifnum #1\expandafter \@firstoftwo
 \else \expandafter \@secondoftwo
 \fi
}%
\providecommand \@ifx [1]{%
 \ifx #1\expandafter \@firstoftwo
 \else \expandafter \@secondoftwo
 \fi
}%
\providecommand \natexlab [1]{#1}%
\providecommand \enquote  [1]{``#1''}%
\providecommand \bibnamefont  [1]{#1}%
\providecommand \bibfnamefont [1]{#1}%
\providecommand \citenamefont [1]{#1}%
\providecommand \href@noop [0]{\@secondoftwo}%
\providecommand \href [0]{\begingroup \@sanitize@url \@href}%
\providecommand \@href[1]{\@@startlink{#1}\@@href}%
\providecommand \@@href[1]{\endgroup#1\@@endlink}%
\providecommand \@sanitize@url [0]{\catcode `\\12\catcode `\$12\catcode
  `\&12\catcode `\#12\catcode `\^12\catcode `\_12\catcode `\%12\relax}%
\providecommand \@@startlink[1]{}%
\providecommand \@@endlink[0]{}%
\providecommand \url  [0]{\begingroup\@sanitize@url \@url }%
\providecommand \@url [1]{\endgroup\@href {#1}{\urlprefix }}%
\providecommand \urlprefix  [0]{URL }%
\providecommand \Eprint [0]{\href }%
\providecommand \doibase [0]{http://dx.doi.org/}%
\providecommand \selectlanguage [0]{\@gobble}%
\providecommand \bibinfo  [0]{\@secondoftwo}%
\providecommand \bibfield  [0]{\@secondoftwo}%
\providecommand \translation [1]{[#1]}%
\providecommand \BibitemOpen [0]{}%
\providecommand \bibitemStop [0]{}%
\providecommand \bibitemNoStop [0]{.\EOS\space}%
\providecommand \EOS [0]{\spacefactor3000\relax}%
\providecommand \BibitemShut  [1]{\csname bibitem#1\endcsname}%
\let\auto@bib@innerbib\@empty
\bibitem [{\citenamefont {Sjostrand}\ \emph {et~al.}(2008)\citenamefont
  {Sjostrand}, \citenamefont {Mrenna},\ and\ \citenamefont
  {Skands}}]{Sjostrand:2007gs}%
  \BibitemOpen
  \bibfield  {author} {\bibinfo {author} {\bibfnamefont {T.}~\bibnamefont
  {Sjostrand}}, \bibinfo {author} {\bibfnamefont {S.}~\bibnamefont {Mrenna}}, \
  and\ \bibinfo {author} {\bibfnamefont {P.~Z.}\ \bibnamefont {Skands}},\
  }\href {\doibase 10.1016/j.cpc.2008.01.036} {\bibfield  {journal} {\bibinfo
  {journal} {Comput. Phys. Commun.}\ }\textbf {\bibinfo {volume} {178}},\
  \bibinfo {pages} {852} (\bibinfo {year} {2008})},\ \Eprint
  {http://arxiv.org/abs/0710.3820} {arXiv:0710.3820} \BibitemShut {NoStop}%
\bibitem [{\citenamefont {Agostinelli}\ \emph {et~al.}(2003)\citenamefont
  {Agostinelli} \emph {et~al.}}]{Agostinelli:2002hh}%
  \BibitemOpen
  \bibfield  {author} {\bibinfo {author} {\bibfnamefont {S.}~\bibnamefont
  {Agostinelli}} \emph {et~al.} (\bibinfo {collaboration} {GEANT4}),\ }\href
  {\doibase 10.1016/S0168-9002(03)01368-8} {\bibfield  {journal} {\bibinfo
  {journal} {Nucl. Instrum. Meth.}\ }\textbf {\bibinfo {volume} {A506}},\
  \bibinfo {pages} {250} (\bibinfo {year} {2003})}\BibitemShut {NoStop}%
\bibitem [{\citenamefont {Cranmer}(2001)}]{Cranmer:2000du}%
  \BibitemOpen
  \bibfield  {author} {\bibinfo {author} {\bibfnamefont {K.~S.}\ \bibnamefont
  {Cranmer}},\ }\href {\doibase 10.1016/S0010-4655(00)00243-5} {\bibfield
  {journal} {\bibinfo  {journal} {Comput. Phys. Commun.}\ }\textbf {\bibinfo
  {volume} {136}},\ \bibinfo {pages} {198} (\bibinfo {year} {2001})},\ \Eprint
  {http://arxiv.org/abs/hep-ex/0011057} {arXiv:hep-ex/0011057 [hep-ex]}
  \BibitemShut {NoStop}%
\bibitem [{\citenamefont {Cranmer}\ \emph {et~al.}(2012)\citenamefont
  {Cranmer}, \citenamefont {Lewis}, \citenamefont {Moneta}, \citenamefont
  {Shibata},\ and\ \citenamefont {Verkerke}}]{Cranmer:2012sba}%
  \BibitemOpen
  \bibfield  {author} {\bibinfo {author} {\bibfnamefont {K.}~\bibnamefont
  {Cranmer}}, \bibinfo {author} {\bibfnamefont {G.}~\bibnamefont {Lewis}},
  \bibinfo {author} {\bibfnamefont {L.}~\bibnamefont {Moneta}}, \bibinfo
  {author} {\bibfnamefont {A.}~\bibnamefont {Shibata}}, \ and\ \bibinfo
  {author} {\bibfnamefont {W.}~\bibnamefont {Verkerke}} (\bibinfo
  {collaboration} {ROOT}),\ }\href@noop {} {\  (\bibinfo {year}
  {2012})}\BibitemShut {NoStop}%
\bibitem [{\citenamefont {Frate}\ \emph {et~al.}(2017)\citenamefont {Frate},
  \citenamefont {Cranmer}, \citenamefont {Kalia}, \citenamefont
  {Vandenberg-Rodes},\ and\ \citenamefont {Whiteson}}]{Frate:2017mai}%
  \BibitemOpen
  \bibfield  {author} {\bibinfo {author} {\bibfnamefont {M.}~\bibnamefont
  {Frate}}, \bibinfo {author} {\bibfnamefont {K.}~\bibnamefont {Cranmer}},
  \bibinfo {author} {\bibfnamefont {S.}~\bibnamefont {Kalia}}, \bibinfo
  {author} {\bibfnamefont {A.}~\bibnamefont {Vandenberg-Rodes}}, \ and\
  \bibinfo {author} {\bibfnamefont {D.}~\bibnamefont {Whiteson}},\ }\href@noop
  {} {\  (\bibinfo {year} {2017})},\ \Eprint {http://arxiv.org/abs/1709.05681}
  {arXiv:1709.05681 [physics.data-an]} \BibitemShut {NoStop}%
\bibitem [{\citenamefont {Brehmer}\ \emph
  {et~al.}(2017{\natexlab{a}})\citenamefont {Brehmer}, \citenamefont {Cranmer},
  \citenamefont {Kling},\ and\ \citenamefont {Plehn}}]{Brehmer:2016nyr}%
  \BibitemOpen
  \bibfield  {author} {\bibinfo {author} {\bibfnamefont {J.}~\bibnamefont
  {Brehmer}}, \bibinfo {author} {\bibfnamefont {K.}~\bibnamefont {Cranmer}},
  \bibinfo {author} {\bibfnamefont {F.}~\bibnamefont {Kling}}, \ and\ \bibinfo
  {author} {\bibfnamefont {T.}~\bibnamefont {Plehn}},\ }\href {\doibase
  10.1103/PhysRevD.95.073002} {\bibfield  {journal} {\bibinfo  {journal} {Phys.
  Rev.}\ }\textbf {\bibinfo {volume} {D95}},\ \bibinfo {pages} {073002}
  (\bibinfo {year} {2017}{\natexlab{a}})},\ \Eprint
  {http://arxiv.org/abs/1612.05261} {arXiv:1612.05261 [hep-ph]} \BibitemShut
  {NoStop}%
\bibitem [{\citenamefont {Kondo}(1988)}]{Kondo:1988yd}%
  \BibitemOpen
  \bibfield  {author} {\bibinfo {author} {\bibfnamefont {K.}~\bibnamefont
  {Kondo}},\ }\href {\doibase 10.1143/JPSJ.57.4126} {\bibfield  {journal}
  {\bibinfo  {journal} {J. Phys. Soc. Jap.}\ }\textbf {\bibinfo {volume}
  {57}},\ \bibinfo {pages} {4126} (\bibinfo {year} {1988})}\BibitemShut
  {NoStop}%
\bibitem [{\citenamefont {Abazov}\ \emph {et~al.}(2004)\citenamefont {Abazov}
  \emph {et~al.}}]{Abazov:2004cs}%
  \BibitemOpen
  \bibfield  {author} {\bibinfo {author} {\bibfnamefont {V.~M.}\ \bibnamefont
  {Abazov}} \emph {et~al.} (\bibinfo {collaboration} {D0}),\ }\href {\doibase
  10.1038/nature02589} {\bibfield  {journal} {\bibinfo  {journal} {Nature}\
  }\textbf {\bibinfo {volume} {429}},\ \bibinfo {pages} {638} (\bibinfo {year}
  {2004})},\ \Eprint {http://arxiv.org/abs/hep-ex/0406031}
  {arXiv:hep-ex/0406031 [hep-ex]} \BibitemShut {NoStop}%
\bibitem [{\citenamefont {Artoisenet}\ and\ \citenamefont
  {Mattelaer}(2008)}]{Artoisenet:2008zz}%
  \BibitemOpen
  \bibfield  {author} {\bibinfo {author} {\bibfnamefont {P.}~\bibnamefont
  {Artoisenet}}\ and\ \bibinfo {author} {\bibfnamefont {O.}~\bibnamefont
  {Mattelaer}},\ }\bibfield  {booktitle} {\emph {\bibinfo {booktitle}
  {{Proceedings, 2nd International Workshop on Prospects for charged Higgs
  discovery at colliders (CHARGED 2008): Uppsala, Sweden, September 16-19,
  2008}}},\ }\href@noop {} {\bibfield  {journal} {\bibinfo  {journal} {PoS}\
  }\textbf {\bibinfo {volume} {CHARGED2008}},\ \bibinfo {pages} {025} (\bibinfo
  {year} {2008})}\BibitemShut {NoStop}%
\bibitem [{\citenamefont {Gao}\ \emph {et~al.}(2010)\citenamefont {Gao},
  \citenamefont {Gritsan}, \citenamefont {Guo}, \citenamefont {Melnikov},
  \citenamefont {Schulze},\ and\ \citenamefont {Tran}}]{Gao:2010qx}%
  \BibitemOpen
  \bibfield  {author} {\bibinfo {author} {\bibfnamefont {Y.}~\bibnamefont
  {Gao}}, \bibinfo {author} {\bibfnamefont {A.~V.}\ \bibnamefont {Gritsan}},
  \bibinfo {author} {\bibfnamefont {Z.}~\bibnamefont {Guo}}, \bibinfo {author}
  {\bibfnamefont {K.}~\bibnamefont {Melnikov}}, \bibinfo {author}
  {\bibfnamefont {M.}~\bibnamefont {Schulze}}, \ and\ \bibinfo {author}
  {\bibfnamefont {N.~V.}\ \bibnamefont {Tran}},\ }\href {\doibase
  10.1103/PhysRevD.81.075022} {\bibfield  {journal} {\bibinfo  {journal} {Phys.
  Rev.}\ }\textbf {\bibinfo {volume} {D81}},\ \bibinfo {pages} {075022}
  (\bibinfo {year} {2010})},\ \Eprint {http://arxiv.org/abs/1001.3396}
  {arXiv:1001.3396 [hep-ph]} \BibitemShut {NoStop}%
\bibitem [{\citenamefont {Alwall}\ \emph {et~al.}(2011)\citenamefont {Alwall},
  \citenamefont {Freitas},\ and\ \citenamefont {Mattelaer}}]{Alwall:2010cq}%
  \BibitemOpen
  \bibfield  {author} {\bibinfo {author} {\bibfnamefont {J.}~\bibnamefont
  {Alwall}}, \bibinfo {author} {\bibfnamefont {A.}~\bibnamefont {Freitas}}, \
  and\ \bibinfo {author} {\bibfnamefont {O.}~\bibnamefont {Mattelaer}},\ }\href
  {\doibase 10.1103/PhysRevD.83.074010} {\bibfield  {journal} {\bibinfo
  {journal} {Phys. Rev.}\ }\textbf {\bibinfo {volume} {D83}},\ \bibinfo {pages}
  {074010} (\bibinfo {year} {2011})},\ \Eprint {http://arxiv.org/abs/1010.2263}
  {arXiv:1010.2263 [hep-ph]} \BibitemShut {NoStop}%
\bibitem [{\citenamefont {Bolognesi}\ \emph {et~al.}(2012)\citenamefont
  {Bolognesi}, \citenamefont {Gao}, \citenamefont {Gritsan}, \citenamefont
  {Melnikov}, \citenamefont {Schulze}, \citenamefont {Tran},\ and\
  \citenamefont {Whitbeck}}]{Bolognesi:2012mm}%
  \BibitemOpen
  \bibfield  {author} {\bibinfo {author} {\bibfnamefont {S.}~\bibnamefont
  {Bolognesi}}, \bibinfo {author} {\bibfnamefont {Y.}~\bibnamefont {Gao}},
  \bibinfo {author} {\bibfnamefont {A.~V.}\ \bibnamefont {Gritsan}}, \bibinfo
  {author} {\bibfnamefont {K.}~\bibnamefont {Melnikov}}, \bibinfo {author}
  {\bibfnamefont {M.}~\bibnamefont {Schulze}}, \bibinfo {author} {\bibfnamefont
  {N.~V.}\ \bibnamefont {Tran}}, \ and\ \bibinfo {author} {\bibfnamefont
  {A.}~\bibnamefont {Whitbeck}},\ }\href {\doibase 10.1103/PhysRevD.86.095031}
  {\bibfield  {journal} {\bibinfo  {journal} {Phys. Rev.}\ }\textbf {\bibinfo
  {volume} {D86}},\ \bibinfo {pages} {095031} (\bibinfo {year} {2012})},\
  \Eprint {http://arxiv.org/abs/1208.4018} {arXiv:1208.4018 [hep-ph]}
  \BibitemShut {NoStop}%
\bibitem [{\citenamefont {Avery}\ \emph {et~al.}(2013)\citenamefont {Avery}
  \emph {et~al.}}]{Avery:2012um}%
  \BibitemOpen
  \bibfield  {author} {\bibinfo {author} {\bibfnamefont {P.}~\bibnamefont
  {Avery}} \emph {et~al.},\ }\href {\doibase 10.1103/PhysRevD.87.055006}
  {\bibfield  {journal} {\bibinfo  {journal} {Phys. Rev.}\ }\textbf {\bibinfo
  {volume} {D87}},\ \bibinfo {pages} {055006} (\bibinfo {year} {2013})},\
  \Eprint {http://arxiv.org/abs/1210.0896} {arXiv:1210.0896 [hep-ph]}
  \BibitemShut {NoStop}%
\bibitem [{\citenamefont {Andersen}\ \emph {et~al.}(2013)\citenamefont
  {Andersen}, \citenamefont {Englert},\ and\ \citenamefont
  {Spannowsky}}]{Andersen:2012kn}%
  \BibitemOpen
  \bibfield  {author} {\bibinfo {author} {\bibfnamefont {J.~R.}\ \bibnamefont
  {Andersen}}, \bibinfo {author} {\bibfnamefont {C.}~\bibnamefont {Englert}}, \
  and\ \bibinfo {author} {\bibfnamefont {M.}~\bibnamefont {Spannowsky}},\
  }\href {\doibase 10.1103/PhysRevD.87.015019} {\bibfield  {journal} {\bibinfo
  {journal} {Phys. Rev.}\ }\textbf {\bibinfo {volume} {D87}},\ \bibinfo {pages}
  {015019} (\bibinfo {year} {2013})},\ \Eprint {http://arxiv.org/abs/1211.3011}
  {arXiv:1211.3011 [hep-ph]} \BibitemShut {NoStop}%
\bibitem [{\citenamefont {Campbell}\ \emph {et~al.}(2013)\citenamefont
  {Campbell}, \citenamefont {Ellis}, \citenamefont {Giele},\ and\ \citenamefont
  {Williams}}]{Campbell:2013hz}%
  \BibitemOpen
  \bibfield  {author} {\bibinfo {author} {\bibfnamefont {J.~M.}\ \bibnamefont
  {Campbell}}, \bibinfo {author} {\bibfnamefont {R.~K.}\ \bibnamefont {Ellis}},
  \bibinfo {author} {\bibfnamefont {W.~T.}\ \bibnamefont {Giele}}, \ and\
  \bibinfo {author} {\bibfnamefont {C.}~\bibnamefont {Williams}},\ }\href
  {\doibase 10.1103/PhysRevD.87.073005} {\bibfield  {journal} {\bibinfo
  {journal} {Phys. Rev.}\ }\textbf {\bibinfo {volume} {D87}},\ \bibinfo {pages}
  {073005} (\bibinfo {year} {2013})},\ \Eprint {http://arxiv.org/abs/1301.7086}
  {arXiv:1301.7086 [hep-ph]} \BibitemShut {NoStop}%
\bibitem [{\citenamefont {Artoisenet}\ \emph {et~al.}(2013)\citenamefont
  {Artoisenet}, \citenamefont {de~Aquino}, \citenamefont {Maltoni},\ and\
  \citenamefont {Mattelaer}}]{Artoisenet:2013vfa}%
  \BibitemOpen
  \bibfield  {author} {\bibinfo {author} {\bibfnamefont {P.}~\bibnamefont
  {Artoisenet}}, \bibinfo {author} {\bibfnamefont {P.}~\bibnamefont
  {de~Aquino}}, \bibinfo {author} {\bibfnamefont {F.}~\bibnamefont {Maltoni}},
  \ and\ \bibinfo {author} {\bibfnamefont {O.}~\bibnamefont {Mattelaer}},\
  }\href {\doibase 10.1103/PhysRevLett.111.091802} {\bibfield  {journal}
  {\bibinfo  {journal} {Phys. Rev. Lett.}\ }\textbf {\bibinfo {volume} {111}},\
  \bibinfo {pages} {091802} (\bibinfo {year} {2013})},\ \Eprint
  {http://arxiv.org/abs/1304.6414} {arXiv:1304.6414 [hep-ph]} \BibitemShut
  {NoStop}%
\bibitem [{\citenamefont {Gainer}\ \emph {et~al.}(2013)\citenamefont {Gainer},
  \citenamefont {Lykken}, \citenamefont {Matchev}, \citenamefont {Mrenna},\
  and\ \citenamefont {Park}}]{Gainer:2013iya}%
  \BibitemOpen
  \bibfield  {author} {\bibinfo {author} {\bibfnamefont {J.~S.}\ \bibnamefont
  {Gainer}}, \bibinfo {author} {\bibfnamefont {J.}~\bibnamefont {Lykken}},
  \bibinfo {author} {\bibfnamefont {K.~T.}\ \bibnamefont {Matchev}}, \bibinfo
  {author} {\bibfnamefont {S.}~\bibnamefont {Mrenna}}, \ and\ \bibinfo {author}
  {\bibfnamefont {M.}~\bibnamefont {Park}},\ }in\ \href
  {http://inspirehep.net/record/1242444/files/arXiv:1307.3546.pdf} {\emph
  {\bibinfo {booktitle} {{Proceedings, 2013 Community Summer Study on the
  Future of U.S. Particle Physics: Snowmass on the Mississippi (CSS2013):
  Minneapolis, MN, USA, July 29-August 6, 2013}}}}\ (\bibinfo {year} {2013})\
  \Eprint {http://arxiv.org/abs/1307.3546} {arXiv:1307.3546 [hep-ph]}
  \BibitemShut {NoStop}%
\bibitem [{\citenamefont {Schouten}\ \emph {et~al.}(2015)\citenamefont
  {Schouten}, \citenamefont {DeAbreu},\ and\ \citenamefont
  {Stelzer}}]{Schouten:2014yza}%
  \BibitemOpen
  \bibfield  {author} {\bibinfo {author} {\bibfnamefont {D.}~\bibnamefont
  {Schouten}}, \bibinfo {author} {\bibfnamefont {A.}~\bibnamefont {DeAbreu}}, \
  and\ \bibinfo {author} {\bibfnamefont {B.}~\bibnamefont {Stelzer}},\ }\href
  {\doibase 10.1016/j.cpc.2015.02.020} {\bibfield  {journal} {\bibinfo
  {journal} {Comput. Phys. Commun.}\ }\textbf {\bibinfo {volume} {192}},\
  \bibinfo {pages} {54} (\bibinfo {year} {2015})},\ \Eprint
  {http://arxiv.org/abs/1407.7595} {arXiv:1407.7595 [physics.comp-ph]}
  \BibitemShut {NoStop}%
\bibitem [{\citenamefont {Martini}\ and\ \citenamefont
  {Uwer}(2015)}]{Martini:2015fsa}%
  \BibitemOpen
  \bibfield  {author} {\bibinfo {author} {\bibfnamefont {T.}~\bibnamefont
  {Martini}}\ and\ \bibinfo {author} {\bibfnamefont {P.}~\bibnamefont {Uwer}},\
  }\href {\doibase 10.1007/JHEP09(2015)083} {\bibfield  {journal} {\bibinfo
  {journal} {JHEP}\ }\textbf {\bibinfo {volume} {09}},\ \bibinfo {pages} {083}
  (\bibinfo {year} {2015})},\ \Eprint {http://arxiv.org/abs/1506.08798}
  {arXiv:1506.08798 [hep-ph]} \BibitemShut {NoStop}%
\bibitem [{\citenamefont {Gritsan}\ \emph {et~al.}(2016)\citenamefont
  {Gritsan}, \citenamefont {R{\"o}ntsch}, \citenamefont {Schulze},\ and\
  \citenamefont {Xiao}}]{Gritsan:2016hjl}%
  \BibitemOpen
  \bibfield  {author} {\bibinfo {author} {\bibfnamefont {A.~V.}\ \bibnamefont
  {Gritsan}}, \bibinfo {author} {\bibfnamefont {R.}~\bibnamefont
  {R{\"o}ntsch}}, \bibinfo {author} {\bibfnamefont {M.}~\bibnamefont
  {Schulze}}, \ and\ \bibinfo {author} {\bibfnamefont {M.}~\bibnamefont
  {Xiao}},\ }\href {\doibase 10.1103/PhysRevD.94.055023} {\bibfield  {journal}
  {\bibinfo  {journal} {Phys. Rev.}\ }\textbf {\bibinfo {volume} {D94}},\
  \bibinfo {pages} {055023} (\bibinfo {year} {2016})},\ \Eprint
  {http://arxiv.org/abs/1606.03107} {arXiv:1606.03107 [hep-ph]} \BibitemShut
  {NoStop}%
\bibitem [{\citenamefont {Martini}\ and\ \citenamefont
  {Uwer}(2017)}]{Martini:2017ydu}%
  \BibitemOpen
  \bibfield  {author} {\bibinfo {author} {\bibfnamefont {T.}~\bibnamefont
  {Martini}}\ and\ \bibinfo {author} {\bibfnamefont {P.}~\bibnamefont {Uwer}},\
  }\href@noop {} {\  (\bibinfo {year} {2017})},\ \Eprint
  {http://arxiv.org/abs/1712.04527} {arXiv:1712.04527 [hep-ph]} \BibitemShut
  {NoStop}%
\bibitem [{\citenamefont {Atwood}\ and\ \citenamefont
  {Soni}(1992)}]{Atwood:1991ka}%
  \BibitemOpen
  \bibfield  {author} {\bibinfo {author} {\bibfnamefont {D.}~\bibnamefont
  {Atwood}}\ and\ \bibinfo {author} {\bibfnamefont {A.}~\bibnamefont {Soni}},\
  }\href {\doibase 10.1103/PhysRevD.45.2405} {\bibfield  {journal} {\bibinfo
  {journal} {Phys. Rev.}\ }\textbf {\bibinfo {volume} {D45}},\ \bibinfo {pages}
  {2405} (\bibinfo {year} {1992})}\BibitemShut {NoStop}%
\bibitem [{\citenamefont {Davier}\ \emph {et~al.}(1993)\citenamefont {Davier},
  \citenamefont {Duflot}, \citenamefont {Le~Diberder},\ and\ \citenamefont
  {Rouge}}]{Davier:1992nw}%
  \BibitemOpen
  \bibfield  {author} {\bibinfo {author} {\bibfnamefont {M.}~\bibnamefont
  {Davier}}, \bibinfo {author} {\bibfnamefont {L.}~\bibnamefont {Duflot}},
  \bibinfo {author} {\bibfnamefont {F.}~\bibnamefont {Le~Diberder}}, \ and\
  \bibinfo {author} {\bibfnamefont {A.}~\bibnamefont {Rouge}},\ }\href
  {\doibase 10.1016/0370-2693(93)90101-M} {\bibfield  {journal} {\bibinfo
  {journal} {Phys. Lett.}\ }\textbf {\bibinfo {volume} {B306}},\ \bibinfo
  {pages} {411} (\bibinfo {year} {1993})}\BibitemShut {NoStop}%
\bibitem [{\citenamefont {Diehl}\ and\ \citenamefont
  {Nachtmann}(1994)}]{Diehl:1993br}%
  \BibitemOpen
  \bibfield  {author} {\bibinfo {author} {\bibfnamefont {M.}~\bibnamefont
  {Diehl}}\ and\ \bibinfo {author} {\bibfnamefont {O.}~\bibnamefont
  {Nachtmann}},\ }\href {\doibase 10.1007/BF01555899} {\bibfield  {journal}
  {\bibinfo  {journal} {Z. Phys.}\ }\textbf {\bibinfo {volume} {C62}},\
  \bibinfo {pages} {397} (\bibinfo {year} {1994})}\BibitemShut {NoStop}%
\bibitem [{\citenamefont {Soper}\ and\ \citenamefont
  {Spannowsky}(2011)}]{Soper:2011cr}%
  \BibitemOpen
  \bibfield  {author} {\bibinfo {author} {\bibfnamefont {D.~E.}\ \bibnamefont
  {Soper}}\ and\ \bibinfo {author} {\bibfnamefont {M.}~\bibnamefont
  {Spannowsky}},\ }\href {\doibase 10.1103/PhysRevD.84.074002} {\bibfield
  {journal} {\bibinfo  {journal} {Phys. Rev.}\ }\textbf {\bibinfo {volume}
  {D84}},\ \bibinfo {pages} {074002} (\bibinfo {year} {2011})},\ \Eprint
  {http://arxiv.org/abs/1102.3480} {arXiv:1102.3480 [hep-ph]} \BibitemShut
  {NoStop}%
\bibitem [{\citenamefont {Soper}\ and\ \citenamefont
  {Spannowsky}(2013)}]{Soper:2012pb}%
  \BibitemOpen
  \bibfield  {author} {\bibinfo {author} {\bibfnamefont {D.~E.}\ \bibnamefont
  {Soper}}\ and\ \bibinfo {author} {\bibfnamefont {M.}~\bibnamefont
  {Spannowsky}},\ }\href {\doibase 10.1103/PhysRevD.87.054012} {\bibfield
  {journal} {\bibinfo  {journal} {Phys. Rev.}\ }\textbf {\bibinfo {volume}
  {D87}},\ \bibinfo {pages} {054012} (\bibinfo {year} {2013})},\ \Eprint
  {http://arxiv.org/abs/1211.3140} {arXiv:1211.3140 [hep-ph]} \BibitemShut
  {NoStop}%
\bibitem [{\citenamefont {Soper}\ and\ \citenamefont
  {Spannowsky}(2014)}]{Soper:2014rya}%
  \BibitemOpen
  \bibfield  {author} {\bibinfo {author} {\bibfnamefont {D.~E.}\ \bibnamefont
  {Soper}}\ and\ \bibinfo {author} {\bibfnamefont {M.}~\bibnamefont
  {Spannowsky}},\ }\href {\doibase 10.1103/PhysRevD.89.094005} {\bibfield
  {journal} {\bibinfo  {journal} {Phys. Rev.}\ }\textbf {\bibinfo {volume}
  {D89}},\ \bibinfo {pages} {094005} (\bibinfo {year} {2014})},\ \Eprint
  {http://arxiv.org/abs/1402.1189} {arXiv:1402.1189 [hep-ph]} \BibitemShut
  {NoStop}%
\bibitem [{\citenamefont {Englert}\ \emph {et~al.}(2016)\citenamefont
  {Englert}, \citenamefont {Mattelaer},\ and\ \citenamefont
  {Spannowsky}}]{Englert:2015dlp}%
  \BibitemOpen
  \bibfield  {author} {\bibinfo {author} {\bibfnamefont {C.}~\bibnamefont
  {Englert}}, \bibinfo {author} {\bibfnamefont {O.}~\bibnamefont {Mattelaer}},
  \ and\ \bibinfo {author} {\bibfnamefont {M.}~\bibnamefont {Spannowsky}},\
  }\href {\doibase 10.1016/j.physletb.2016.02.074} {\bibfield  {journal}
  {\bibinfo  {journal} {Phys. Lett.}\ }\textbf {\bibinfo {volume} {B756}},\
  \bibinfo {pages} {103} (\bibinfo {year} {2016})},\ \Eprint
  {http://arxiv.org/abs/1512.03429} {arXiv:1512.03429 [hep-ph]} \BibitemShut
  {NoStop}%
\bibitem [{\citenamefont {Rubin}(1984)}]{rubin1984}%
  \BibitemOpen
  \bibfield  {author} {\bibinfo {author} {\bibfnamefont {D.~B.}\ \bibnamefont
  {Rubin}},\ }\href {\doibase 10.1214/aos/1176346785} {\bibfield  {journal}
  {\bibinfo  {journal} {Ann. Statist.}\ }\textbf {\bibinfo {volume} {12}},\
  \bibinfo {pages} {1151} (\bibinfo {year} {1984})}\BibitemShut {NoStop}%
\bibitem [{\citenamefont {Beaumont}\ \emph {et~al.}(2002)\citenamefont
  {Beaumont}, \citenamefont {Zhang},\ and\ \citenamefont
  {Balding}}]{beaumont2002approximate}%
  \BibitemOpen
  \bibfield  {author} {\bibinfo {author} {\bibfnamefont {M.~A.}\ \bibnamefont
  {Beaumont}}, \bibinfo {author} {\bibfnamefont {W.}~\bibnamefont {Zhang}}, \
  and\ \bibinfo {author} {\bibfnamefont {D.~J.}\ \bibnamefont {Balding}},\
  }\href@noop {} {\bibfield  {journal} {\bibinfo  {journal} {Genetics}\
  }\textbf {\bibinfo {volume} {162}},\ \bibinfo {pages} {2025} (\bibinfo {year}
  {2002})}\BibitemShut {NoStop}%
\bibitem [{\citenamefont {Marjoram}\ \emph {et~al.}(2003)\citenamefont
  {Marjoram}, \citenamefont {Molitor}, \citenamefont {Plagnol},\ and\
  \citenamefont {Tavar{\'e}}}]{marjoram2003markov}%
  \BibitemOpen
  \bibfield  {author} {\bibinfo {author} {\bibfnamefont {P.}~\bibnamefont
  {Marjoram}}, \bibinfo {author} {\bibfnamefont {J.}~\bibnamefont {Molitor}},
  \bibinfo {author} {\bibfnamefont {V.}~\bibnamefont {Plagnol}}, \ and\
  \bibinfo {author} {\bibfnamefont {S.}~\bibnamefont {Tavar{\'e}}},\
  }\href@noop {} {\bibfield  {journal} {\bibinfo  {journal} {Proceedings of the
  National Academy of Sciences}\ }\textbf {\bibinfo {volume} {100}},\ \bibinfo
  {pages} {15324} (\bibinfo {year} {2003})}\BibitemShut {NoStop}%
\bibitem [{\citenamefont {Sisson}\ \emph {et~al.}(2007)\citenamefont {Sisson},
  \citenamefont {Fan},\ and\ \citenamefont {Tanaka}}]{sisson2007sequential}%
  \BibitemOpen
  \bibfield  {author} {\bibinfo {author} {\bibfnamefont {S.~A.}\ \bibnamefont
  {Sisson}}, \bibinfo {author} {\bibfnamefont {Y.}~\bibnamefont {Fan}}, \ and\
  \bibinfo {author} {\bibfnamefont {M.~M.}\ \bibnamefont {Tanaka}},\
  }\href@noop {} {\bibfield  {journal} {\bibinfo  {journal} {Proceedings of the
  National Academy of Sciences}\ }\textbf {\bibinfo {volume} {104}},\ \bibinfo
  {pages} {1760} (\bibinfo {year} {2007})}\BibitemShut {NoStop}%
\bibitem [{\citenamefont {Sisson}\ and\ \citenamefont
  {Fan}(2011)}]{sisson2011likelihood}%
  \BibitemOpen
  \bibfield  {author} {\bibinfo {author} {\bibfnamefont {S.~A.}\ \bibnamefont
  {Sisson}}\ and\ \bibinfo {author} {\bibfnamefont {Y.}~\bibnamefont {Fan}},\
  }\href@noop {} {\emph {\bibinfo {title} {Likelihood-free MCMC}}}\ (\bibinfo
  {publisher} {Chapman \& Hall/CRC, New York.[839]},\ \bibinfo {year}
  {2011})\BibitemShut {NoStop}%
\bibitem [{\citenamefont {Marin}\ \emph {et~al.}(2012)\citenamefont {Marin},
  \citenamefont {Pudlo}, \citenamefont {Robert},\ and\ \citenamefont
  {Ryder}}]{marin2012approximate}%
  \BibitemOpen
  \bibfield  {author} {\bibinfo {author} {\bibfnamefont {J.-M.}\ \bibnamefont
  {Marin}}, \bibinfo {author} {\bibfnamefont {P.}~\bibnamefont {Pudlo}},
  \bibinfo {author} {\bibfnamefont {C.~P.}\ \bibnamefont {Robert}}, \ and\
  \bibinfo {author} {\bibfnamefont {R.~J.}\ \bibnamefont {Ryder}},\ }\href@noop
  {} {\bibfield  {journal} {\bibinfo  {journal} {Statistics and Computing}\ ,\
  \bibinfo {pages} {1}} (\bibinfo {year} {2012})}\BibitemShut {NoStop}%
\bibitem [{\citenamefont {Charnock}\ \emph {et~al.}(2018)\citenamefont
  {Charnock}, \citenamefont {Lavaux},\ and\ \citenamefont
  {Wandelt}}]{Charnock:2018ogm}%
  \BibitemOpen
  \bibfield  {author} {\bibinfo {author} {\bibfnamefont {T.}~\bibnamefont
  {Charnock}}, \bibinfo {author} {\bibfnamefont {G.}~\bibnamefont {Lavaux}}, \
  and\ \bibinfo {author} {\bibfnamefont {B.~D.}\ \bibnamefont {Wandelt}},\
  }\href {\doibase 10.1103/PhysRevD.97.083004} {\bibfield  {journal} {\bibinfo
  {journal} {Phys. Rev.}\ }\textbf {\bibinfo {volume} {D97}},\ \bibinfo {pages}
  {083004} (\bibinfo {year} {2018})},\ \Eprint
  {http://arxiv.org/abs/1802.03537} {arXiv:1802.03537 [astro-ph.IM]}
  \BibitemShut {NoStop}%
\bibitem [{\citenamefont {Cranmer}\ \emph {et~al.}(2015)\citenamefont
  {Cranmer}, \citenamefont {Pavez},\ and\ \citenamefont
  {Louppe}}]{Cranmer:2015bka}%
  \BibitemOpen
  \bibfield  {author} {\bibinfo {author} {\bibfnamefont {K.}~\bibnamefont
  {Cranmer}}, \bibinfo {author} {\bibfnamefont {J.}~\bibnamefont {Pavez}}, \
  and\ \bibinfo {author} {\bibfnamefont {G.}~\bibnamefont {Louppe}},\
  }\href@noop {} {\  (\bibinfo {year} {2015})},\ \Eprint
  {http://arxiv.org/abs/1506.02169} {arXiv:1506.02169 [stat.AP]} \BibitemShut
  {NoStop}%
\bibitem [{\citenamefont {Cranmer}\ and\ \citenamefont
  {Louppe}(2016)}]{Cranmer:2016lzt}%
  \BibitemOpen
  \bibfield  {author} {\bibinfo {author} {\bibfnamefont {K.}~\bibnamefont
  {Cranmer}}\ and\ \bibinfo {author} {\bibfnamefont {G.}~\bibnamefont
  {Louppe}},\ }\href {\doibase 10.5281/zenodo.198541} {\bibfield  {journal}
  {\bibinfo  {journal} {J. Brief Ideas}\ } (\bibinfo {year} {2016}),\
  10.5281/zenodo.198541}\BibitemShut {NoStop}%
\bibitem [{\citenamefont {{Fan}}\ \emph {et~al.}(2012)\citenamefont {{Fan}},
  \citenamefont {{Nott}},\ and\ \citenamefont
  {{Sisson}}}]{2012arXiv1212.1479F}%
  \BibitemOpen
  \bibfield  {author} {\bibinfo {author} {\bibfnamefont {Y.}~\bibnamefont
  {{Fan}}}, \bibinfo {author} {\bibfnamefont {D.~J.}\ \bibnamefont {{Nott}}}, \
  and\ \bibinfo {author} {\bibfnamefont {S.~A.}\ \bibnamefont {{Sisson}}},\
  }\href@noop {} {\bibfield  {journal} {\bibinfo  {journal} {ArXiv e-prints}\ }
  (\bibinfo {year} {2012})},\ \Eprint {http://arxiv.org/abs/1212.1479}
  {arXiv:1212.1479 [stat.CO]} \BibitemShut {NoStop}%
\bibitem [{\citenamefont {Papamakarios}\ and\ \citenamefont
  {Murray}(2016)}]{NIPS2016_6084}%
  \BibitemOpen
  \bibfield  {author} {\bibinfo {author} {\bibfnamefont {G.}~\bibnamefont
  {Papamakarios}}\ and\ \bibinfo {author} {\bibfnamefont {I.}~\bibnamefont
  {Murray}},\ }in\ \href
  {http://papers.nips.cc/paper/6084-fast-free-inference-of-simulation-models-with-bayesian-conditional-density-estimation.pdf}
  {\emph {\bibinfo {booktitle} {Advances in Neural Information Processing
  Systems 29}}},\ \bibinfo {editor} {edited by\ \bibinfo {editor}
  {\bibfnamefont {D.~D.}\ \bibnamefont {Lee}}, \bibinfo {editor} {\bibfnamefont
  {M.}~\bibnamefont {Sugiyama}}, \bibinfo {editor} {\bibfnamefont {U.~V.}\
  \bibnamefont {Luxburg}}, \bibinfo {editor} {\bibfnamefont {I.}~\bibnamefont
  {Guyon}}, \ and\ \bibinfo {editor} {\bibfnamefont {R.}~\bibnamefont
  {Garnett}}}\ (\bibinfo  {publisher} {Curran Associates, Inc.},\ \bibinfo
  {year} {2016})\ pp.\ \bibinfo {pages} {1028--1036}\BibitemShut {NoStop}%
\bibitem [{\citenamefont {{Paige}}\ and\ \citenamefont
  {{Wood}}(2016)}]{2016arXiv160206701P}%
  \BibitemOpen
  \bibfield  {author} {\bibinfo {author} {\bibfnamefont {B.}~\bibnamefont
  {{Paige}}}\ and\ \bibinfo {author} {\bibfnamefont {F.}~\bibnamefont
  {{Wood}}},\ }\href@noop {} {\bibfield  {journal} {\bibinfo  {journal} {ArXiv
  e-prints}\ } (\bibinfo {year} {2016})},\ \Eprint
  {http://arxiv.org/abs/1602.06701} {arXiv:1602.06701 [stat.ML]} \BibitemShut
  {NoStop}%
\bibitem [{\citenamefont {{Dutta}}\ \emph {et~al.}(2016)\citenamefont
  {{Dutta}}, \citenamefont {{Corander}}, \citenamefont {{Kaski}},\ and\
  \citenamefont {{Gutmann}}}]{2016arXiv161110242D}%
  \BibitemOpen
  \bibfield  {author} {\bibinfo {author} {\bibfnamefont {R.}~\bibnamefont
  {{Dutta}}}, \bibinfo {author} {\bibfnamefont {J.}~\bibnamefont {{Corander}}},
  \bibinfo {author} {\bibfnamefont {S.}~\bibnamefont {{Kaski}}}, \ and\
  \bibinfo {author} {\bibfnamefont {M.~U.}\ \bibnamefont {{Gutmann}}},\
  }\href@noop {} {\bibfield  {journal} {\bibinfo  {journal} {ArXiv e-prints}\ }
  (\bibinfo {year} {2016})},\ \Eprint {http://arxiv.org/abs/1611.10242}
  {arXiv:1611.10242 [stat.ML]} \BibitemShut {NoStop}%
\bibitem [{\citenamefont {Gutmann}\ \emph {et~al.}(2017)\citenamefont
  {Gutmann}, \citenamefont {Dutta}, \citenamefont {Kaski},\ and\ \citenamefont
  {Corander}}]{gutmann2017likelihood}%
  \BibitemOpen
  \bibfield  {author} {\bibinfo {author} {\bibfnamefont {M.~U.}\ \bibnamefont
  {Gutmann}}, \bibinfo {author} {\bibfnamefont {R.}~\bibnamefont {Dutta}},
  \bibinfo {author} {\bibfnamefont {S.}~\bibnamefont {Kaski}}, \ and\ \bibinfo
  {author} {\bibfnamefont {J.}~\bibnamefont {Corander}},\ }\href@noop {}
  {\bibfield  {journal} {\bibinfo  {journal} {Statistics and Computing}\ ,\
  \bibinfo {pages} {1}} (\bibinfo {year} {2017})}\BibitemShut {NoStop}%
\bibitem [{\citenamefont {{Tran}}\ \emph {et~al.}(2017)\citenamefont {{Tran}},
  \citenamefont {{Ranganath}},\ and\ \citenamefont
  {{Blei}}}]{2017arXiv170208896T}%
  \BibitemOpen
  \bibfield  {author} {\bibinfo {author} {\bibfnamefont {D.}~\bibnamefont
  {{Tran}}}, \bibinfo {author} {\bibfnamefont {R.}~\bibnamefont {{Ranganath}}},
  \ and\ \bibinfo {author} {\bibfnamefont {D.~M.}\ \bibnamefont {{Blei}}},\
  }\href@noop {} {\bibfield  {journal} {\bibinfo  {journal} {ArXiv e-prints}\ }
  (\bibinfo {year} {2017})},\ \Eprint {http://arxiv.org/abs/1702.08896}
  {arXiv:1702.08896 [stat.ML]} \BibitemShut {NoStop}%
\bibitem [{\citenamefont {{Louppe}}\ and\ \citenamefont
  {{Cranmer}}(2017)}]{2017arXiv170707113L}%
  \BibitemOpen
  \bibfield  {author} {\bibinfo {author} {\bibfnamefont {G.}~\bibnamefont
  {{Louppe}}}\ and\ \bibinfo {author} {\bibfnamefont {K.}~\bibnamefont
  {{Cranmer}}},\ }\href@noop {} {\bibfield  {journal} {\bibinfo  {journal}
  {ArXiv e-prints}\ } (\bibinfo {year} {2017})},\ \Eprint
  {http://arxiv.org/abs/1707.07113} {arXiv:1707.07113 [stat.ML]} \BibitemShut
  {NoStop}%
\bibitem [{\citenamefont {{Dinh}}\ \emph {et~al.}(2014)\citenamefont {{Dinh}},
  \citenamefont {{Krueger}},\ and\ \citenamefont
  {{Bengio}}}]{2014arXiv1410.8516D}%
  \BibitemOpen
  \bibfield  {author} {\bibinfo {author} {\bibfnamefont {L.}~\bibnamefont
  {{Dinh}}}, \bibinfo {author} {\bibfnamefont {D.}~\bibnamefont {{Krueger}}}, \
  and\ \bibinfo {author} {\bibfnamefont {Y.}~\bibnamefont {{Bengio}}},\
  }\href@noop {} {\bibfield  {journal} {\bibinfo  {journal} {ArXiv e-prints}\ }
  (\bibinfo {year} {2014})},\ \Eprint {http://arxiv.org/abs/1410.8516}
  {arXiv:1410.8516 [cs.LG]} \BibitemShut {NoStop}%
\bibitem [{\citenamefont {{Jimenez Rezende}}\ and\ \citenamefont
  {{Mohamed}}(2015)}]{2015arXiv150505770J}%
  \BibitemOpen
  \bibfield  {author} {\bibinfo {author} {\bibfnamefont {D.}~\bibnamefont
  {{Jimenez Rezende}}}\ and\ \bibinfo {author} {\bibfnamefont {S.}~\bibnamefont
  {{Mohamed}}},\ }\href@noop {} {\bibfield  {journal} {\bibinfo  {journal}
  {ArXiv e-prints}\ } (\bibinfo {year} {2015})},\ \Eprint
  {http://arxiv.org/abs/1505.05770} {arXiv:1505.05770 [stat.ML]} \BibitemShut
  {NoStop}%
\bibitem [{\citenamefont {{Dinh}}\ \emph {et~al.}(2016)\citenamefont {{Dinh}},
  \citenamefont {{Sohl-Dickstein}},\ and\ \citenamefont
  {{Bengio}}}]{2016arXiv160508803D}%
  \BibitemOpen
  \bibfield  {author} {\bibinfo {author} {\bibfnamefont {L.}~\bibnamefont
  {{Dinh}}}, \bibinfo {author} {\bibfnamefont {J.}~\bibnamefont
  {{Sohl-Dickstein}}}, \ and\ \bibinfo {author} {\bibfnamefont
  {S.}~\bibnamefont {{Bengio}}},\ }\href@noop {} {\bibfield  {journal}
  {\bibinfo  {journal} {ArXiv e-prints}\ } (\bibinfo {year} {2016})},\ \Eprint
  {http://arxiv.org/abs/1605.08803} {arXiv:1605.08803 [cs.LG]} \BibitemShut
  {NoStop}%
\bibitem [{\citenamefont {{Papamakarios}}\ \emph {et~al.}(2017)\citenamefont
  {{Papamakarios}}, \citenamefont {{Pavlakou}},\ and\ \citenamefont
  {{Murray}}}]{2017arXiv170507057P}%
  \BibitemOpen
  \bibfield  {author} {\bibinfo {author} {\bibfnamefont {G.}~\bibnamefont
  {{Papamakarios}}}, \bibinfo {author} {\bibfnamefont {T.}~\bibnamefont
  {{Pavlakou}}}, \ and\ \bibinfo {author} {\bibfnamefont {I.}~\bibnamefont
  {{Murray}}},\ }\href@noop {} {\bibfield  {journal} {\bibinfo  {journal}
  {ArXiv e-prints}\ } (\bibinfo {year} {2017})},\ \Eprint
  {http://arxiv.org/abs/1705.07057} {arXiv:1705.07057 [stat.ML]} \BibitemShut
  {NoStop}%
\bibitem [{\citenamefont {{Uria}}\ \emph {et~al.}(2016)\citenamefont {{Uria}},
  \citenamefont {{C{\^o}t{\'e}}}, \citenamefont {{Gregor}}, \citenamefont
  {{Murray}},\ and\ \citenamefont {{Larochelle}}}]{2016arXiv160502226U}%
  \BibitemOpen
  \bibfield  {author} {\bibinfo {author} {\bibfnamefont {B.}~\bibnamefont
  {{Uria}}}, \bibinfo {author} {\bibfnamefont {M.-A.}\ \bibnamefont
  {{C{\^o}t{\'e}}}}, \bibinfo {author} {\bibfnamefont {K.}~\bibnamefont
  {{Gregor}}}, \bibinfo {author} {\bibfnamefont {I.}~\bibnamefont {{Murray}}},
  \ and\ \bibinfo {author} {\bibfnamefont {H.}~\bibnamefont {{Larochelle}}},\
  }\href@noop {} {\bibfield  {journal} {\bibinfo  {journal} {ArXiv e-prints}\ }
  (\bibinfo {year} {2016})},\ \Eprint {http://arxiv.org/abs/1605.02226}
  {arXiv:1605.02226 [cs.LG]} \BibitemShut {NoStop}%
\bibitem [{\citenamefont {{van den Oord}}\ \emph
  {et~al.}(2016{\natexlab{a}})\citenamefont {{van den Oord}}, \citenamefont
  {{Dieleman}}, \citenamefont {{Zen}}, \citenamefont {{Simonyan}},
  \citenamefont {{Vinyals}}, \citenamefont {{Graves}}, \citenamefont
  {{Kalchbrenner}}, \citenamefont {{Senior}},\ and\ \citenamefont
  {{Kavukcuoglu}}}]{2016arXiv160903499V}%
  \BibitemOpen
  \bibfield  {author} {\bibinfo {author} {\bibfnamefont {A.}~\bibnamefont {{van
  den Oord}}}, \bibinfo {author} {\bibfnamefont {S.}~\bibnamefont
  {{Dieleman}}}, \bibinfo {author} {\bibfnamefont {H.}~\bibnamefont {{Zen}}},
  \bibinfo {author} {\bibfnamefont {K.}~\bibnamefont {{Simonyan}}}, \bibinfo
  {author} {\bibfnamefont {O.}~\bibnamefont {{Vinyals}}}, \bibinfo {author}
  {\bibfnamefont {A.}~\bibnamefont {{Graves}}}, \bibinfo {author}
  {\bibfnamefont {N.}~\bibnamefont {{Kalchbrenner}}}, \bibinfo {author}
  {\bibfnamefont {A.}~\bibnamefont {{Senior}}}, \ and\ \bibinfo {author}
  {\bibfnamefont {K.}~\bibnamefont {{Kavukcuoglu}}},\ }\href@noop {} {\bibfield
   {journal} {\bibinfo  {journal} {ArXiv e-prints}\ } (\bibinfo {year}
  {2016}{\natexlab{a}})},\ \Eprint {http://arxiv.org/abs/1609.03499}
  {arXiv:1609.03499 [cs.SD]} \BibitemShut {NoStop}%
\bibitem [{\citenamefont {{van den Oord}}\ \emph
  {et~al.}(2016{\natexlab{b}})\citenamefont {{van den Oord}}, \citenamefont
  {{Kalchbrenner}}, \citenamefont {{Vinyals}}, \citenamefont {{Espeholt}},
  \citenamefont {{Graves}},\ and\ \citenamefont
  {{Kavukcuoglu}}}]{2016arXiv160605328V}%
  \BibitemOpen
  \bibfield  {author} {\bibinfo {author} {\bibfnamefont {A.}~\bibnamefont {{van
  den Oord}}}, \bibinfo {author} {\bibfnamefont {N.}~\bibnamefont
  {{Kalchbrenner}}}, \bibinfo {author} {\bibfnamefont {O.}~\bibnamefont
  {{Vinyals}}}, \bibinfo {author} {\bibfnamefont {L.}~\bibnamefont
  {{Espeholt}}}, \bibinfo {author} {\bibfnamefont {A.}~\bibnamefont
  {{Graves}}}, \ and\ \bibinfo {author} {\bibfnamefont {K.}~\bibnamefont
  {{Kavukcuoglu}}},\ }\href@noop {} {\bibfield  {journal} {\bibinfo  {journal}
  {ArXiv e-prints}\ } (\bibinfo {year} {2016}{\natexlab{b}})},\ \Eprint
  {http://arxiv.org/abs/1606.05328} {arXiv:1606.05328 [cs.CV]} \BibitemShut
  {NoStop}%
\bibitem [{\citenamefont {{van den Oord}}\ \emph
  {et~al.}(2016{\natexlab{c}})\citenamefont {{van den Oord}}, \citenamefont
  {{Kalchbrenner}},\ and\ \citenamefont {{Kavukcuoglu}}}]{2016arXiv160106759V}%
  \BibitemOpen
  \bibfield  {author} {\bibinfo {author} {\bibfnamefont {A.}~\bibnamefont {{van
  den Oord}}}, \bibinfo {author} {\bibfnamefont {N.}~\bibnamefont
  {{Kalchbrenner}}}, \ and\ \bibinfo {author} {\bibfnamefont {K.}~\bibnamefont
  {{Kavukcuoglu}}},\ }\href@noop {} {\bibfield  {journal} {\bibinfo  {journal}
  {ArXiv e-prints}\ } (\bibinfo {year} {2016}{\natexlab{c}})},\ \Eprint
  {http://arxiv.org/abs/1601.06759} {arXiv:1601.06759 [cs.CV]} \BibitemShut
  {NoStop}%
\bibitem [{\citenamefont {{Papamakarios}}\ \emph {et~al.}(2018)\citenamefont
  {{Papamakarios}}, \citenamefont {{Sterratt}},\ and\ \citenamefont
  {{Murray}}}]{2018arXiv180507226P}%
  \BibitemOpen
  \bibfield  {author} {\bibinfo {author} {\bibfnamefont {G.}~\bibnamefont
  {{Papamakarios}}}, \bibinfo {author} {\bibfnamefont {D.~C.}\ \bibnamefont
  {{Sterratt}}}, \ and\ \bibinfo {author} {\bibfnamefont {I.}~\bibnamefont
  {{Murray}}},\ }\href@noop {} {\bibfield  {journal} {\bibinfo  {journal}
  {ArXiv e-prints}\ } (\bibinfo {year} {2018})},\ \Eprint
  {http://arxiv.org/abs/1805.07226} {arXiv:1805.07226 [stat.ML]} \BibitemShut
  {NoStop}%
\bibitem [{\citenamefont {Brehmer}\ \emph
  {et~al.}(2018{\natexlab{a}})\citenamefont {Brehmer}, \citenamefont {Cranmer},
  \citenamefont {Louppe},\ and\ \citenamefont {Pavez}}]{companion_short}%
  \BibitemOpen
  \bibfield  {author} {\bibinfo {author} {\bibfnamefont {J.}~\bibnamefont
  {Brehmer}}, \bibinfo {author} {\bibfnamefont {K.}~\bibnamefont {Cranmer}},
  \bibinfo {author} {\bibfnamefont {G.}~\bibnamefont {Louppe}}, \ and\ \bibinfo
  {author} {\bibfnamefont {J.}~\bibnamefont {Pavez}},\ }\href@noop {} {\
  (\bibinfo {year} {2018}{\natexlab{a}})},\ \Eprint
  {http://arxiv.org/abs/1805.00013} {arXiv:1805.00013 [hep-ph]} \BibitemShut
  {NoStop}%
\bibitem [{\citenamefont {Brehmer}\ \emph
  {et~al.}(2018{\natexlab{b}})\citenamefont {Brehmer}, \citenamefont {Louppe},
  \citenamefont {Pavez},\ and\ \citenamefont {Cranmer}}]{companion_nips}%
  \BibitemOpen
  \bibfield  {author} {\bibinfo {author} {\bibfnamefont {J.}~\bibnamefont
  {Brehmer}}, \bibinfo {author} {\bibfnamefont {G.}~\bibnamefont {Louppe}},
  \bibinfo {author} {\bibfnamefont {J.}~\bibnamefont {Pavez}}, \ and\ \bibinfo
  {author} {\bibfnamefont {K.}~\bibnamefont {Cranmer}},\ }\href@noop {} {\
  (\bibinfo {year} {2018}{\natexlab{b}})},\ \Eprint
  {http://arxiv.org/abs/1805.12244} {arXiv:1805.12244 [stat.ML]} \BibitemShut
  {NoStop}%
\bibitem [{\citenamefont {Brehmer}\ \emph
  {et~al.}(2018{\natexlab{c}})\citenamefont {Brehmer}, \citenamefont {Cranmer},
  \citenamefont {Louppe},\ and\ \citenamefont {Pavez}}]{repository}%
  \BibitemOpen
  \bibfield  {author} {\bibinfo {author} {\bibfnamefont {J.}~\bibnamefont
  {Brehmer}}, \bibinfo {author} {\bibfnamefont {K.}~\bibnamefont {Cranmer}},
  \bibinfo {author} {\bibfnamefont {G.}~\bibnamefont {Louppe}}, \ and\ \bibinfo
  {author} {\bibfnamefont {J.}~\bibnamefont {Pavez}},\ }\href@noop {} {\enquote
  {\bibinfo {title} {{Code repository for the paper ``Constraining Effective
  Field Theories with Machine Learning''}},}\ }\bibinfo {howpublished}
  {\url{http://github.com/johannbrehmer/higgs_inference}} (\bibinfo {year}
  {2018}{\natexlab{c}})\BibitemShut {NoStop}%
\bibitem [{\citenamefont {Coleman}\ \emph {et~al.}(1969)\citenamefont
  {Coleman}, \citenamefont {Wess},\ and\ \citenamefont
  {Zumino}}]{Coleman:1969sm}%
  \BibitemOpen
  \bibfield  {author} {\bibinfo {author} {\bibfnamefont {S.~R.}\ \bibnamefont
  {Coleman}}, \bibinfo {author} {\bibfnamefont {J.}~\bibnamefont {Wess}}, \
  and\ \bibinfo {author} {\bibfnamefont {B.}~\bibnamefont {Zumino}},\ }\href
  {\doibase 10.1103/PhysRev.177.2239} {\bibfield  {journal} {\bibinfo
  {journal} {Phys. Rev.}\ }\textbf {\bibinfo {volume} {177}},\ \bibinfo {pages}
  {2239} (\bibinfo {year} {1969})}\BibitemShut {NoStop}%
\bibitem [{\citenamefont {Callan}\ \emph {et~al.}(1969)\citenamefont {Callan},
  \citenamefont {Coleman}, \citenamefont {Wess},\ and\ \citenamefont
  {Zumino}}]{Callan:1969sn}%
  \BibitemOpen
  \bibfield  {author} {\bibinfo {author} {\bibfnamefont {C.~G.}\ \bibnamefont
  {Callan}, \bibfnamefont {Jr.}}, \bibinfo {author} {\bibfnamefont {S.~R.}\
  \bibnamefont {Coleman}}, \bibinfo {author} {\bibfnamefont {J.}~\bibnamefont
  {Wess}}, \ and\ \bibinfo {author} {\bibfnamefont {B.}~\bibnamefont
  {Zumino}},\ }\href {\doibase 10.1103/PhysRev.177.2247} {\bibfield  {journal}
  {\bibinfo  {journal} {Phys. Rev.}\ }\textbf {\bibinfo {volume} {177}},\
  \bibinfo {pages} {2247} (\bibinfo {year} {1969})}\BibitemShut {NoStop}%
\bibitem [{\citenamefont {Weinberg}(1980)}]{Weinberg:1980wa}%
  \BibitemOpen
  \bibfield  {author} {\bibinfo {author} {\bibfnamefont {S.}~\bibnamefont
  {Weinberg}},\ }\href {\doibase 10.1016/0370-2693(80)90660-7} {\bibfield
  {journal} {\bibinfo  {journal} {Phys. Lett.}\ }\textbf {\bibinfo {volume}
  {B91}},\ \bibinfo {pages} {51} (\bibinfo {year} {1980})}\BibitemShut
  {NoStop}%
\bibitem [{\citenamefont {Burges}\ and\ \citenamefont
  {Schnitzer}(1983)}]{Burges:1983zg}%
  \BibitemOpen
  \bibfield  {author} {\bibinfo {author} {\bibfnamefont {C.~J.~C.}\
  \bibnamefont {Burges}}\ and\ \bibinfo {author} {\bibfnamefont {H.~J.}\
  \bibnamefont {Schnitzer}},\ }\href {\doibase 10.1016/0550-3213(83)90555-2}
  {\bibfield  {journal} {\bibinfo  {journal} {Nucl. Phys.}\ }\textbf {\bibinfo
  {volume} {B228}},\ \bibinfo {pages} {464} (\bibinfo {year}
  {1983})}\BibitemShut {NoStop}%
\bibitem [{\citenamefont {Leung}\ \emph {et~al.}(1986)\citenamefont {Leung},
  \citenamefont {Love},\ and\ \citenamefont {Rao}}]{Leung:1984ni}%
  \BibitemOpen
  \bibfield  {author} {\bibinfo {author} {\bibfnamefont {C.~N.}\ \bibnamefont
  {Leung}}, \bibinfo {author} {\bibfnamefont {S.~T.}\ \bibnamefont {Love}}, \
  and\ \bibinfo {author} {\bibfnamefont {S.}~\bibnamefont {Rao}},\ }\href
  {\doibase 10.1007/BF01588041} {\bibfield  {journal} {\bibinfo  {journal} {Z.
  Phys.}\ }\textbf {\bibinfo {volume} {C31}},\ \bibinfo {pages} {433} (\bibinfo
  {year} {1986})}\BibitemShut {NoStop}%
\bibitem [{\citenamefont {Buchmuller}\ and\ \citenamefont
  {Wyler}(1986)}]{Buchmuller:1985jz}%
  \BibitemOpen
  \bibfield  {author} {\bibinfo {author} {\bibfnamefont {W.}~\bibnamefont
  {Buchmuller}}\ and\ \bibinfo {author} {\bibfnamefont {D.}~\bibnamefont
  {Wyler}},\ }\href {\doibase 10.1016/0550-3213(86)90262-2} {\bibfield
  {journal} {\bibinfo  {journal} {Nucl. Phys.}\ }\textbf {\bibinfo {volume}
  {B268}},\ \bibinfo {pages} {621} (\bibinfo {year} {1986})}\BibitemShut
  {NoStop}%
\bibitem [{\citenamefont {Arzt}\ \emph {et~al.}(1995)\citenamefont {Arzt},
  \citenamefont {Einhorn},\ and\ \citenamefont {Wudka}}]{Arzt:1994gp}%
  \BibitemOpen
  \bibfield  {author} {\bibinfo {author} {\bibfnamefont {C.}~\bibnamefont
  {Arzt}}, \bibinfo {author} {\bibfnamefont {M.~B.}\ \bibnamefont {Einhorn}}, \
  and\ \bibinfo {author} {\bibfnamefont {J.}~\bibnamefont {Wudka}},\ }\href
  {\doibase 10.1016/0550-3213(94)00336-D} {\bibfield  {journal} {\bibinfo
  {journal} {Nucl. Phys.}\ }\textbf {\bibinfo {volume} {B433}},\ \bibinfo
  {pages} {41} (\bibinfo {year} {1995})},\ \Eprint
  {http://arxiv.org/abs/hep-ph/9405214} {arXiv:hep-ph/9405214 [hep-ph]}
  \BibitemShut {NoStop}%
\bibitem [{\citenamefont {Hagiwara}\ \emph {et~al.}(1993)\citenamefont
  {Hagiwara}, \citenamefont {Ishihara}, \citenamefont {Szalapski},\ and\
  \citenamefont {Zeppenfeld}}]{Hagiwara:1993ck}%
  \BibitemOpen
  \bibfield  {author} {\bibinfo {author} {\bibfnamefont {K.}~\bibnamefont
  {Hagiwara}}, \bibinfo {author} {\bibfnamefont {S.}~\bibnamefont {Ishihara}},
  \bibinfo {author} {\bibfnamefont {R.}~\bibnamefont {Szalapski}}, \ and\
  \bibinfo {author} {\bibfnamefont {D.}~\bibnamefont {Zeppenfeld}},\ }\href
  {\doibase 10.1103/PhysRevD.48.2182} {\bibfield  {journal} {\bibinfo
  {journal} {Phys. Rev.}\ }\textbf {\bibinfo {volume} {D48}},\ \bibinfo {pages}
  {2182} (\bibinfo {year} {1993})}\BibitemShut {NoStop}%
\bibitem [{\citenamefont {Grzadkowski}\ \emph {et~al.}(2010)\citenamefont
  {Grzadkowski}, \citenamefont {Iskrzynski}, \citenamefont {Misiak},\ and\
  \citenamefont {Rosiek}}]{Grzadkowski:2010es}%
  \BibitemOpen
  \bibfield  {author} {\bibinfo {author} {\bibfnamefont {B.}~\bibnamefont
  {Grzadkowski}}, \bibinfo {author} {\bibfnamefont {M.}~\bibnamefont
  {Iskrzynski}}, \bibinfo {author} {\bibfnamefont {M.}~\bibnamefont {Misiak}},
  \ and\ \bibinfo {author} {\bibfnamefont {J.}~\bibnamefont {Rosiek}},\ }\href
  {\doibase 10.1007/JHEP10(2010)085} {\bibfield  {journal} {\bibinfo  {journal}
  {JHEP}\ }\textbf {\bibinfo {volume} {10}},\ \bibinfo {pages} {085} (\bibinfo
  {year} {2010})},\ \Eprint {http://arxiv.org/abs/1008.4884} {arXiv:1008.4884
  [hep-ph]} \BibitemShut {NoStop}%
\bibitem [{\citenamefont {Alwall}\ \emph {et~al.}(2014)\citenamefont {Alwall},
  \citenamefont {Frederix}, \citenamefont {Frixione}, \citenamefont {Hirschi},
  \citenamefont {Maltoni}, \citenamefont {Mattelaer}, \citenamefont {Shao},
  \citenamefont {Stelzer}, \citenamefont {Torrielli},\ and\ \citenamefont
  {Zaro}}]{Alwall:2014hca}%
  \BibitemOpen
  \bibfield  {author} {\bibinfo {author} {\bibfnamefont {J.}~\bibnamefont
  {Alwall}}, \bibinfo {author} {\bibfnamefont {R.}~\bibnamefont {Frederix}},
  \bibinfo {author} {\bibfnamefont {S.}~\bibnamefont {Frixione}}, \bibinfo
  {author} {\bibfnamefont {V.}~\bibnamefont {Hirschi}}, \bibinfo {author}
  {\bibfnamefont {F.}~\bibnamefont {Maltoni}}, \bibinfo {author} {\bibfnamefont
  {O.}~\bibnamefont {Mattelaer}}, \bibinfo {author} {\bibfnamefont {H.~S.}\
  \bibnamefont {Shao}}, \bibinfo {author} {\bibfnamefont {T.}~\bibnamefont
  {Stelzer}}, \bibinfo {author} {\bibfnamefont {P.}~\bibnamefont {Torrielli}},
  \ and\ \bibinfo {author} {\bibfnamefont {M.}~\bibnamefont {Zaro}},\ }\href
  {\doibase 10.1007/JHEP07(2014)079} {\bibfield  {journal} {\bibinfo  {journal}
  {JHEP}\ }\textbf {\bibinfo {volume} {07}},\ \bibinfo {pages} {079} (\bibinfo
  {year} {2014})},\ \Eprint {http://arxiv.org/abs/1405.0301} {arXiv:1405.0301
  [hep-ph]} \BibitemShut {NoStop}%
\bibitem [{\citenamefont {Henning}\ \emph {et~al.}(2016)\citenamefont
  {Henning}, \citenamefont {Lu},\ and\ \citenamefont
  {Murayama}}]{Henning:2014wua}%
  \BibitemOpen
  \bibfield  {author} {\bibinfo {author} {\bibfnamefont {B.}~\bibnamefont
  {Henning}}, \bibinfo {author} {\bibfnamefont {X.}~\bibnamefont {Lu}}, \ and\
  \bibinfo {author} {\bibfnamefont {H.}~\bibnamefont {Murayama}},\ }\href
  {\doibase 10.1007/JHEP01(2016)023} {\bibfield  {journal} {\bibinfo  {journal}
  {JHEP}\ }\textbf {\bibinfo {volume} {01}},\ \bibinfo {pages} {023} (\bibinfo
  {year} {2016})},\ \Eprint {http://arxiv.org/abs/1412.1837} {arXiv:1412.1837
  [hep-ph]} \BibitemShut {NoStop}%
\bibitem [{\citenamefont {de~Florian}\ \emph {et~al.}(2016)\citenamefont
  {de~Florian} \emph {et~al.}}]{deFlorian:2016spz}%
  \BibitemOpen
  \bibfield  {author} {\bibinfo {author} {\bibfnamefont {D.}~\bibnamefont
  {de~Florian}} \emph {et~al.} (\bibinfo {collaboration} {LHC Higgs Cross
  Section Working Group}),\ }\href {\doibase 10.23731/CYRM-2017-002} {\
  (\bibinfo {year} {2016}),\ 10.23731/CYRM-2017-002},\ \Eprint
  {http://arxiv.org/abs/1610.07922} {arXiv:1610.07922 [hep-ph]} \BibitemShut
  {NoStop}%
\bibitem [{\citenamefont {Brehmer}\ \emph {et~al.}(2016)\citenamefont
  {Brehmer}, \citenamefont {Freitas}, \citenamefont {Lopez-Val},\ and\
  \citenamefont {Plehn}}]{Brehmer:2015rna}%
  \BibitemOpen
  \bibfield  {author} {\bibinfo {author} {\bibfnamefont {J.}~\bibnamefont
  {Brehmer}}, \bibinfo {author} {\bibfnamefont {A.}~\bibnamefont {Freitas}},
  \bibinfo {author} {\bibfnamefont {D.}~\bibnamefont {Lopez-Val}}, \ and\
  \bibinfo {author} {\bibfnamefont {T.}~\bibnamefont {Plehn}},\ }\href
  {\doibase 10.1103/PhysRevD.93.075014} {\bibfield  {journal} {\bibinfo
  {journal} {Phys. Rev.}\ }\textbf {\bibinfo {volume} {D93}},\ \bibinfo {pages}
  {075014} (\bibinfo {year} {2016})},\ \Eprint
  {http://arxiv.org/abs/1510.03443} {arXiv:1510.03443 [hep-ph]} \BibitemShut
  {NoStop}%
\bibitem [{\citenamefont {Brehmer}\ \emph
  {et~al.}(2017{\natexlab{b}})\citenamefont {Brehmer}, \citenamefont {Kling},
  \citenamefont {Plehn},\ and\ \citenamefont {Tait}}]{Brehmer:2017lrt}%
  \BibitemOpen
  \bibfield  {author} {\bibinfo {author} {\bibfnamefont {J.}~\bibnamefont
  {Brehmer}}, \bibinfo {author} {\bibfnamefont {F.}~\bibnamefont {Kling}},
  \bibinfo {author} {\bibfnamefont {T.}~\bibnamefont {Plehn}}, \ and\ \bibinfo
  {author} {\bibfnamefont {T.~M.~P.}\ \bibnamefont {Tait}},\ }\href@noop {} {\
  (\bibinfo {year} {2017}{\natexlab{b}})},\ \Eprint
  {http://arxiv.org/abs/1712.02350} {arXiv:1712.02350 [hep-ph]} \BibitemShut
  {NoStop}%
\bibitem [{\citenamefont {de~Favereau}\ \emph {et~al.}(2014)\citenamefont
  {de~Favereau}, \citenamefont {Delaere}, \citenamefont {Demin}, \citenamefont
  {Giammanco}, \citenamefont {Lema\^{i}tre}, \citenamefont {Mertens},\ and\
  \citenamefont {Selvaggi}}]{deFavereau:2013fsa}%
  \BibitemOpen
  \bibfield  {author} {\bibinfo {author} {\bibfnamefont {J.}~\bibnamefont
  {de~Favereau}}, \bibinfo {author} {\bibfnamefont {C.}~\bibnamefont
  {Delaere}}, \bibinfo {author} {\bibfnamefont {P.}~\bibnamefont {Demin}},
  \bibinfo {author} {\bibfnamefont {A.}~\bibnamefont {Giammanco}}, \bibinfo
  {author} {\bibfnamefont {V.}~\bibnamefont {Lema\^{i}tre}}, \bibinfo {author}
  {\bibfnamefont {A.}~\bibnamefont {Mertens}}, \ and\ \bibinfo {author}
  {\bibfnamefont {M.}~\bibnamefont {Selvaggi}} (\bibinfo {collaboration}
  {DELPHES 3}),\ }\href {\doibase 10.1007/JHEP02(2014)057} {\bibfield
  {journal} {\bibinfo  {journal} {JHEP}\ }\textbf {\bibinfo {volume} {02}},\
  \bibinfo {pages} {057} (\bibinfo {year} {2014})},\ \Eprint
  {http://arxiv.org/abs/1307.6346} {arXiv:1307.6346 [hep-ex]} \BibitemShut
  {NoStop}%
\bibitem [{\citenamefont {Aad}\ \emph {et~al.}(2015)\citenamefont {Aad} \emph
  {et~al.}}]{ATLAS:morphing}%
  \BibitemOpen
  \bibfield  {author} {\bibinfo {author} {\bibfnamefont {G.}~\bibnamefont
  {Aad}} \emph {et~al.} (\bibinfo {collaboration} {{ATLAS}}),\ }\href
  {http://cds.cern.ch/record/2066980} {\enquote {\bibinfo {title} {{A morphing
  technique for signal modelling in a multidimensional space of coupling
  parameters}},}\ } (\bibinfo {year} {2015}),\ \bibinfo {note} {{Physics note
  ATL-PHYS-PUB-2015-047}}\BibitemShut {NoStop}%
\bibitem [{\citenamefont {Dell'Aquila}\ and\ \citenamefont
  {Nelson}(1986)}]{DellAquila:1985mtb}%
  \BibitemOpen
  \bibfield  {author} {\bibinfo {author} {\bibfnamefont {J.~R.}\ \bibnamefont
  {Dell'Aquila}}\ and\ \bibinfo {author} {\bibfnamefont {C.~A.}\ \bibnamefont
  {Nelson}},\ }\href {\doibase 10.1103/PhysRevD.33.80} {\bibfield  {journal}
  {\bibinfo  {journal} {Phys. Rev.}\ }\textbf {\bibinfo {volume} {D33}},\
  \bibinfo {pages} {80} (\bibinfo {year} {1986})}\BibitemShut {NoStop}%
\bibitem [{\citenamefont {Plehn}\ \emph {et~al.}(2002)\citenamefont {Plehn},
  \citenamefont {Rainwater},\ and\ \citenamefont {Zeppenfeld}}]{Plehn:2001nj}%
  \BibitemOpen
  \bibfield  {author} {\bibinfo {author} {\bibfnamefont {T.}~\bibnamefont
  {Plehn}}, \bibinfo {author} {\bibfnamefont {D.~L.}\ \bibnamefont
  {Rainwater}}, \ and\ \bibinfo {author} {\bibfnamefont {D.}~\bibnamefont
  {Zeppenfeld}},\ }\href {\doibase 10.1103/PhysRevLett.88.051801} {\bibfield
  {journal} {\bibinfo  {journal} {Phys. Rev. Lett.}\ }\textbf {\bibinfo
  {volume} {88}},\ \bibinfo {pages} {051801} (\bibinfo {year} {2002})},\
  \Eprint {http://arxiv.org/abs/hep-ph/0105325} {arXiv:hep-ph/0105325 [hep-ph]}
  \BibitemShut {NoStop}%
\bibitem [{\citenamefont {Hankele}\ \emph {et~al.}(2006)\citenamefont
  {Hankele}, \citenamefont {Klamke}, \citenamefont {Zeppenfeld},\ and\
  \citenamefont {Figy}}]{Hankele:2006ma}%
  \BibitemOpen
  \bibfield  {author} {\bibinfo {author} {\bibfnamefont {V.}~\bibnamefont
  {Hankele}}, \bibinfo {author} {\bibfnamefont {G.}~\bibnamefont {Klamke}},
  \bibinfo {author} {\bibfnamefont {D.}~\bibnamefont {Zeppenfeld}}, \ and\
  \bibinfo {author} {\bibfnamefont {T.}~\bibnamefont {Figy}},\ }\href {\doibase
  10.1103/PhysRevD.74.095001} {\bibfield  {journal} {\bibinfo  {journal} {Phys.
  Rev.}\ }\textbf {\bibinfo {volume} {D74}},\ \bibinfo {pages} {095001}
  (\bibinfo {year} {2006})},\ \Eprint {http://arxiv.org/abs/hep-ph/0609075}
  {arXiv:hep-ph/0609075 [hep-ph]} \BibitemShut {NoStop}%
\bibitem [{\citenamefont {Hagiwara}\ \emph {et~al.}(2009)\citenamefont
  {Hagiwara}, \citenamefont {Li},\ and\ \citenamefont
  {Mawatari}}]{Hagiwara:2009wt}%
  \BibitemOpen
  \bibfield  {author} {\bibinfo {author} {\bibfnamefont {K.}~\bibnamefont
  {Hagiwara}}, \bibinfo {author} {\bibfnamefont {Q.}~\bibnamefont {Li}}, \ and\
  \bibinfo {author} {\bibfnamefont {K.}~\bibnamefont {Mawatari}},\ }\href
  {\doibase 10.1088/1126-6708/2009/07/101} {\bibfield  {journal} {\bibinfo
  {journal} {JHEP}\ }\textbf {\bibinfo {volume} {07}},\ \bibinfo {pages} {101}
  (\bibinfo {year} {2009})},\ \Eprint {http://arxiv.org/abs/0905.4314}
  {arXiv:0905.4314 [hep-ph]} \BibitemShut {NoStop}%
\bibitem [{\citenamefont {Englert}\ \emph {et~al.}(2013)\citenamefont
  {Englert}, \citenamefont {Goncalves-Netto}, \citenamefont {Mawatari},\ and\
  \citenamefont {Plehn}}]{Englert:2012xt}%
  \BibitemOpen
  \bibfield  {author} {\bibinfo {author} {\bibfnamefont {C.}~\bibnamefont
  {Englert}}, \bibinfo {author} {\bibfnamefont {D.}~\bibnamefont
  {Goncalves-Netto}}, \bibinfo {author} {\bibfnamefont {K.}~\bibnamefont
  {Mawatari}}, \ and\ \bibinfo {author} {\bibfnamefont {T.}~\bibnamefont
  {Plehn}},\ }\href {\doibase 10.1007/JHEP01(2013)148} {\bibfield  {journal}
  {\bibinfo  {journal} {JHEP}\ }\textbf {\bibinfo {volume} {01}},\ \bibinfo
  {pages} {148} (\bibinfo {year} {2013})},\ \Eprint
  {http://arxiv.org/abs/1212.0843} {arXiv:1212.0843 [hep-ph]} \BibitemShut
  {NoStop}%
\bibitem [{\citenamefont {Cranmer}\ and\ \citenamefont
  {Plehn}(2007)}]{Cranmer:2006zs}%
  \BibitemOpen
  \bibfield  {author} {\bibinfo {author} {\bibfnamefont {K.}~\bibnamefont
  {Cranmer}}\ and\ \bibinfo {author} {\bibfnamefont {T.}~\bibnamefont
  {Plehn}},\ }\href {\doibase 10.1140/epjc/s10052-007-0309-4} {\bibfield
  {journal} {\bibinfo  {journal} {Eur. Phys. J.}\ }\textbf {\bibinfo {volume}
  {C51}},\ \bibinfo {pages} {415} (\bibinfo {year} {2007})},\ \Eprint
  {http://arxiv.org/abs/hep-ph/0605268} {arXiv:hep-ph/0605268 [hep-ph]}
  \BibitemShut {NoStop}%
\bibitem [{\citenamefont {Plehn}\ \emph {et~al.}(2014)\citenamefont {Plehn},
  \citenamefont {Schichtel},\ and\ \citenamefont {Wiegand}}]{Plehn:2013paa}%
  \BibitemOpen
  \bibfield  {author} {\bibinfo {author} {\bibfnamefont {T.}~\bibnamefont
  {Plehn}}, \bibinfo {author} {\bibfnamefont {P.}~\bibnamefont {Schichtel}}, \
  and\ \bibinfo {author} {\bibfnamefont {D.}~\bibnamefont {Wiegand}},\ }\href
  {\doibase 10.1103/PhysRevD.89.054002} {\bibfield  {journal} {\bibinfo
  {journal} {Phys. Rev.}\ }\textbf {\bibinfo {volume} {D89}},\ \bibinfo {pages}
  {054002} (\bibinfo {year} {2014})},\ \Eprint {http://arxiv.org/abs/1311.2591}
  {arXiv:1311.2591 [hep-ph]} \BibitemShut {NoStop}%
\bibitem [{\citenamefont {Kling}\ \emph {et~al.}(2017)\citenamefont {Kling},
  \citenamefont {Plehn},\ and\ \citenamefont {Schichtel}}]{Kling:2016lay}%
  \BibitemOpen
  \bibfield  {author} {\bibinfo {author} {\bibfnamefont {F.}~\bibnamefont
  {Kling}}, \bibinfo {author} {\bibfnamefont {T.}~\bibnamefont {Plehn}}, \ and\
  \bibinfo {author} {\bibfnamefont {P.}~\bibnamefont {Schichtel}},\ }\href
  {\doibase 10.1103/PhysRevD.95.035026} {\bibfield  {journal} {\bibinfo
  {journal} {Phys. Rev.}\ }\textbf {\bibinfo {volume} {D95}},\ \bibinfo {pages}
  {035026} (\bibinfo {year} {2017})},\ \Eprint
  {http://arxiv.org/abs/1607.07441} {arXiv:1607.07441 [hep-ph]} \BibitemShut
  {NoStop}%
\bibitem [{\citenamefont {Baldi}\ \emph {et~al.}(2016)\citenamefont {Baldi},
  \citenamefont {Cranmer}, \citenamefont {Faucett}, \citenamefont {Sadowski},\
  and\ \citenamefont {Whiteson}}]{Baldi:2016fzo}%
  \BibitemOpen
  \bibfield  {author} {\bibinfo {author} {\bibfnamefont {P.}~\bibnamefont
  {Baldi}}, \bibinfo {author} {\bibfnamefont {K.}~\bibnamefont {Cranmer}},
  \bibinfo {author} {\bibfnamefont {T.}~\bibnamefont {Faucett}}, \bibinfo
  {author} {\bibfnamefont {P.}~\bibnamefont {Sadowski}}, \ and\ \bibinfo
  {author} {\bibfnamefont {D.}~\bibnamefont {Whiteson}},\ }\href {\doibase
  10.1140/epjc/s10052-016-4099-4} {\bibfield  {journal} {\bibinfo  {journal}
  {Eur. Phys. J.}\ }\textbf {\bibinfo {volume} {C76}},\ \bibinfo {pages} {235}
  (\bibinfo {year} {2016})},\ \Eprint {http://arxiv.org/abs/1601.07913}
  {arXiv:1601.07913 [hep-ex]} \BibitemShut {NoStop}%
\bibitem [{\citenamefont {Cranmer}\ \emph {et~al.}(2016)\citenamefont
  {Cranmer}, \citenamefont {Pavez}, \citenamefont {Louppe},\ and\ \citenamefont
  {Brooks}}]{Cranmer:2016swd}%
  \BibitemOpen
  \bibfield  {author} {\bibinfo {author} {\bibfnamefont {K.}~\bibnamefont
  {Cranmer}}, \bibinfo {author} {\bibfnamefont {J.}~\bibnamefont {Pavez}},
  \bibinfo {author} {\bibfnamefont {G.}~\bibnamefont {Louppe}}, \ and\ \bibinfo
  {author} {\bibfnamefont {W.~K.}\ \bibnamefont {Brooks}},\ }\bibfield
  {booktitle} {\emph {\bibinfo {booktitle} {{Proceedings, 17th International
  Workshop on Advanced Computing and Analysis Techniques in Physics Research
  (ACAT 2016): Valparaiso, Chile, January 18-22, 2016}}},\ }\href {\doibase
  10.1088/1742-6596/762/1/012034} {\bibfield  {journal} {\bibinfo  {journal}
  {J. Phys. Conf. Ser.}\ }\textbf {\bibinfo {volume} {762}},\ \bibinfo {pages}
  {012034} (\bibinfo {year} {2016})}\BibitemShut {NoStop}%
\bibitem [{\citenamefont {Alsing}\ \emph {et~al.}(2018)\citenamefont {Alsing},
  \citenamefont {Wandelt},\ and\ \citenamefont {Feeney}}]{Alsing:2018eau}%
  \BibitemOpen
  \bibfield  {author} {\bibinfo {author} {\bibfnamefont {J.}~\bibnamefont
  {Alsing}}, \bibinfo {author} {\bibfnamefont {B.}~\bibnamefont {Wandelt}}, \
  and\ \bibinfo {author} {\bibfnamefont {S.}~\bibnamefont {Feeney}},\
  }\href@noop {} {\  (\bibinfo {year} {2018})},\ \Eprint
  {http://arxiv.org/abs/1801.01497} {arXiv:1801.01497 [astro-ph.CO]}
  \BibitemShut {NoStop}%
\bibitem [{\citenamefont {Hyv{\"a}rinen}(2005)}]{hyvarinen2005estimation}%
  \BibitemOpen
  \bibfield  {author} {\bibinfo {author} {\bibfnamefont {A.}~\bibnamefont
  {Hyv{\"a}rinen}},\ }\href@noop {} {\bibfield  {journal} {\bibinfo  {journal}
  {Journal of Machine Learning Research}\ }\textbf {\bibinfo {volume} {6}},\
  \bibinfo {pages} {695} (\bibinfo {year} {2005})}\BibitemShut {NoStop}%
\bibitem [{\citenamefont {{Kingma}}\ and\ \citenamefont {{Ba}}(2014)}]{adam}%
  \BibitemOpen
  \bibfield  {author} {\bibinfo {author} {\bibfnamefont {D.~P.}\ \bibnamefont
  {{Kingma}}}\ and\ \bibinfo {author} {\bibfnamefont {J.}~\bibnamefont
  {{Ba}}},\ }\href@noop {} {\bibfield  {journal} {\bibinfo  {journal} {ArXiv
  e-prints}\ } (\bibinfo {year} {2014})},\ \Eprint
  {http://arxiv.org/abs/1412.6980} {arXiv:1412.6980 [cs.LG]} \BibitemShut
  {NoStop}%
\bibitem [{\citenamefont {Chollet}\ \emph {et~al.}(2015)\citenamefont {Chollet}
  \emph {et~al.}}]{chollet2015keras}%
  \BibitemOpen
  \bibfield  {author} {\bibinfo {author} {\bibfnamefont {F.}~\bibnamefont
  {Chollet}} \emph {et~al.},\ }\href@noop {} {\enquote {\bibinfo {title}
  {Keras},}\ }\bibinfo {howpublished}
  {\url{https://github.com/keras-team/keras}} (\bibinfo {year}
  {2015})\BibitemShut {NoStop}%
\bibitem [{\citenamefont {{Abadi}}\ \emph {et~al.}(2016)\citenamefont
  {{Abadi}}, \citenamefont {{Agarwal}}, \citenamefont {{Barham}}, \citenamefont
  {{Brevdo}}, \citenamefont {{Chen}}, \citenamefont {{Citro}}, \citenamefont
  {{Corrado}}, \citenamefont {{Davis}}, \citenamefont {{Dean}}, \citenamefont
  {{Devin}}, \citenamefont {{Ghemawat}}, \citenamefont {{Goodfellow}},
  \citenamefont {{Harp}}, \citenamefont {{Irving}}, \citenamefont {{Isard}},
  \citenamefont {{Jia}}, \citenamefont {{Jozefowicz}}, \citenamefont
  {{Kaiser}}, \citenamefont {{Kudlur}}, \citenamefont {{Levenberg}},
  \citenamefont {{Mane}}, \citenamefont {{Monga}}, \citenamefont {{Moore}},
  \citenamefont {{Murray}}, \citenamefont {{Olah}}, \citenamefont {{Schuster}},
  \citenamefont {{Shlens}}, \citenamefont {{Steiner}}, \citenamefont
  {{Sutskever}}, \citenamefont {{Talwar}}, \citenamefont {{Tucker}},
  \citenamefont {{Vanhoucke}}, \citenamefont {{Vasudevan}}, \citenamefont
  {{Viegas}}, \citenamefont {{Vinyals}}, \citenamefont {{Warden}},
  \citenamefont {{Wattenberg}}, \citenamefont {{Wicke}}, \citenamefont {{Yu}},\
  and\ \citenamefont {{Zheng}}}]{tensorflow}%
  \BibitemOpen
  \bibfield  {author} {\bibinfo {author} {\bibfnamefont {M.}~\bibnamefont
  {{Abadi}}}, \bibinfo {author} {\bibfnamefont {A.}~\bibnamefont {{Agarwal}}},
  \bibinfo {author} {\bibfnamefont {P.}~\bibnamefont {{Barham}}}, \bibinfo
  {author} {\bibfnamefont {E.}~\bibnamefont {{Brevdo}}}, \bibinfo {author}
  {\bibfnamefont {Z.}~\bibnamefont {{Chen}}}, \bibinfo {author} {\bibfnamefont
  {C.}~\bibnamefont {{Citro}}}, \bibinfo {author} {\bibfnamefont {G.~S.}\
  \bibnamefont {{Corrado}}}, \bibinfo {author} {\bibfnamefont {A.}~\bibnamefont
  {{Davis}}}, \bibinfo {author} {\bibfnamefont {J.}~\bibnamefont {{Dean}}},
  \bibinfo {author} {\bibfnamefont {M.}~\bibnamefont {{Devin}}}, \bibinfo
  {author} {\bibfnamefont {S.}~\bibnamefont {{Ghemawat}}}, \bibinfo {author}
  {\bibfnamefont {I.}~\bibnamefont {{Goodfellow}}}, \bibinfo {author}
  {\bibfnamefont {A.}~\bibnamefont {{Harp}}}, \bibinfo {author} {\bibfnamefont
  {G.}~\bibnamefont {{Irving}}}, \bibinfo {author} {\bibfnamefont
  {M.}~\bibnamefont {{Isard}}}, \bibinfo {author} {\bibfnamefont
  {Y.}~\bibnamefont {{Jia}}}, \bibinfo {author} {\bibfnamefont
  {R.}~\bibnamefont {{Jozefowicz}}}, \bibinfo {author} {\bibfnamefont
  {L.}~\bibnamefont {{Kaiser}}}, \bibinfo {author} {\bibfnamefont
  {M.}~\bibnamefont {{Kudlur}}}, \bibinfo {author} {\bibfnamefont
  {J.}~\bibnamefont {{Levenberg}}}, \bibinfo {author} {\bibfnamefont
  {D.}~\bibnamefont {{Mane}}}, \bibinfo {author} {\bibfnamefont
  {R.}~\bibnamefont {{Monga}}}, \bibinfo {author} {\bibfnamefont
  {S.}~\bibnamefont {{Moore}}}, \bibinfo {author} {\bibfnamefont
  {D.}~\bibnamefont {{Murray}}}, \bibinfo {author} {\bibfnamefont
  {C.}~\bibnamefont {{Olah}}}, \bibinfo {author} {\bibfnamefont
  {M.}~\bibnamefont {{Schuster}}}, \bibinfo {author} {\bibfnamefont
  {J.}~\bibnamefont {{Shlens}}}, \bibinfo {author} {\bibfnamefont
  {B.}~\bibnamefont {{Steiner}}}, \bibinfo {author} {\bibfnamefont
  {I.}~\bibnamefont {{Sutskever}}}, \bibinfo {author} {\bibfnamefont
  {K.}~\bibnamefont {{Talwar}}}, \bibinfo {author} {\bibfnamefont
  {P.}~\bibnamefont {{Tucker}}}, \bibinfo {author} {\bibfnamefont
  {V.}~\bibnamefont {{Vanhoucke}}}, \bibinfo {author} {\bibfnamefont
  {V.}~\bibnamefont {{Vasudevan}}}, \bibinfo {author} {\bibfnamefont
  {F.}~\bibnamefont {{Viegas}}}, \bibinfo {author} {\bibfnamefont
  {O.}~\bibnamefont {{Vinyals}}}, \bibinfo {author} {\bibfnamefont
  {P.}~\bibnamefont {{Warden}}}, \bibinfo {author} {\bibfnamefont
  {M.}~\bibnamefont {{Wattenberg}}}, \bibinfo {author} {\bibfnamefont
  {M.}~\bibnamefont {{Wicke}}}, \bibinfo {author} {\bibfnamefont
  {Y.}~\bibnamefont {{Yu}}}, \ and\ \bibinfo {author} {\bibfnamefont
  {X.}~\bibnamefont {{Zheng}}},\ }\href@noop {} {\bibfield  {journal} {\bibinfo
   {journal} {ArXiv e-prints}\ } (\bibinfo {year} {2016})},\ \Eprint
  {http://arxiv.org/abs/1603.04467} {arXiv:1603.04467 [cs.DC]} \BibitemShut
  {NoStop}%
\bibitem [{\citenamefont {Kruskal}(1964)}]{Kruskal1964}%
  \BibitemOpen
  \bibfield  {author} {\bibinfo {author} {\bibfnamefont {J.~B.}\ \bibnamefont
  {Kruskal}},\ }\href {\doibase 10.1007/BF02289694} {\bibfield  {journal}
  {\bibinfo  {journal} {Psychometrika}\ }\textbf {\bibinfo {volume} {29}},\
  \bibinfo {pages} {115} (\bibinfo {year} {1964})}\BibitemShut {NoStop}%
\bibitem [{\citenamefont {Wilks}(1938)}]{Wilks:1938dza}%
  \BibitemOpen
  \bibfield  {author} {\bibinfo {author} {\bibfnamefont {S.~S.}\ \bibnamefont
  {Wilks}},\ }\href {\doibase 10.1214/aoms/1177732360} {\bibfield  {journal}
  {\bibinfo  {journal} {Annals Math. Statist.}\ }\textbf {\bibinfo {volume}
  {9}},\ \bibinfo {pages} {60} (\bibinfo {year} {1938})}\BibitemShut {NoStop}%
\bibitem [{\citenamefont {Cowan}\ \emph {et~al.}(2011)\citenamefont {Cowan},
  \citenamefont {Cranmer}, \citenamefont {Gross},\ and\ \citenamefont
  {Vitells}}]{Cowan:2010js}%
  \BibitemOpen
  \bibfield  {author} {\bibinfo {author} {\bibfnamefont {G.}~\bibnamefont
  {Cowan}}, \bibinfo {author} {\bibfnamefont {K.}~\bibnamefont {Cranmer}},
  \bibinfo {author} {\bibfnamefont {E.}~\bibnamefont {Gross}}, \ and\ \bibinfo
  {author} {\bibfnamefont {O.}~\bibnamefont {Vitells}},\ }\href {\doibase
  10.1140/epjc/s10052-011-1554-0, 10.1140/epjc/s10052-013-2501-z} {\bibfield
  {journal} {\bibinfo  {journal} {Eur. Phys. J.}\ }\textbf {\bibinfo {volume}
  {C71}},\ \bibinfo {pages} {1554} (\bibinfo {year} {2011})},\ \bibinfo {note}
  {[Erratum: Eur.\ Phys.\ J.\ C73, p.\ 2501, 2013]},\ \Eprint
  {http://arxiv.org/abs/1007.1727} {arXiv:1007.1727 [physics.data-an]}
  \BibitemShut {NoStop}%
\bibitem [{\citenamefont {Wald}(1943)}]{Wald}%
  \BibitemOpen
  \bibfield  {author} {\bibinfo {author} {\bibfnamefont {A.}~\bibnamefont
  {Wald}},\ }\href@noop {} {\bibfield  {journal} {\bibinfo  {journal}
  {Transactions of the American Mathematical Society}\ }\textbf {\bibinfo
  {volume} {54}},\ \bibinfo {pages} {426} (\bibinfo {year} {1943})}\BibitemShut
  {NoStop}%
\bibitem [{\citenamefont {Louppe}\ \emph {et~al.}(2016)\citenamefont {Louppe},
  \citenamefont {Kagan},\ and\ \citenamefont {Cranmer}}]{Louppe:2016ylz}%
  \BibitemOpen
  \bibfield  {author} {\bibinfo {author} {\bibfnamefont {G.}~\bibnamefont
  {Louppe}}, \bibinfo {author} {\bibfnamefont {M.}~\bibnamefont {Kagan}}, \
  and\ \bibinfo {author} {\bibfnamefont {K.}~\bibnamefont {Cranmer}},\
  }\href@noop {} {\  (\bibinfo {year} {2016})},\ \Eprint
  {http://arxiv.org/abs/1611.01046} {arXiv:1611.01046 [stat.ME]} \BibitemShut
  {NoStop}%
\bibitem [{\citenamefont {Mertens}(2014)}]{Mertens:2014iya}%
  \BibitemOpen
  \bibfield  {author} {\bibinfo {author} {\bibfnamefont {A.}~\bibnamefont
  {Mertens}},\ }in\ \href {\doibase 10.1088/1742-6596/523/1/012028} {\emph
  {\bibinfo {booktitle} {{Proceedings, 15th International Workshop on Advanced
  Computing and Analysis Techniques in Physics Research (ACAT 2013)}}}},\ Vol.\
  \bibinfo {volume} {523}\ (\bibinfo {year} {2014})\ p.\ \bibinfo {pages}
  {012028}\BibitemShut {NoStop}%
\end{thebibliography}%

\end{document}